\documentclass[prd,superscriptaddress,nofootinbib,preprintnumbers]{revtex4-2}
\usepackage{amsmath,amsfonts,amssymb}
\usepackage{graphicx}
\usepackage{dcolumn}
\usepackage[mathlines]{lineno}
\usepackage{float}
\usepackage{hyperref}
\usepackage[usenames,dvipsnames]{color}
\usepackage{etex}
\usepackage{booktabs}
\usepackage{adjustbox}

\begin{document}

\title{A Novel One-loop Model for Majorana Neutrino Mass and Dark Matter}

\author{Mohamed Belfkir}
\email{m_belfkir@uaeu.ac.ae}
\affiliation{College of General Education, University of Doha for Science and Technology, Qatar}

\author{Mohamed Amin Loualidi}
\email{ma.loualidi@uaeu.ac.ae}
\affiliation{Department of physics, United Arab Emirates University, Al-Ain, UAE}

\author{Salah Nasri}
\email{snasri@uaeu.ac.ae, salah.nasri@cern.ch}
\affiliation{Department of physics, United Arab Emirates University, Al-Ain, UAE}

\begin{abstract}
We present the first complete field-theoretic realization of the finite one-loop \texttt{T4-3-i} topology for Majorana neutrino mass. Here, \texttt{T4-3-i} denotes a one-loop realization of the Weinberg operator in which a fermion links the two external lepton--Higgs pairs. If this fermion is a Majorana singlet or triplet, the same interactions generate a tree-level type-I or type-III seesaw contribution, respectively, so
that the loop is not the leading source of neutrino mass. This lower-order contribution is removed by taking the mediator to be a Dirac fermion $N$, placing lepton-number violation in a separate Majorana fermion $\psi$ inside the loop, and imposing an exact $Z_2$ symmetry that keeps the new scalars inert and stabilizes the lightest odd state.
We classify the allowed electroweak charge assignments and study the minimal singlet-doublet realization, denoted \texttt{T4-3-i-B1}, which contains one Dirac fermion, one Majorana fermion, an inert scalar doublet, and an inert scalar singlet. The resulting rank-two neutrino mass matrix predicts one massless neutrino. We confront both normal and inverted neutrino-mass orderings with neutrino-oscillation and cosmological data, charged-lepton flavor violation, including $\mu-e$ conversion, electroweak precision observables, $h\to\gamma\gamma$, theoretical consistency conditions, the relic abundance, and direct-detection limits. Both fermionic and scalar dark matter are viable. The fermionic candidate has only a loop-induced Higgs coupling and consequently a
strongly suppressed spin-independent scattering rate, whereas the scalar candidate couples through a tree-level Higgs portal and can lie above the neutrino floor while remaining compatible with current limits. In both cases, coannihilation with inert scalars is essential for reproducing the observed relic abundance.
\end{abstract}

%\keywords{Topology T4-3-i, radiative neutrino mass, dark matter, cLFV.}

\maketitle
%\tableofcontents

%%%%%%%%%%%%%%%%%%%%%%%%%%%%%%%%%%%%%%%%%%%%%%%%%%%%%%%%%
\section{Introduction}
\label{sec:intro}
%%%%%%%%%%%%%%%%%%%%%%%%%%%%%%%%%%%%%%%%%%%%%%%%%%%%%%%%%
Radiative generation of neutrino mass at the one-loop level offers a well-motivated framework that simultaneously explains the smallness of neutrino masses and provides viable dark matter (DM) candidates. A broad class of such models can be interpreted as $n$-loop realizations of the dimension-5 Weinberg operator $O_W = LLHH$, where $L$ and $H$ denote the Standard Model (SM) lepton and Higgs doublets. The complete set of one-loop diagrams that realize $O_W$ is identified in Ref.~\cite{Bonnet:2012kz}, while systematic explorations of two- and three-loop realizations are presented in Refs.~\cite{AristizabalSierra:2014wal,Cao:2017xgk} and Ref.~\cite{Cepedello:2018rfh}, respectively. A topological classification shows that one-loop realizations of $O_W$ reduce to six four-point diagrams, labeled $T_1,\ldots,T_6$~\cite{Bonnet:2012kz}. As stressed in Refs.~\cite{Bonnet:2012kz,Cai:2017jrq}, a finite one-loop diagram is not yet a complete radiative model. The loop must also be the leading source of neutrino mass. If a diagram requires a Weinberg-operator counterterm, the neutrino mass is not predicted by the loop alone. After removing such cases, the finite one-loop candidates are \textit{T1-i}, \textit{T1-ii}, \textit{T1-iii}, \textit{T3}, \textit{T4-2-i}, and \textit{T4-3-i}. The first four topologies have many known realizations. The most familiar example is the scotogenic model, which realizes \textit{T3} by adding an inert electroweak doublet $\eta$ and the coupling $\lambda_{5} (H^{\dagger}\eta)^{2}$~\cite{Ma:2006km}; for a recent review of scotogenic schemes and their experimental prospects, see Ref.~\cite{Avila:2025qsc}. The \textit{T1} box topologies lead to different classes of models. For example, if this $\lambda_5$ coupling is forbidden by a discrete symmetry, neutrino masses can arise through \textit{T1-i}~\cite{Budhi:2014gxa,Kashiwase:2015yla,Budhi:2015sha}\textit{T1-i}, the Zee model is a well-known \textit{T1-ii} realization without a stable $Z_2$-odd DM candidate in its minimal form~\cite{Zee:1980ai}, and \textit{T1-iii} was discussed in Ref.~\cite{Ma:1998dn}. The broader question of which one-loop neutrino mass models can also contain a stable neutral $Z_2$-odd particle was studied systematically in Ref.~\cite{Restrepo:2013aga}. This includes box-type scotogenic models, which are distinct from the original \textit{T3} scotogenic model. Recent charged lepton flavor violation (cLFV) studies of such box-type scotogenic constructions show that three-body charged-lepton decays can provide strong phenomenological tests~\cite{Darricau:2026iwg}. \\

The two \textit{T4} cases are different from these scotogenic examples and need extra care. They are one-loop versions of the usual seesaws. In \textit{T4-2-i}, the loop contains the scalar structure of the type-II seesaw. If the corresponding tree-level term is allowed, it dominates over the loop. In \textit{T4-3-i}, the loop contains the fermion structure of the type-I or type-III seesaw. If that fermion is a Majorana singlet or triplet with the usual $LH$ coupling, the tree-level type-I or type-III contribution is generated. Thus, for the two \textit{T4} topologies, the main task is not only to draw a finite loop. One must also choose the fields and symmetries so that the tree-level seesaw is absent, the one-loop diagram remains active and a DM candidate remains stable. For topology \textit{T4-2-i}, the authors in Ref.~\cite{Loualidi:2020jlj} used the dihedral $D_4$ symmetry and extended the particle content of the SM to prevent the tree-level contribution of the type-II seesaw mechanism. For the same topology, the authors in Ref.~\cite{Kashav:2022kpk} employed the modular $A_4$ symmetry, which by construction required the supersymmetrization of the SM. More recently, a nonholomorphic modular-invariant realization based on the double-cover group $T'$ was shown to forbid the dangerous tree-level terms while providing a residual stabilizing symmetry for DM~\cite{Loualidi:2026pld}. For \textit{T4-3-i}, the radiative linear-seesaw model with a $U(1)_{B-L}$ symmetry proposed in Ref.~\cite{Wang:2015saa} reduces to this topology after $U(1)_{B-L}$ breaking. To the best of our knowledge, however, a fully symmetry invariant field-theoretic realization of \textit{T4-3-i} has not yet been constructed. \\

In this paper, we build a complete \textit{T4-3-i} model with these requirements. We restrict the analysis to the type-I scenario of the topology, where the fermion attached to the external lepton-Higgs legs is an $SU(2)_L$ singlet, and classify the electroweak charge assignments of the new particles running in the loop. Among the allowed cases, we focus on the minimal singlet-doublet model we denote by \texttt{T4-3-i-B1}. It contains one Dirac fermion $N$, one Majorana fermion $\psi$, one inert scalar doublet and one inert scalar singlet. This field content is small, has no large electroweak multiplets, and contains neutral odd states that can be DM. The Dirac nature of $N$ removes the ordinary type-I seesaw term, while the Majorana mass of $\psi$ provides lepton number violation inside the loop. The exact $Z_2$ symmetry forbids vacuum expectation values (VEVs) for the new scalars and stabilizes the lightest odd particle. The model predicts a rank-two neutrino mass matrix, and therefore one massless neutrino at leading order. It also allows two different DM pictures: a fermionic one with $\psi$ as DM, and a scalar one with $H_1^0$ as DM. We study both normal and inverted neutrino mass orderings and test the model against neutrino data, cLFV processes including $\mu-e$ conversion, oblique parameters, $h\to\gamma\gamma$, theoretical constraints, relic density and direct detection (DD). \\

The rest of the paper is organized as follows. In Sec.~\ref{SecII}, we present the \texttt{T4-3-i} setup, classify the possible variants and define the scalar sector of the \texttt{T4-3-i-B1} model. In Sec.~\ref{SecIII}, we derive the one-loop neutrino mass matrix and the cLFV observables. In Sec.~\ref{SecIV}, we summarize the theoretical and experimental constraints used in the analysis. In Sec.~\ref{SecV}, we discuss the relic density and direct detection for fermionic and scalar DM. In Sec.~\ref{SecVI}, we present the numerical scan and the results. In Sec.~\ref{SecVII}, we give our conclusions.
%%%%%%%%%%%%%%%%%%%%%%%%%%%%%%%%%%%%%%%%%%%%%%%%%%%%%%%%%%%%%%%%%
\section{Theoretical framework}
\label{SecII}
%%%%%%%%%%%%%%%%%%%%%%%%%%%%%%%%%%%%%%%%%%%%%%%%%%%%%%%%%%%%%%%%%
\subsection{Topology \texttt{T4-3-i}}
\label{SecII-A}
\begin{figure}[H]
    \centering
    \includegraphics[scale=0.4]{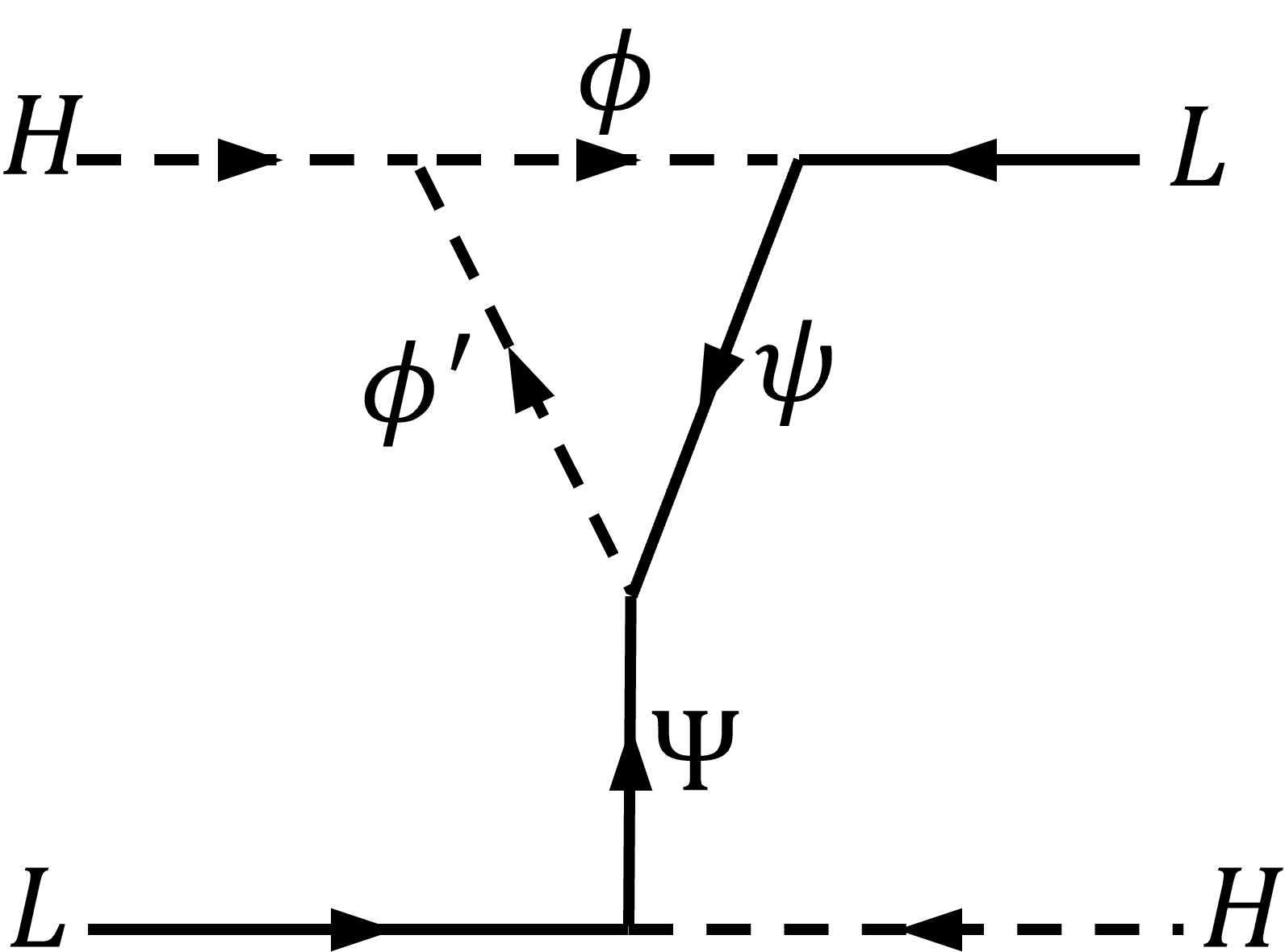}
    \hspace{2.5cm}
    \includegraphics[scale=0.4]{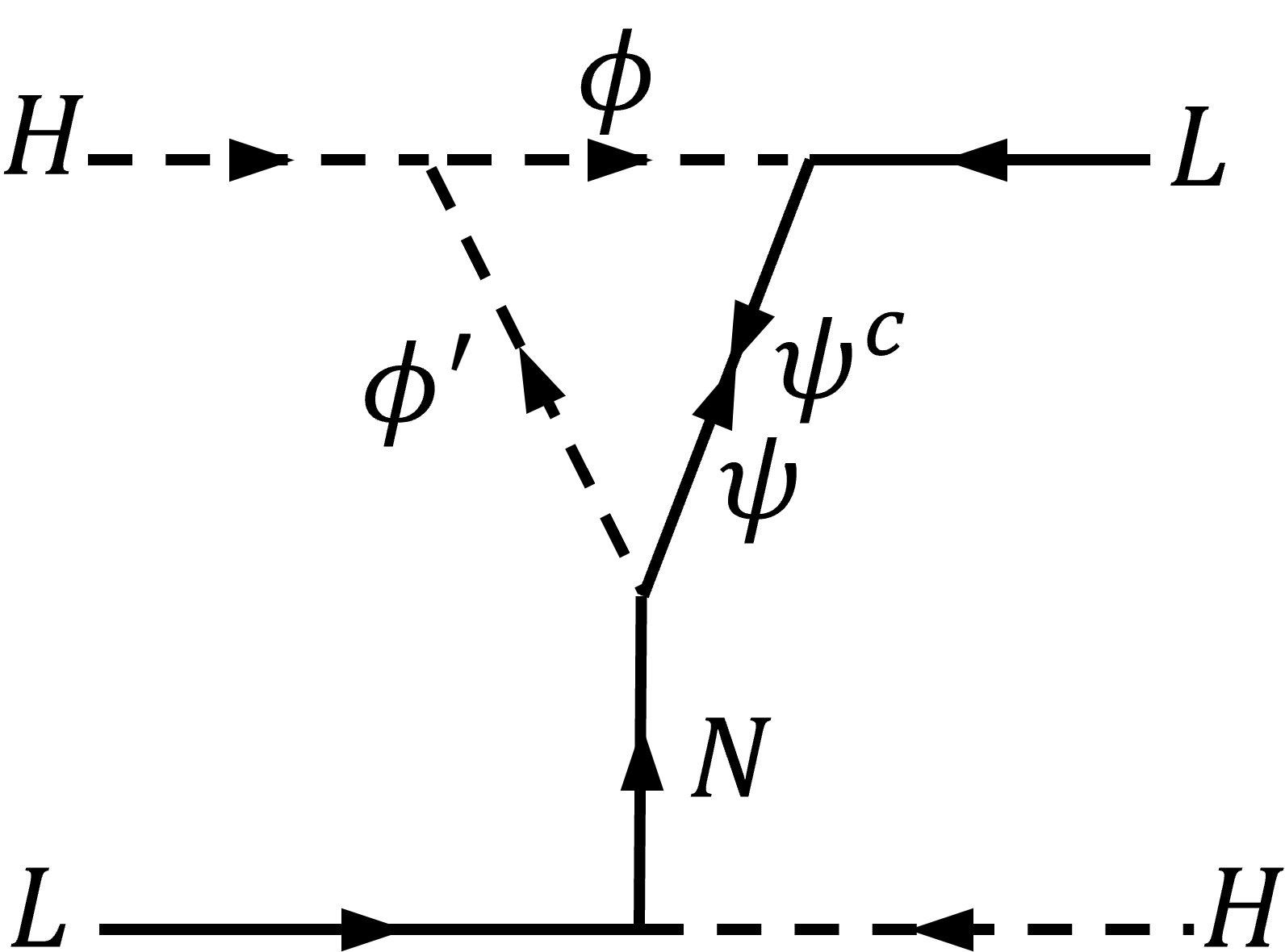}
    \caption{Left: Topology T4-3-i accompanied by the tree level type-I or type-III seesaw contributions when $\Psi$ is a Majorana fermion which transforms under $SU(2)_L$ as $\Psi \sim 1$ or $\Psi \sim 3$, respectively. Right: Genuine T4-3-i topology obtained by promoting $\Psi$ to a Dirac spinor, $\psi$ to Majorana fermion while imposing lepton number conservation for all the couplings.}
    \label{f1}
\end{figure}

%%%%%%%%%%%%%%%%%%%%%%%%%%%%%%%%%%%%%%%%%%%%%%%%%%%%%%%%%%%%%%%%%
Opening up the Weinberg operator $O_{W}$ at the 1-loop level using scalars and fermions provides the basis for the one-loop topology T4-3-i as depicted in the left panel of Fig. \ref{f1}. Depending on the $SU(2)_L$ representation of the fermion $\Psi$, this topology is always accompanied by the tree level type-I seesaw contribution (if $\Psi$ is an $SU(2)_L$ Majorana singlet) or the tree level type-III seesaw contribution (if $\Psi$ is an $SU(2)_L$ Majorana triplet) which are generically dominant. 
Before we delve into the requirements to make this topology the leading contribution to neutrino masses, let us consider the case where $\Psi$ is an $SU(2)_L$ singlet and classify the models according to the $SU(2)_L \times U(1)_Y$ charge assignments of the particles running in the loop. Following the notations of \cite{Bonnet:2012kz}, the transformations of the new fields are denoted as $X_Y^{\mathcal{L}}$ where $X$ corresponds to their $SU(2)_L$ representations, $Y$ to their hypercharges, and $\mathcal{L}$ to their Lorentz structures. Initially, for arbitrary hypercharge values, we identify four possible neutrino mass models. This is based on our understanding that the external legs of $O_W$ correspond to the lepton and Higgs doublets of the SM and that the tensor product of two $SU(2)_L$ doublets follows the decomposition $\mathbf{2 \otimes 2 = 1 \oplus 3}$. These models are denoted in Table \ref{t1} as T4-3-i-A, T4-3-i-B, T4-3-i-C, and T4-3-i-D. \newline
To eliminate the tree-level type-I seesaw contribution, we adopt the approach outlined in Ref. \cite{Bonnet:2012kz} and redefine the properties of new fields as follows: {\it (1)} to avoid a Majorana mass term for the field $\Psi$ and subsequently the ordinary tree-level type-I seesaw contribution, we promote this field to a Dirac spinor which we denote from now on by $N=\Psi_L + \Psi_R$. {\it (2)} we promote the field $\psi$ to a Majorana fermion and assume that all couplings conserve lepton number while the Majorana mass term for $\psi$ is allowed to break the lepton number. This way, a Dirac mass term $\Bar{L}\Tilde{H}\psi$ is absent in this setup. {\it (3)} to maintain the loop in topology T4-3-i closed, we need to impose an exact symmetry that prevents the scalars $\phi$ and $\phi^{\prime}$ to acquire VEVs. The simplest symmetry for this requirement is a $Z_2$ symmetry, wherein $\phi$, $\phi^{\prime}$, and $\psi$ are odd, while the remaining fields are even. After these modifications, we obtain the Feynman diagram in the right panel of Fig \ref{f1}.
\begin{table}[H]
    \centering
    \begin{tabular}{c|c|c|c|c}
         \hline
    Models & \texttt{T4-3-i-A}  & \texttt{T4-3-i-B} & \texttt{T4-3-i-C} & \texttt{T4-3-i-D}  \\
         \hline \hline
    $\phi$ & $1_{1+\alpha}^S$ & $2_{1+\alpha}^S$ & $2_{1+\alpha}^S$ & $3_{1+\alpha}^S$  \\
         \hline
    $\phi^{\prime}$ & $2_{\alpha}^S$ & $1_{\alpha}^S$ & $3_{\alpha}^S$ & $2_{\alpha}^S$  \\
         \hline
    $\psi$ & $2_{\alpha}^F$ & $1_{\alpha}^F$ & $3_{\alpha}^F$ & $2_{\alpha}^F$   \\
         \hline
     
    \end{tabular}
    \caption{Gauge quantum number assignments for the four singlet-fermion realizations of the \texttt{T4-3-i} topology. The parameter $\alpha$ fixes the hypercharge of the loop fields.}
    \label{t1}
\end{table}
In Table \ref{t1}, the $\alpha$ index within $SU(2)_L$ representations determines the hypercharges for the sub-models within the initial four. Here, we provide the sub-models with $\alpha = 0,\pm 1, \pm 2, \pm 3$, while consider that at least one of the new particles must be color and electrically neutral to potentially serve as a candidate for DM. Let us now explore the components of the fields running in the loop for different values of the parameter $\alpha$ that are consistent with DM for each model.
\begin{itemize}
    \item \textbf{T4-3-i-A model:} In this model, we find two values of $\alpha$ that are consistent with DM candidates. For $\alpha=-1$, let us call this model \texttt{T4-3-i-A1}. The components of the fields running in the loop are given explicitly as
    \begin{equation}
        \phi = \phi^0 \sim 1_0^S, \quad \phi^{\prime} = \begin{pmatrix} \phi^{\prime 0} \\ \phi^{\prime -} \end{pmatrix} \sim 2_{-1}^S , \quad \psi = \begin{pmatrix} \psi^{0} \\ \psi^{-} \end{pmatrix} \sim 2_{-1}^F.
    \end{equation}
    For $\alpha=+1$, let us call this model \texttt{T4-3-i-A2}. The components of the fields running in the loop are given by
    \begin{equation}
        \phi = \phi^+ \sim 1_2^S, \quad \phi^{\prime} = \begin{pmatrix} \phi^{\prime +} \\ \phi^{\prime 0} \end{pmatrix} \sim 2_{1}^S , \quad \psi = \begin{pmatrix} \psi^{+} \\ \psi^{0} \end{pmatrix} \sim 2_{1}^F.
    \end{equation}
    We eliminate the cases where $\alpha=0$ and $\alpha = \pm 2$ since they result in fractional electric charges. Furthermore, we omit the case where $\alpha = \pm 3$ as it does not give rise to neutral odd particles under the discrete $Z_2$ symmetry and thus no DM candidate can be accommodated with this choice of $\alpha$.

    \item \textbf{T4-3-i-B model:} In this model, we find also two values of $\alpha$ that are consistent with DM candidates. For $\alpha=0$, let us call this model \texttt{T4-3-i-B1}. The components of the fields running in the loop are given by
    \begin{equation}
        \phi = \begin{pmatrix} \phi^{+} \\ \phi^{0} \end{pmatrix} \sim 2_1^S, \quad \phi^{\prime} = \phi^{\prime0} \sim 1_{0}^S , \quad \psi = \psi^0 \sim 1_{0}^F. 
    \label{B1}
    \end{equation}
    For $\alpha=-2$, let us call this model \texttt{T4-3-i-B2}. The components of the fields running in the loop are given by
    \begin{equation}
        \phi = \begin{pmatrix} \phi^{0} \\ \phi^{-} \end{pmatrix} \sim 2_{-1}^S, \quad \phi^{\prime} = \phi^{\prime -} \sim 1_{-2}^S , \quad \psi = \psi^{-} \sim 1_{-2}^F.
    \label{B2}    
    \end{equation}
    The cases where $\alpha=\pm 1$ and $\alpha = \pm 3$ are discarded since they give rise to fractional electric charges. The remaining case $\alpha = +2$ is also omitted because it does not contain a neutral $Z_2$-odd state.

    \item \textbf{T4-3-i-C model:} In this model, we find three values of $\alpha$ with potential DM candidates. For $\alpha=0$, let us call this model \texttt{T4-3-i-C1}. The fields running in the loop are expressed in this case as
    \begin{equation}
        \phi = \begin{pmatrix} \phi^{+} \\ \phi^{0} \end{pmatrix} \sim 2_1^S, \quad \phi^{\prime} = \begin{pmatrix} \phi^{\prime +} \\ \phi^{\prime 0} \\ \phi^{\prime -} \end{pmatrix} \sim 3_{0}^S , \quad \psi = \begin{pmatrix} \psi^+ \\ \psi^0 \\ \psi^- \end{pmatrix} \sim 3_{0}^F.
    \end{equation}
    For $\alpha=-2$, let us call this model \texttt{T4-3-i-C2}. The components of the fields running in the loop are given by
    \begin{equation}
        \phi = \begin{pmatrix} \phi^{0} \\ \phi^{-} \end{pmatrix} \sim 2_{-1}^S, \quad \phi^{\prime} = \begin{pmatrix} \phi^{\prime 0} \\ \phi^{\prime -} \\ \phi^{\prime --} \end{pmatrix} \sim 3_{-2}^S , \quad \psi = \begin{pmatrix} \psi^0 \\ \psi^- \\ \psi^{--} \end{pmatrix} \sim 3_{-2}^F.
    \end{equation}
    For $\alpha=+2$, let us call this model \texttt{T4-3-i-C3}. The components of the fields running in the loop are given by
    \begin{equation}
        \phi = \begin{pmatrix} \phi^{++} \\ \phi^{+} \end{pmatrix} \sim 2_3^S, \quad \phi^{\prime} = \begin{pmatrix} \phi^{\prime ++} \\ \phi^{\prime +} \\ \phi^{\prime 0} \end{pmatrix} \sim 3_{2}^S , \quad \psi = \begin{pmatrix} \psi^{++} \\ \psi^+ \\ \psi^0 \end{pmatrix} \sim 3_{2}^F.
    \end{equation}
    The cases where $\alpha=\pm 1$ and $\alpha = \pm 3$ are discarded since they give rise to fractional electric charges.

    \item \textbf{T4-3-i-D model:} In this model, we find two values of $\alpha$ that are compatible with DM candidates. For $\alpha=-1$, let us call this model \texttt{T4-3-i-D1}. The field representations are expressed in this case as
    \begin{equation}
        \phi = \begin{pmatrix} \phi^{+} \\ \phi^{0} \\ \phi^- \end{pmatrix} \sim 3_0^S, \quad \phi^{\prime} = \begin{pmatrix} \phi^{\prime 0} \\ \phi^{\prime -} \end{pmatrix} \sim 2_{-1}^S , \quad \psi = \begin{pmatrix} \psi^0 \\ \psi^- \end{pmatrix} \sim 2_{-1}^F.
    \end{equation}
    For $\alpha=+1$, let us call this model \texttt{T4-3-i-D2}. The components of the fields running in the loop are given by
    \begin{equation}
        \phi = \begin{pmatrix} \phi^{++} \\ \phi^{+} \\ \phi^0 \end{pmatrix} \sim 3_2^S, \quad \phi^{\prime} = \begin{pmatrix} \phi^{\prime +} \\ \phi^{\prime 0} \end{pmatrix} \sim 2_{1}^S , \quad \psi = \begin{pmatrix} \psi^+ \\ \psi^0 \end{pmatrix} \sim 2_{1}^F.
    \end{equation}
    For $\alpha=-3$, we call this model \texttt{T4-3-i-D3}. The components of the fields running in the loop are given by
    \begin{equation}
        \phi = \begin{pmatrix} \phi^{0} \\ \phi^{-} \\ \phi^{--} \end{pmatrix} \sim 3_{-2}^S, \quad \phi^{\prime} = \begin{pmatrix} \phi^{\prime -} \\ \phi^{\prime --} \end{pmatrix} \sim 2_{-3}^S , \quad \psi = \begin{pmatrix} \psi^- \\ \psi^{--} \end{pmatrix} \sim 2_{-3}^F.
    \end{equation}
    The cases where $\alpha=0$ and $\alpha = \pm 2$ are omitted since they give rise to fractional electric charges. For simplicity, we restrict our analysis to models whose multiplets do not contain components with electric charges greater than two; for example, the \texttt{T4-3-i-D} model with $\alpha = +3$ is not included in this list.
\end{itemize}

Notice that models \texttt{T4-3-i-A} and \texttt{T4-3-i-B} share the same scalar potential, with the roles of $\phi$ and $\phi^{\prime}$ swapped: in \texttt{T4-3-i-A}, $\phi$ is an $SU(2)_{L}$ singlet and $\phi^{\prime}$ is an $SU(2)_{L}$ doublet, while in \texttt{T4-3-i-B}, $\phi$ is a doublet and $\phi^{\prime}$ is a singlet. The electric charge of each scalar multiplet depends on the value of $\alpha $, which determines their hypercharge. Similarly, in models \texttt{T4-3-i-C} and \texttt{T4-3-i-D}, the scalar potential is the same, but the roles of $\phi$ and $\phi^{\prime }$ are reversed. In this work, we focus on the \texttt{T4-3-i-B1} model, notable for its simplicity, as it involves only singlet and doublet fields while accommodating both scalar and Majorana DM candidates. This makes it one of the most minimal realizations of radiative neutrino mass generation within the \texttt{T4-3-i} class. Notably, when the DM is identified as a neutral Majorana fermion--whose mass induces lepton number violation--only a single generation of this fermion is required to fit the six neutrino oscillation observables, highlighting the minimal and predictive power of the model.
%%%%%%%%%%%%%%%%%%%%%%%%%%%%%%%%%%%%%%%%%%%%%%%%%%%%%%%%%%%%%%%%%
\subsection{Scalar potential}
%%%%%%%%%%%%%%%%%%%%%%%%%%%%%%%%%%%%%%%%%%%%%%%%%%%%%%%%%%%%%%%%%
The most general gauge invariant, $Z_{2}$-invariant, and renormalizable potential for model T4-3-i-B1 is given by 
\begin{eqnarray}
\mathcal{V}_{B} &=& -\mu_H^2 H^{\dagger} H + m_{\phi}^2\phi^{\dagger}\phi + m_{\phi^{\prime}}^2 \phi^{\prime \ast}\phi^{\prime} + \left(\mu_{\phi}\phi^{\dagger} H \phi^{\prime } + h.c.\right) +\lambda_1 \left( H^{\dagger} H \right)^2 + \lambda_2\left( \phi^{\dagger }\phi \right)^{2} +\lambda_{3}\left( \phi^{\prime \ast }\phi^{\prime}\right)
^{2} 
\\
&& + \kappa_1 \left(H^{\dagger} H \right) \left( \phi^{\dagger }\phi \right) + \kappa_{2} \left( H^{\dagger}\phi \right) \left( \phi^{\dagger} H \right) +\frac{\kappa_{3}}{2} \left[ \left( H^{\dagger}\phi \right)^{2} + h.c. \right] + \kappa_{4}\left( H^{\dagger} H \right) \left( \phi^{\prime \ast}\phi^{\prime}\right) + \kappa_5 \left( \phi^{\dagger}\phi \right) \left( \phi^{\prime \ast }\phi^{\prime} \right)  \nonumber
\label{vb}
\end{eqnarray}
The scalar fields in this model are shown in Eq. \ref{B1}. After electroweak symmetry breaking (EWSB), only the neutral the {\it CP}-even component of SM Higgs doublet obtains a vacuum expectation value (VEV), $\upsilon = 246GeV$, while the neutral components of $\phi$ and $\phi^{\prime}$ remain inert. These scalar fields can be parametrized as follows 
\begin{equation}
H=\left( 
\begin{array}{c}
G^{+} \\ 
\frac{1}{\sqrt{2}}(\upsilon + h + iG_{0})
\end{array}
\right) ,\quad \phi =\left( 
\begin{array}{c}
\phi^{+} \\ 
\frac{1}{\sqrt{2}}(h_{1} + i\omega_{1})
\end{array}
\right) ,\quad \phi^{\prime} = \frac{1}{\sqrt{2}}(h_{2} + i\omega_{2})
\end{equation}
To avoid explicit {\it CP} violation in the scalar sector, all parameters in $\mathcal{V}_{B}$ are taken to be real. The resulting physical spectrum contains the SM-like Higgs boson $h$, two {\it CP}-even insert scalars $\left\{H_1^0,H_2^0\right\}$, two {\it CP}-odd scalars $\left\{ A_1^0,A_2^0\right\} $ and one charged scalar pair $\phi^{\pm}$. The trilinear term proportional to $\mu_{\phi}$ mixes the doublet-like and the singlet-like fields separately in the {\it CP}-even and {\it CP}-odd neutral sectors. The physical neutral inert states are therefore obtained from 
\begin{equation}
\left( 
\begin{array}{c}
H_1^0 \\ 
H_2^0
\end{array}
\right) =\left( 
\begin{array}{cc}
c_{\theta_H} & s_{\theta_H} \\ 
-s_{\theta_H} & c_{\theta_H}
\end{array}
\right) \left( 
\begin{array}{c}
h_1 \\ 
h_2
\end{array}
\right) ,\quad \left( 
\begin{array}{c}
A_1^0 \\ 
A_2^0
\end{array}
\right) =\left( 
\begin{array}{cc}
c_{\theta_A} & s_{\theta_A} \\ 
-s_{\theta_A} & c_{\theta_A}
\end{array}
\right) \left( 
\begin{array}{c}
\omega_1 \\ 
\omega_2
\end{array}
\right) 
\end{equation}
where we have used the shorthand $c_{\theta_x}=\cos \theta_x$ and $s_{\theta_x}=\sin \theta_x$ with $\theta_H$ and $\theta_A$
representing the mixing angles that diagonalize the CP-even and CP-odd scalar mass matrices, respectively. After EWSB, the physical masses of the Higgs boson $h$ and the charged scalars $\phi^{\pm}$ are given by
\begin{equation}
m_h^{2} = 2\lambda_1 \upsilon^2,\quad m_{\phi^{\pm}}^2=m_{\phi
}^{2} + \frac{\kappa_1}{2}\upsilon^{2}
\end{equation}
The CP-even and CP-odd scalar squared mass matrices in the bases $\{h_1,h_2\}$ and $\{\omega_1,\omega_2\}$, respectively, are given by
\begin{equation}
M_H^2 = \left( 
\begin{array}{cc}
\left[ m_H^2\right]_{11} & \left[ m_H^2\right]_{12} \\ 
\left[ m_H^2\right]_{12} & \left[ m_H^2\right]_{22}
\end{array}
\right) ,\qquad 
M_A^2 = \left( 
\begin{array}{cc}
\left[ m_A^2\right]_{11} & \left[ m_A^2\right]_{12} \\ 
\left[ m_A^2\right]_{12} & \left[ m_A^2\right]_{22}
\end{array}
\right),
\end{equation}
with
\begin{eqnarray}
\left[ m_H^2\right]_{11} &=& m_{\phi}^2+\left( \kappa_1 + \kappa_2 + \kappa_3 \right)\frac{\upsilon^2}{2}, \qquad 
\left[ m_H^2 \right]_{12} = \frac{\mu_{\phi} \upsilon}{\sqrt{2}}, \qquad 
\left[ m_H^2 \right]_{22} = m_{\phi^{\prime}}^2+\frac{\kappa_4 \upsilon^2}{2}, \notag\\
\left[ m_A^2 \right]_{11} &=& m_{\phi}^2+\left( \kappa_1 + \kappa_2 - \kappa_3 \right)\frac{\upsilon^2}{2}, \qquad 
\left[ m_A^2 \right]_{12} = \frac{\mu_{\phi}\upsilon}{\sqrt{2}}, \qquad 
\left[ m_A^2 \right]_{22} = m_{\phi^{\prime}}^2+\frac{\kappa_4 \upsilon^2}{2}.
\label{mHmA}
\end{eqnarray}
The mass eigenvalues and mixing angles are then given by
\begin{equation}
m_{F_{1,2}}^{2}=\frac{1}{2}\left( \left[ m_{F}^{2}\right] _{11} + \left[ m_{F}^{2}\right]_{22}\mp \sqrt{\left( \left[ m_{F}^{2}\right]_{22} - \left[ m_{F}^{2}\right]_{11}\right) ^{2} + 4\left[ m_{F}^{2}\right]_{12}^{2}} \right),
\qquad 
\tan \left( 2\theta_{F}\right) = \frac{2\left[ m_{F}^{2} \right]_{12}}{\left[ m_{F}^{2}\right]_{22}-\left[ m_{F}^{2}\right]_{11}},
\label{ev}
\end{equation}
where $[F_{1,2},\theta_F]=[H_{1,2}^0,\theta_H]$ for the {\it CP}-even sector and $[F_{1,2},\theta_F]=[A_{1,2}^0,\theta_A]$ for the {\it CP}-odd one. It is important to note that the {\it CP}-even and {\it CP}-odd neutral sectors are not independent. Indeed, from Eq.~\eqref{mHmA}, both matrices share the same off-diagonal entry induced by the trilinear term $\mu_\phi \phi^\dagger H\phi'$ and the same singlet-like diagonal entry. Defining
\begin{equation}
a_A \equiv \left[m_A^2\right]_{11}, \qquad 
b \equiv \left[m_A^2\right]_{12}=\frac{\mu_\phi \upsilon}{\sqrt{2}}, \qquad 
c \equiv \left[m_A^2\right]_{22}=m_{\phi'}^2+\frac{\kappa_4 \upsilon^2}{2},
\end{equation}
the {\it CP}-odd mass matrix can be written as
\begin{equation}
M_A^2=
\begin{pmatrix}
a_A & b\\
b & c
\end{pmatrix}.
\end{equation}
Using the {\it CP}-even sector as input, the common entries $b$ and $c$ may be expressed as
\begin{equation}
b=s_{\theta_H}c_{\theta_H}\left(m_{H_1^0}^2-m_{H_2^0}^2\right),
\qquad
c=s_{\theta_H}^2\,m_{H_1^0}^2+c_{\theta_H}^2\,m_{H_2^0}^2.
\label{bccommon}
\end{equation}

Assuming that $m_{A_1^0}$ is taken as an input parameter, the remaining {\it CP}-odd quantities can be reconstructed from the characteristic equation $ \det\!\left(M_A^2-m_{A_1^0}^2\,\mathbf{1}\right)=0$; namely $(a_A-m_{A_1^0}^2)(c-m_{A_1^0}^2)-b^2=0$.
Solving for $a_A$, one obtains
\begin{equation}
a_A=m_{A_1^0}^2+\frac{b^2}{c-m_{A_1^0}^2}.
\end{equation}
Using next the trace relation $m_{A_1^0}^2+m_{A_2^0}^2=a_A+c$, the second {\it CP}-odd mass eigenvalue is found to be
\begin{equation}
m_{A_2^0}^2
=
c+\frac{b^2}{c-m_{A_1^0}^2}.
\label{eq:MA02derived}
\end{equation}
Likewise, the {\it CP}-odd mixing angle is determined through
\begin{equation}
\tan(2\theta_A)=\frac{2b}{c-a_A}
=
\frac{2b\left(c-m_{A_1^0}^2\right)}
{\left(c-m_{A_1^0}^2\right)^2-b^2}.
\label{eq:thetaAderived}
\end{equation}
Therefore, once $m_h$ is fixed, the scalar sector is fully specified by the ten independent parameters
\begin{equation}
\left\{
m_{\phi^\pm}, m_{H_1^0}, m_{H_2^0}, \theta_H, m_{A_1^0}, \kappa_1, \kappa_4, \kappa_5, \lambda_2, \lambda_3 \right\},
\end{equation}
while $m_{A_2^0}$, $\theta_A$, $\mu_\phi$, $\kappa_2$, $\kappa_3$, $m_\phi^2$, and $m_{\phi'}^2$ are dependent quantities. In the numerical implementation, one may equivalently use $s_{\theta_H}^2$ instead of $\theta_H$ as an external input. The remaining parameters of the scalar potential can be expressed as
\begin{eqnarray}
\mu _{\phi } &=& \frac{\sqrt{2}}{\upsilon }s_{\theta _{H}}c_{\theta _{H}}\left( m_{H_2^0}^2 - m_{H_1^0}^2\right), \qquad
\kappa_2 = \frac{1}{\upsilon ^{2}}\left( c_{\theta_H}^{2} m_{H_1^0}^2 + s_{\theta_H}^2 m_{H_2^0}^2 + c_{\theta_A}^2 m_{A_1^0}^2 + s_{\theta_A}^2 m_{A_2^0}^2 - 2 m_{\phi^{\pm }}^2\right), \nonumber \\
\kappa_3 &=& \frac{1}{\upsilon^2}\left( c_{\theta_H}^2 m_{H_1^0}^2 + s_{\theta_H}^2 m_{H_2^0}^2 - c_{\theta_A}^2 m_{A_1^0}^2 - s_{\theta_A}^2 m_{A_2^0}^2 \right), \qquad
m_{\phi^{\prime }}^{2} = s_{\theta_H}^2 m_{H_1^0}^2 + c_{\theta_H}^2 m_{H_2^0}^2 - \frac{\kappa_4 \upsilon^2}{2}.
\label{dependentpars}
\end{eqnarray}
This parameter choice is particularly convenient for a numerical scan because all physical scalar masses entering DM relic density, DM direct detection, Higgs diphoton and oblique parameter observables are specified directly.
%%%%%%%%%%%%%%%%%%%%%%%%%%%%%%%%%%%%%%%%%%%%%%%%%%%%%%%%%%%%%%%%
\section{Neutrino mass and cLFV}
\label{SecIII}
%%%%%%%%%%%%%%%%%%%%%%%%%%%%%%%%%%%%%%%%%%%%%%%%%%%%%%%%%%%%%%%%%
\subsection{Neutrino masses and mixing}
%%%%%%%%%%%%%%%%%%%%%%%%%%%%%%%%%%%%%%%%%%%%%%%%%%%%%%%%%%%%%%%%%
In the present realization of the \texttt{T4-3-i-B} topology, light neutrino masses are generated radiatively at one loop through the simultaneous presence of the Yukawa interactions $Y_N$, $y$, $y'$, the trilinear scalar coupling $\mu_\phi$, and the Majorana mass insertion of the singlet fermion $\psi$. From the right panel of Fig. \ref{f1}, the invariant couplings under gauge and $Z_2$ symmetries relevant for the neutrino mass are given by the following Lagrangian
\begin{eqnarray}
    \mathcal{L} &=& [ (Y_N)_\beta^{\alpha} \overline{N^\beta} P_L L_\alpha H + (y^{\prime})^{a \alpha} \overline{\psi_a^c} P_L N_\alpha \phi^{\prime \ast} + (y)_a^{\alpha} \overline{\psi^a} P_L L_\alpha \phi + \mu_\phi \phi^\ast \phi^{\prime} H + h.c.] \nonumber \\ 
    &+& M_N \overline{N} N + M_\psi \overline{\psi} \psi + m^2_\phi \phi^\ast \phi + m^2_{\phi^{\prime}} \phi^{\prime \ast} \phi^{\prime}
    \label{newYuk}
\end{eqnarray}
After electroweak symmetry breaking, the neutrino mass matrix can be written as
\begin{equation}
(M_\nu)_{\alpha\beta} = -\mu_\phi \langle H^0\rangle^2
\left[ (Y_N)_\alpha \frac{1}{M_N}\,y'_{11}\,M_\psi\,y_\beta + y_\alpha\,M_\psi\,y'_{11}\,\frac{1}{M_N}(Y_N)_\beta \right] I_3(m_\phi^2,m_{\phi'}^2,M_\psi^2),
\label{eq:Mnu_general}
\end{equation}
where $\langle H^0\rangle=v/\sqrt{2}$ and $I_3$ denotes the loop function expressed as 
\begin{equation}
    I_3(m_\phi^2,m^2_{\phi^{\prime}},M_\psi^2) = -\frac{1}{(4 \pi)^2} \left[ \frac{m_\phi^2 \ln\frac{M_\psi^2}{m_\phi^2}}{(m_\phi^2 - m^2_{\phi^{\prime}}) (m_\phi^2 - M_\psi^2)} + \frac{m^2_{\phi^{\prime}} \ln\frac{M_\psi^2}{m^2_{\phi^{\prime}}}}{(m^2_{\phi^{\prime}} - m_\phi^2) (m^2_{\phi^{\prime}} - M_\psi^2)} \right]
\end{equation}
For the minimal setup considered in this work, with one Dirac fermion $N$ and one Majorana fermion $\psi$, it is convenient to introduce the flavor vectors $Y \equiv (Y_e,Y_\mu,Y_\tau)^T$ and $y\equiv (y_e,y_\mu,y_\tau)^T$, so that Eq.~\eqref{eq:Mnu_general} takes the compact form
\begin{equation}
M_\nu=\Lambda\left(Yy^T+yY^T\right) \qquad \text{with} \qquad \Lambda=
-\mu_\phi \langle H^0\rangle^2
\frac{M_\psi}{M_N}\,
y'_{11}\,
I_3(m_\phi^2,m_{\phi'}^2,M_\psi^2).
\label{eq:Mnu_compact}
\end{equation}
Explicitly, the neutrino mass matrix reads
\begin{equation}
M_\nu= \Lambda
\begin{pmatrix}
2Y_e y_e &
Y_e y_\mu + Y_\mu y_e &
Y_e y_\tau + Y_\tau y_e \\[2mm]
Y_\mu y_e + Y_e y_\mu &
2Y_\mu y_\mu &
Y_\mu y_\tau + Y_\tau y_\mu \\[2mm]
Y_\tau y_e + Y_e y_\tau &
Y_\tau y_\mu + Y_\mu y_\tau &
2Y_\tau y_\tau
\end{pmatrix}.
\label{eq:Mnu_explicit}
\end{equation}
A key feature of this minimal realization is that the mass matrix in Eq.~\eqref{eq:Mnu_compact} is the sum of two outer products, and therefore satisfies
$ {\rm rank}(M_\nu)\leq 2$. For generic flavor vectors $Y$ and $y$, the light-neutrino mass matrix has rank two, and therefore $\det M_\nu = 0$. Consequently, one light neutrino is massless at this order. The mass matrix is diagonalized by the PMNS matrix as
\begin{equation}
U_{\rm PMNS}^T M_\nu U_{\rm PMNS}
=
{\rm diag}(m_1,m_2,m_3).
\end{equation}
Since $M_\nu$ is complex symmetric, the appropriate numerical procedure for diagonalization is an Autonne-Takagi factorization \cite{autonne1915matrices,takagi1924algebraic}. We adopt the standard parameterization
\begin{equation}
U_{\rm PMNS} = V(\theta_{12},\theta_{13},\theta_{23},\delta_{\rm CP})~{\rm diag}\left(1,e^{i\alpha_{21}/2},e^{i\alpha_{31}/2}\right),
\end{equation}
where
\begin{equation}
V = \begin{pmatrix}
c_{12}c_{13} & s_{12}c_{13} & s_{13}e^{-i\delta_{\rm CP}} \\
-s_{12}c_{23}-c_{12}s_{23}s_{13}e^{i\delta_{\rm CP}} &
c_{12}c_{23}-s_{12}s_{23}s_{13}e^{i\delta_{\rm CP}} &
s_{23}c_{13} \\
s_{12}s_{23}-c_{12}c_{23}s_{13}e^{i\delta_{\rm CP}} &
-c_{12}s_{23}-s_{12}c_{23}s_{13}e^{i\delta_{\rm CP}} &
c_{23}c_{13}
\end{pmatrix},
\end{equation}
with $s_{ij}\equiv \sin\theta_{ij}$ and $c_{ij}\equiv \cos\theta_{ij}$. Because one neutrino is massless, only a single Majorana-phase combination is physical. The minimal setup therefore admits two possible spectra: normal ordering (NO) with $m_1=0$, or inverted ordering (IO) with $m_3=0$. In each case, the remaining two masses are determined by the measured mass-squared differences, making the low-energy neutrino spectrum strongly constrained by oscillation data.

In the numerical analysis, the Yukawa couplings $Y_\alpha$, $y_\alpha$, and $y'_{11}$ are treated as complex quantities and scanned in terms of their real and imaginary parts. For each point in parameter space, the one-loop mass matrix in Eq.~\eqref{eq:Mnu_explicit} is diagonalized numerically, yielding the oscillation observables
$\{\sin^2\theta_{12},\sin^2\theta_{13},\sin^2\theta_{23},\delta_{\rm CP},\Delta m_{21}^2,\Delta m_{3\ell}^2 \}$. In this way, the observed lepton mixing pattern directly constrains the flavor structure of the radiative neutrino mass mechanism. Besides oscillation observables, the quantities relevant for non-oscillation probes of neutrino mass are considered. In particular, the effective Majorana mass governing neutrinoless double beta decay is defined as $m_{\beta\beta} = \left| \sum_{i=1}^3 U_{ei}^2 m_i \right|$. Since one neutrino is massless in the minimal setup, this expression simplifies substantially, leading to stronger correlations than in a general three-massive-neutrino scenario. The effective electron neutrino mass relevant for beta decay endpoint measurements is $m_\beta = \left( \sum_{i=1}^3 |U_{ei}|^2 m_i^2 \right)^{1/2}$, while the cosmological mass sum is $\Sigma~m_i = m_1+m_2+m_3$. In the present model, $m_\beta$, $m_{\beta\beta}$, and $\Sigma$ are not independent from the oscillation fit: once the mass ordering and the two measured mass-squared splittings are specified, they are sharply constrained. This makes them particularly useful as complementary probes of the construction.
%%%%%%%%%%%%%%%%%%%%%%%%%%%%%%%%%%%%%%%%%%%%%%%%%%%%%%%%%%%%%%%%%
\subsection{Lepton flavor violation}
%%%%%%%%%%%%%%%%%%%%%%%%%%%%%%%%%%%%%%%%%%%%%%%%%%%%%%%%%%%%%%%%%
The search for cLFV provides a particularly clean window into physics BSM. While neutrino oscillation experiments have established that lepton flavor is not an exact symmetry of nature, the corresponding cLFV rates predicted by the SM are suppressed to an experimentally inaccessible level. As a result, the observation of any cLFV process would represent a clear manifestation of new physics, placing these observables among the most sensitive tests of BSM scenarios \cite{deGouvea:2013zba,Bernstein:2013hba,Calibbi:2017uvl,Ardu:2022sbt,Davidson:2022jai}. In our model, the interaction Lagrangian presented in Eq.~\ref{newYuk} contains a term of the form $y^\alpha~\bar{\psi}_R \phi^+ l_{L \alpha}$ which induces cLFV processes at the one-loop level. These processes are mediated by the charged scalar singlet $\phi^+$ and the Majorana fermion $\psi$. In this work, we focus on the cLFV decay channels $l_\alpha \rightarrow l_\beta \gamma$, $l_\alpha \rightarrow l_\beta l_\beta \bar{l}_\beta$ and the coherent $\mu-e$ conversion in nuclei for which the current experimental upper limits on the branching ratios and conversion rates (CR) are summarized in Table~\ref{tab:clfv}. These three classes of observables are complementary; radiative decays are controlled mainly by the dipole form factor, three-body decays also test nondipole, box and interference terms, and conversion rates are enhanced by coherent nuclear effects.
\begin{table}[H]
\centering
\renewcommand{\arraystretch}{1.15}
\setlength{\tabcolsep}{4pt}
\begin{tabular}{lll}
\hline
\textbf{Observable} & \textbf{Current limit} & \textbf{Future sensitivity} \\
\hline
$\mathrm{Br}(\mu\to e\gamma)$
& $<1.5\times10^{-13}$ \cite{MEGII:2025muegamma}
& $6.0\times10^{-14}$ \cite{Baldini:2018uhj} \\
$\mathrm{Br}(\tau\to e\gamma)$
& $<3.3\times10^{-8}$ \cite{BaBar:2010taulgamma}
& $3.0\times10^{-9}$ \cite{Belle-II:2018jsg,Aoki:2025clfv} \\
$\mathrm{Br}(\tau\to \mu\gamma)$
& $<4.2\times10^{-8}$ \cite{Belle:2021taugamma}
& $1.0\times10^{-9}$ \cite{Belle-II:2018jsg,Aoki:2025clfv} \\
$\mathrm{Br}(\mu\to 3e)$
& $<1.0\times10^{-12}$ \cite{SINDRUM:1988mu3e}
& $2\times10^{-15}$, Phase I; $10^{-16}$, Phase II
  \cite{Blondel:2021fji,Aoki:2025clfv} \\
$\mathrm{Br}(\tau\to 3e)$
& $<2.7\times10^{-8}$ \cite{Belle:2010tau3l}
& $5.0\times10^{-10}$ \cite{Belle-II:2018jsg} \\
$\mathrm{Br}(\tau\to 3\mu)$
& $<2.1\times10^{-8}$ \cite{Belle:2010tau3l}
& $4.0\times10^{-10}$ \cite{Belle-II:2018jsg} \\
${\rm CR}(\mu\!-\!e,\mathrm{Ti})$
& $<4.3\times10^{-12}$ \cite{ParticleDataGroup:2024cfk}
& $10^{-18}$, PRISM/PRIME benchmark \cite{Aoki:2025clfv} \\
${\rm CR}(\mu\!-\!e,\mathrm{Au})$
& $<7.0\times10^{-13}$ \cite{ParticleDataGroup:2024cfk}
& -- \\
${\rm CR}(\mu\!-\!e,\mathrm{Pb})$
& $<4.6\times10^{-11}$ \cite{ParticleDataGroup:2024cfk}
& -- \\
\hline
\end{tabular}
\caption{Current limits and projected sensitivities for the cLFV observables. All present bounds are quoted at $90\%$ C.L.; future sensitivities correspond to the design or benchmark reaches reported in the cited references.}
\label{tab:clfv}
\end{table}
Let us first discuss radiative lepton decays. In our analysis, we considered a single generation of the Majorana fermion $\psi$. Consequently, the radiative decay $l_\alpha \rightarrow l_\beta \gamma$ receives one-loop contributions via diagrams mediated by $\phi^\pm$ and $\psi$. In these diagrams, the photon is attached to the charged scalar or to the external charged-lepton legs; no photon is emitted from the neutral fermion $\psi$. Evaluating the one-loop integrals yields the dipole form factor $A_D$, given by
\begin{equation}
    A_D = \frac{y_{l_\beta\psi}^\ast y_{l_\alpha\psi}}{2(4\pi)^2} \frac{1}{m_{\phi^\pm}^2} F_2(x) \quad \text{with~}x = \frac{M_\psi^2}{m_{\phi^\pm}^2} \quad \text{and~} \quad F_2(x) = \frac{1 - 6x + 3x^2 + 2x^3 - 6x^2\log x}{6(1-x)^4}.
    \label{AD}
\end{equation}
The resulting branching ratio for the cLFV decay $l_\alpha \rightarrow l_\beta \gamma$ is then expressed as
\begin{equation}
    BR(l_\alpha \rightarrow l_\beta \gamma) = \frac{3(4\pi)^3 \alpha_{em}}{4G_F^2} |A_D|^2~BR(l_\alpha \rightarrow l_\beta \nu_\alpha \bar{\nu}_\beta)
\end{equation}
where $G_F$ is the Fermi constant, $\alpha_{em} = e^2 / 4\pi$ is the electromagnetic fine structure constant, and $BR(l_\alpha \rightarrow l_\beta \nu_\alpha \bar{\nu}_\beta)$ is the SM branching ratio.

Next we consider the 3-body decay processes $l_\alpha \rightarrow l_\beta l_\beta \bar{l}_\beta$ which receive four contributions from four types of 1-loop diagrams: $\gamma$-penguins, $Z$-penguins, Higgs-penguins and box diagrams. As in radiative decays, the $\phi^\pm-\psi$ loop generates the off-shell photon and $Z$ penguins, while direct and crossed box diagrams contain two internal $\phi^\pm$ lines and two internal $\psi$ lines. Higgs-penguin effects are charged-lepton-Yukawa suppressed and are not kept in the leading expression. The branching ratio for these decay processes is given by \cite{Toma:2013zsa}
\begin{eqnarray}
    BR(l_\alpha \rightarrow l_\beta \bar{l}_\beta l_\beta) &=& \frac{3(4\pi)^2 \alpha_{em}^2}{8 G_F^2} \biggl[ \overbrace{ |A_{ND}|^2 + |A_D|^2 \left( \frac{16}{3} \log\left(\frac{m_\alpha}{m_\beta}   \right) - \frac{22}{3}  \right) }^{\gamma~\text{penguin}} + \overbrace{ \frac{1}{3} \left( 2|Z_{RR}|^2 + |Z_{RL}|^2 \right) }^{Z~\text{penguin}}  \nonumber \\
     &+& \underbrace{ \frac{1}{6} |B|^2 }_{\text{Box}} + \underbrace{\left( -2 A_{ND} A_D^{\ast} + \frac{1}{3} A_{ND} B^\ast - \frac{2}{3} A_D B^\ast + h.c.  \right) }_{\text{Interference}} \biggr] \times BR(l_\alpha \rightarrow l_\beta \nu_\alpha \bar{\nu}_\beta )
\end{eqnarray}
The $\gamma$-penguin contribution involves the dipole form factor $A_D$, defined in Eq.~\ref{AD}, and the coefficient $A_{ND}$, which accounts for the photonic nondipole contributions and is given by
\begin{equation}
    A_{ND} = \frac{y_{l_\beta\psi}^\ast y_{l_\alpha\psi}}{6(4\pi)^2} \frac{1}{m_{\phi^\pm}^2} G_2(x) \quad \text{with~}\quad G_2(x) = \frac{2 - 9x + 18 x^2 - 11 x^3 + 6 x^3 \log x}{6(1-x)^4}.
    \label{AND}
\end{equation}
where $G_2(x)$ is the one-loop function for the non-dipole $\gamma$-penguin, satisfying $G_2(0) = 1/3$ and $G_2(1) = 1/4$. The $Z$ penguin contribution reads
\begin{equation}
    Z_{RR} = \frac{Z_{ND}g_R^l}{g_2^2 \sin^2\theta_W m_Z^2}, \quad Z_{RL} = \frac{Z_{ND}g_L^l}{g_2^2 \sin^2\theta_W m_Z^2}\quad \text{where~} \quad Z_{ND} = \frac{y_{l_\beta\psi}^\ast y_{l_\alpha\psi}}{2(4\pi)^2} \frac{ g_2 m_\alpha m_\beta}{ \cos\theta_W~m_{\phi^{\pm}}^2} F_2(x)
\end{equation}
is the momentum-independent Z-boson form factor, $g_2$ is the $SU(2)_L$ gauge coupling constant, while $g_R^l$ and $g_L^l$ are the right- and left-handed components of the tree-level $Z$-boson couplings to charged leptons defined as: $g_R^l = -g_2 \sin^2\theta_W/\cos\theta_W$ and $g_L^l = (g_2 / \cos\theta_W) (\frac{1}{2} - \sin^2\theta_W)$. Finally, the box contribution is
given by the coefficient $B$ as follows
\begin{equation}
    B = \frac{y_{l_\beta \psi}^{\ast 2} y_{l_\beta \psi} y_{l_\alpha \psi}}{2(4\pi)^3 \alpha_{em}m_{\phi^{\pm}}^2} \left[ D_1(x) + 2 x D_2(x) \right]
\end{equation}
where the 1-loop box functions $D_{1,2}(x)$ are given by $D_1(x) = (-1 + x^2 - 2x \log x)/(1-x)^3$ and $D_2(x) = (-2 + 2x - (1+x) \log x) / (1-x)^3$.

Let us now consider coherent $\mu-e$ conversion in nuclei. When a negative muon is stopped in matter, it is captured into an atomic orbit and cascades to the $1s$ state of a muonic atom. From this bound state the muon can undergo ordinary nuclear capture, $\mu^-+{\cal N}(A,Z)\to\nu_\mu+{\cal N}(A,Z-1)$, or it can convert into an electron through a cLFV interaction while leaving the nucleus in its ground state, $\mu^- + {\cal N} \rightarrow e^- + {\cal N}$. The coherent channel is particularly important because the amplitudes induced by vector quark currents add over the nucleons in the target. The experimental observable is conventionally normalized to the ordinary muon capture rate,
\begin{equation}
    CR(\mu {\cal N} \rightarrow e {\cal N}) \equiv \frac{\Gamma(\mu^- + {\cal N}(A,Z) \rightarrow e^- + {\cal N}(A,Z))}{\Gamma(\mu^- + {\cal N}(A,Z) \rightarrow \text{muon captures})} \equiv \frac{\Gamma(\mu-e~\text{conversion})}{\Gamma_{\text{capt}}^{\cal N}},
\end{equation}
where ${\cal N}(A,Z)$ denotes a nucleus with mass number $A$, proton number $Z$, and neutron number $N=A-Z$. This normalization is the one used in the experimental literature and in the standard theoretical treatment of coherent conversion \cite{Kuno:1999jp,Kitano:2002mt,deGouvea:2013zba,Bernstein:2013hba,Calibbi:2017uvl,Ardu:2022sbt,Davidson:2022jai}. This process is complementary to $\mu\to e\gamma$ and $\mu\to 3e$ since it probes the same flavor-changing $\mu-e$ transition, but at nonzero momentum transfer and in the presence of a nuclear source. Consequently, off-shell photon penguins, $Z$ penguins and, in more general theories, scalar or four-fermion contact operators can all be tested. In the present model there are no tree-level lepton-quark interactions, the $Z_2$ symmetry forbids box diagrams with quarks, and Higgs-mediated scalar quark operators are negligible. The conversion amplitude is therefore controlled by the photonic dipole and non-dipole form factors, $A_D$ and $A_{ND}$, and by the momentum-independent $Z$-penguin form factor $Z_{ND}$ introduced above. Diagrammatically, coherent conversion is described here by the same $\phi^\pm-\psi$ loop attached to an off-shell photon or
$Z$ boson, which then couples to the nuclear quark current. Since the $Z_2$-odd fields do not couple to quarks, quark-box diagrams are absent, and Higgs-penguin scalar terms are Yukawa suppressed.
\newline
At the quark level, the relevant vector coefficients are obtained by matching the off-shell photon and $Z$ penguin amplitudes onto the low-energy operators $\left(\bar e\gamma_\mu P_X\mu\right)\left(\bar q\gamma^\mu q\right)$, with $X=L,R$. With our form-factor normalization, the non-negligible pieces are \cite{Kitano:2002mt,Toma:2013zsa}
\begin{align}
    g_{LV,\gamma}^{(q)} = \frac{\sqrt{2}}{G_F}\,e^2 Q_q\left(A_{ND}-A_D\right), \qquad
    g_{RV,Z}^{(q)} = -\frac{\sqrt{2}}{G_F}\,\frac{g_L^q+g_R^q}{2}\,\frac{Z_{ND}}{m_Z^2},
    \label{mue-quark}
\end{align}
with
\begin{equation}
    g_L^q = \frac{g_2}{\cos\theta_W}\left(Q_q\sin^2\theta_W-T_3^q\right),
    \qquad
    g_R^q = \frac{g_2}{\cos\theta_W}Q_q\sin^2\theta_W .
    \label{eq:mue-z-quark-couplings}
\end{equation}
The superscript $(q)$ labels the quark current on which the effective operator acts. In a completely general effective Lagrangian the vector coefficient $g_{XV}^{(q)}$ may receive both photon- and Z-penguin pieces. In the present charged-scalar realization, however, the leading
contribution in the external charged-lepton masses has the chiral structure shown in Eq. \ref{mue-quark}. The coefficients $g_{LV,\gamma}^{(q)}$ and $g_{RV,Z}^{(q)}$ require the opposite charged-lepton chirality and vanish in this leading chiral limit, and thus $g_{LV}^{(q)}=g_{LV,\gamma}^{(q)}$ and $
g_{RV}^{(q)}=g_{RV,Z}^{(q)}$.
For coherent conversion it is sufficient to match the quark-level vector coefficients onto proton and neutron vector currents. Following the standard convention used in Refs.~\cite{Kitano:2002mt,Kuno:1999jp,Toma:2013zsa,Calibbi:2017uvl}, this matching is written in terms of the nucleon vector charges $G_V^{(q,B)}$, where $q=u,d,s$ labels the quark current and $B=p,n$ denotes the proton or neutron. The required charges are
\begin{equation}
    G_V^{(u,p)}=2,\quad G_V^{(u,n)}=1,\qquad
    G_V^{(d,p)}=1,\quad G_V^{(d,n)}=2,\qquad
    G_V^{(s,p)}=G_V^{(s,n)}=0 .
    \label{eq:mue-nucleon-vector-charges}
\end{equation}
The corresponding proton and neutron coefficients are then
\begin{equation}
    g_{XV}^{(p)}=\sum_{q=u,d,s}g_{XV}^{(q)}G_V^{(q,p)},\qquad
    g_{XV}^{(n)}=\sum_{q=u,d,s}g_{XV}^{(q)}G_V^{(q,n)} .
    \label{eq:mue-proton-neutron-couplings}
\end{equation}
Here $X=L,R$ specifies the chirality of the charged-lepton current in the effective operator. It is convenient to rewrite the same information in terms of isoscalar and isovector combinations,
\begin{equation}
    g_{XV}^{(0)}=\frac{1}{2}\sum_{q=u,d,s}g_{XV}^{(q)}
    \left(G_V^{(q,p)}+G_V^{(q,n)}\right),\qquad
    g_{XV}^{(1)}=\frac{1}{2}\sum_{q=u,d,s}g_{XV}^{(q)}
    \left(G_V^{(q,p)}-G_V^{(q,n)}\right),
    \label{eq:mue-isospin-couplings}
\end{equation}
which are equivalently $g_{XV}^{(0)}=(g_{XV}^{(p)}+g_{XV}^{(n)})/2$ and $g_{XV}^{(1)}=(g_{XV}^{(p)}-g_{XV}^{(n)})/2$. Therefore the coherent sum over $Z$ protons and $N$ neutrons, $Zg_{XV}^{(p)}+Ng_{XV}^{(n)}$, can be written as $(Z+N)g_{XV}^{(0)}+(Z-N)g_{XV}^{(1)}$. The amplitudes entering the rate are then
\begin{align}
    {\cal A}_L({\cal N}) = (Z+N) g_{LV}^{(0)}+(Z-N) g_{LV}^{(1)}, \qquad
    {\cal A}_R({\cal N}) = (Z+N) g_{RV}^{(0)}+(Z-N) g_{RV}^{(1)} .
    \label{eq:mue-coherent-amplitudes}
\end{align}
With the normalization convention of Refs.~\cite{Kitano:2002mt,Toma:2013zsa}, and neglecting scalar operators consistently with the discussion above, the conversion rate is \cite{Kitano:2002mt,Abada:2014kba,Lindner:2016bgg,Toma:2013zsa}
\begin{equation}
    CR(\mu-e,{\cal N}) = \frac{G_F^2\alpha_{\rm em}^3m_\mu^5}{8\pi^2 Z ~\Gamma_{\rm capt}^{\cal N}} Z_{\rm eff}^4 ~ \left|F_p\right|^2
    \left[ \left|{\cal A}_L({\cal N})\right|^2 +  \left|{\cal A}_R({\cal N})\right|^2 \right].
    \label{eq:mue-conversion-rate}
\end{equation}
This expression is the $p_e\simeq E_e\simeq m_\mu$ limit of the Kitano--Koike--Okada convention, in which the prefactor contains $p_eE_e m_\mu^3/(8\pi^2 Z\Gamma_{\rm capt}^{\cal N})$ before this approximation is made \cite{Kitano:2002mt,Toma:2013zsa}. The quantities $Z_{\rm eff}$ and $F_p$ encode, respectively, the effective atomic charge felt by the bound muon and the nuclear proton form factor. The corresponding values of $Z_{\rm eff}$, $F_p$ and $\Gamma_{\rm capt}^{\cal N}$ for the nuclei commonly used in $\mu-e$ conversion searches are tabulated in the coherent-conversion literature \cite{Kitano:2002mt,Toma:2013zsa}.
%%%%%%%%%%%%%%%%%%%%%%%%%%%%%%%%%%%%%%%%%%%%%%%%%%%%%%%%%%%%%%%%
\section{Theoretical and experimental constraints}
\label{SecIV}
%%%%%%%%%%%%%%%%%%%%%%%%%%%%%%%%%%%%%%%%%%%%%%%%%%%%%%%%%%%%%%%%%
In this section, we outline the main theoretical and experimental constraints applied to the model. Theoretical consistency is ensured by requiring perturbativity, perturbative unitarity, and vacuum stability of the scalar potential \cite{Akeroyd:2000wc}, while experimental bounds further restrict the viable parameter space.
%%%%%%%%%%%%%%%%%%%%%%%%%%%%%%%%%%%%%%%%%%%%%%%%%%%%%%%%%%%%%%%%%
\subsection{Theoretical constraints}
\label{sub.A}
%%%%%%%%%%%%%%%%%%%%%%%%%%%%%%%%%%%%%%%%%%%%%%%%%%%%%%%%%%%%%%%%%
\textbf{Boundedness-from-below:} For the potential to have a stable minimum, it must not go to negative infinity in any direction of the field space--that is, it must be bounded from below. This large-field behavior is mainly controlled by the quartic terms, so cubic terms can be ignored here. Using co-positivity conditions \cite{Kannike:2012pe}, the following criteria ensure the potential remains bounded from below in all directions
\begin{align}
    \lambda_1 \geq 0, \quad \lambda_2 \geq 0, \quad \lambda_3 \geq 0, \quad \kappa_1 + \kappa_2 - |\kappa_3| \geq -2\sqrt{\lambda_1\lambda_2}, \quad \kappa_4 \geq -2\sqrt{\lambda_1\lambda_3}, \quad \kappa_5 \geq -2\sqrt{\lambda_2\lambda_3}.
\end{align}
If any of the mixed conditions is close to being saturated or negative, the following bound is needed
\begin{equation}
    \sqrt{\lambda_1 \lambda_2 \lambda_3} + \frac{1}{2} [(\kappa_1+\kappa_2-|\kappa_3|)\sqrt{\lambda_3} + \kappa_4 \sqrt{\lambda_2} + \kappa_5 \sqrt{\lambda_1}] \geq 0.
\end{equation}

\textbf{Perturbativity:} To maintain the validity of perturbative calculations up to the TeV scale, all scalar quartic and Yukawa couplings are required to lie within the perturbative regime, specifically obeying the condition
\begin{equation}
    |\lambda_i|,|\kappa_j|,|Y_N|^2,|y'|^2,|y|^2 < 4\pi, \quad \text{where i=1,2,3 }\text{and }j=1,...,5.
\end{equation}
In the numerical study presented later, these theoretical requirements are imposed before the DM observables are evaluated. This step ensures that theoretically inconsistent parameter points, such as those with non-perturbative reconstructed quartic couplings or an unstable scalar potential, are excluded from the analysis. As a result, such points will not affect the phenomenological interpretation.

\textbf{Perturbative unitarity:} To ensure the consistency of the model at high energies, perturbative unitarity must be preserved in all scattering processes involving scalar and gauge bosons \cite{Lee:1977yc}. According to the equivalence theorem, at very high energies, gauge bosons can be replaced by their corresponding Goldstone bosons, simplifying the analysis to purely scalar interactions. As a result, the scattering amplitudes for processes like $S_i S_j \rightarrow S_k S_l$ can be computed more straightforwardly, with dominant contributions arising from the quartic scalar couplings \cite{Arhrib:2000is}. To satisfy perturbative unitarity, the eigenvalues of the scattering amplitude matrix are required to remain below the bound $|e_i|< 8 \pi$ \cite{Arhrib:2012ia,Arhrib:2013ela,Belyaev:2016lok}. In our model, the structure of the scattering amplitude matrix is tightly constrained by three exact symmetries: electric charge conservation, {\it CP} invariance, and a global $Z_2$ symmetry. These symmetries play a critical role in simplifying the matrix by enforcing selection rules that allow it to be block-diagonalized. As a result, the full scattering matrix decomposes into a set of independent sub-matrices, each corresponding to distinct quantum number combinations of the initial and final particle states.\\
The classification of these sub-matrices is based on the quantum characteristics of the initial state $S_i S_j$. Specifically, we distinguish the sectors according to three criteria: whether the state is {\it CP}-even or {\it CP}-odd, electrically neutral or charged, and $Z_2$-even or $Z_2$-odd. This systematic organization leads to six distinct sub-matrices, each with its own set of basis states. The explicit bases for these sectors are listed in Appendix~\ref{app.A}.
%%%%%%%%%%%%%%%%%%%%%%%%%%%%%%%%%%%%%%%%%%%%%%%%%%%%%%%%%%%%%%%%%
\subsection{$h\to \gamma\gamma$ in in the minimal \texttt{T4-3-i} model}
%%%%%%%%%%%%%%%%%%%%%%%%%%%%%%%%%%%%%%%%%%%%%%%%%%%%%%%%%%%%%%%%%
The Higgs couplings to two photons ($H\gamma\gamma$) is generated at the one-loop level, primarily through contributions from $W$ bosons and charged fermions in the SM. In the model under consideration, additional contributions arise from the new charged scalar $\phi^\pm$, which modify the amplitudes for $H\gamma\gamma$ interaction. This arise from the $\kappa_1$ term in the scalar potential which leads eventually to $g_{h \phi^- \phi^+} \sim \kappa_1 \upsilon$. Therefore, the partial decay widths for this process, incorporating the contributions from $\phi^\pm$, is given by
\begin{equation}
\Gamma(h \rightarrow \gamma\gamma)^{\texttt{T4-3-i}} = \frac{G_F \alpha^2 m_h^3}{128 \sqrt{2} \pi^3} 
\left| \sum_f N_c Q_f^2 A_{1/2}^{\gamma\gamma}(\tau_f) + A_1^{\gamma\gamma}(\tau_W) 
+ \frac{\kappa_1 v^2}{2 m_{\phi^\pm}^2} A_0^{\gamma\gamma}(\tau_{\phi^\pm}) \right|^2.
\end{equation}
Here, $G_F$ and $\alpha$ denote the Fermi constant and the fine-structure constant, respectively. The color factor $N_c$ equals 3 for quarks and 1 for leptons. $Q_f$ is the electric charge of the fermion, while $c_W \equiv \cos\theta_W = \sqrt{1 - \sin^2\theta_W}$, with $\theta_W$ being the electroweak mixing angle. The loop functions $A_i^{\gamma\gamma}$ encode the spin-dependent form factors of the particles running in the loop and are explicitly given in Refs. \cite{Gunion:1989we, Djouadi:2005gi, Chen:2013vi}. For phenomenological studies, it is convenient to define the ratio of the partial decay width to its SM value. Denoted as $R_{\gamma\gamma}$, this ratio quantify the relative enhancement or suppression due to the presence of the new charged scalar. Including only the dominant contribution from the top quark in the SM loop, the ratio is given by 
\begin{equation}
    R_{\gamma\gamma} = \frac{\Gamma(h \rightarrow\gamma\gamma)^{\texttt{T4-3-i}}}{\Gamma(h \rightarrow\gamma\gamma)^{\text{SM}}} = \left| 1 + \frac{\kappa_1 \upsilon^2}{2 m_{\phi^\pm}^2} \frac{A_0^{\gamma\gamma}(\tau_{\phi^\pm})}{ \frac{4}{3} A_{1/2}^{\gamma\gamma}(\tau_f) + A_1^{\gamma\gamma}(\tau_W)} \right|^2.
\end{equation}
Recent measurements from the ATLAS collaboration provide the experimental values of this ratio as follows \cite{ATLAS:2022vkf}: $R_{\gamma\gamma} = 1.088_{-0.089}^{+0.095}$. In the numerical analysis, the predicted diphoton rate is required to be compatible with this measurement. Since the charged scalar only modifies the loop-induced partial width and no sizable new Higgs decay mode is opened, we use the SM-like total-width approximation, so that ${\rm BR}(h\to\gamma\gamma)^{\texttt{T4-3-i}}\simeq R_{\gamma\gamma}~{\rm BR}(h\to\gamma\gamma)^{\rm SM}$. This constraint is directly linked to the dark sector because the same coupling $\kappa_1$ that controls the $h\phi^+\phi^-$ loop amplitude, contributes to the charged scalar mass, and enters the Higgs portal interactions relevant for scalar DM.
%%%%%%%%%%%%%%%%%%%%%%%%%%%%%%%%%%%%%%%%%%%%%%%%%%%%%%%%%%%%%%%%
\subsection{Oblique parameters}
%%%%%%%%%%%%%%%%%%%%%%%%%%%%%%%%%%%%%%%%%%%%%%%%%%%%%%%%%%%%%%%%
In the present model, the inert scalars introduce quantum corrections to the self-energies of the $W^\pm$ and $Z$ bosons, which can be systematically captured using the oblique parameters $S$, $T$, and $U$ \cite{Peskin:1990zt,Peskin:1991sw}. These parameters serve as a powerful, model-independent framework to probe indirect effects of new physics through EW precision observables, particularly via the universal two-point functions of the EW gauge bosons. The parameters $S$ and $T$ are especially sensitive to mass splittings among the scalar states, and under the assumption $\Delta U = 0$, current global fits to EW data impose the following constraints: $\Delta S = -0.05 \pm 0.07$ and $\Delta T = 0.0 \pm 0.06$ \cite{ParticleDataGroup:2024cfk}. The analytical expression of $\Delta T$ and $\Delta S$ in our model are given by
\begin{eqnarray}
    \Delta T &=& \frac{1}{16\pi s_W^2 m_W^2} \{ c_{\theta_H}^2 F(m_{H_1^0}^2,~m_{\phi^\pm}^2) + s_{\theta_H}^2 F(m_{H_2^0}^2,~m_{\phi^\pm}^2) + c_{\theta_A}^2 F(m_{A_1^0}^2,~m_{\phi^\pm}^2) + s_{\theta_A}^2 F(m_{A_2^0}^2,~m_{\phi^\pm}^2) \nonumber \\
    &-& c_{\theta_H}^2 c_{\theta_A}^2 F(m_{H_1^0}^2,~m_{A_1^0}^2) - c_{\theta_H}^2 s_{\theta_A}^2 F(m_{H_1^0}^2,~m_{A_2^0}^2) - s_{\theta_H}^2 c_{\theta_A}^2 F(m_{H_2^0}^2,~m_{A_1^0}^2) - s_{\theta_H}^2 s_{\theta_A}^2 F(m_{H_2^0}^2,~m_{A_2^0}^2) \} \\
    \Delta S &=& \frac{1}{24\pi} \{ (2 s_W^2 - 1) G(m_{\phi^\pm}^2,~m_{\phi^\pm}^2,~m_{Z}^2) + c_{\theta_H}^2 c_{\theta_A}^2 G(m_{H_1^0}^2,~m_{A_1^0}^2,~m_{Z}^2) + c_{\theta_H}^2 s_{\theta_A}^2 G(m_{H_1^0}^2,~m_{A_2^0}^2,~m_{Z}^2) \nonumber \\
    &+& s_{\theta_H}^2 c_{\theta_A}^2 G(m_{H_2^0}^2,~m_{A_1^0}^2,~m_{Z}^2) + s_{\theta_H}^2 s_{\theta_A}^2 G(m_{H_2^0}^2,~m_{A_2^0}^2,~m_{Z}^2) + c_{\theta_H}^2 \log\left(\frac{m_{H_1^0}^2}{m_{\phi^\pm}^2}\right) + s_{\theta_H}^2 \log\left(\frac{m_{H_2^0}^2}{m_{\phi^\pm}^2}\right) \nonumber \\
    &+& c_{\theta_A}^2 \log\left(\frac{m_{A_1^0}^2}{m_{\phi^\pm}^2}\right) + s_{\theta_A}^2 \log\left(\frac{m_{A_2^0}^2}{m_{\phi^\pm}^2}\right),
\end{eqnarray}
where $F(x, y)$ and $G(x, y, z)$ are one-loop functions that can be found in Ref. \cite{Grimus:2008nb}.
%%%%%%%%%%%%%%%%%%%%%%%%%%%%%%%%%%%%%%%%%%%%%%%%%%%%%%%%%%%%%%%%
\section{Dark matter constraints}
\label{SecV}
%%%%%%%%%%%%%%%%%%%%%%%%%%%%%%%%%%%%%%%%%%%%%%%%%%%%%%%%%%%%%%%%%
The $Z_2$ symmetry forbids terms linear in the inert fields and prevents mixing between the SM-like Higgs boson and the $Z_2$-odd neutral states. As a result, the lightest $Z_2$-odd state is stable. Depending on the spectrum, this state can be either the Majorana fermion $\psi$, or one of the neutral inert scalars. These two possibilities probe the same neutrino mass and flavor structure but lead to different relic density and direct detection mechanisms.
%%%%%%%%%%%%%%%%%%%%%%%%%%%%%%%%%%%%%%%%%%%%%%%%%%%%%%%%%%%%%%%%%
\subsection{Thermal relic abundance}
%%%%%%%%%%%%%%%%%%%%%%%%%%%%%%%%%%%%%%%%%%%%%%%%%%%%%%%%%%%%%%%%
We assume the standard thermal freeze-out picture in a radiation-dominated Universe. The odd particles are initially in thermal equilibrium with the SM plasma and then depart from equilibrium once the interaction rate becomes smaller than the Hubble expansion rate. Since viable points often contain several $Z_2$-odd states with mass splittings comparable to the freeze-out temperature, the relevant quantity is not only the annihilation rate of the lightest particle, but the effective annihilation rate of the whole thermally populated odd sector \cite{Gondolo:1990dk,Griest:1990kh}. Defining $n=\sum_i n_i$, where $i$ runs over all odd states that remain populated around freeze-out, the total number density satisfies
\begin{equation}
\frac{dn}{dt}+3Hn
=-\langle\sigma_{\rm eff}v\rangle\left(n^2-n_{\rm eq}^2\right).
\label{eq:dm-boltzmann-total}
\end{equation}
Here $H$ is the Hubble parameter, $n_{\rm eq}=\sum_i n_i^{\rm eq}$ is the total equilibrium density, and $\langle\sigma_{\rm eff}v\rangle$ is thermally averaged annihilation rate factor which contains all annihilation and coannihilation processes among the thermally accessible odd states. It is often useful to rewrite Eq.~\eqref{eq:dm-boltzmann-total} in terms of the yield $Y=n/s$, with $s$ the entropy density and $x=m_{\rm DM}/T$ \cite{Gondolo:1990dk}
\begin{equation}
\frac{dY}{dx}
=-\frac{s}{Hx}\,\langle\sigma_{\rm eff}v\rangle
\left(Y^2-Y_{\rm eq}^2\right).
\label{eq:dm-yield}
\end{equation}
This form makes clear that increasing the effective annihilation rate decreases the final relic abundance, while suppressed annihilation tends to overclose the Universe.

For a set of coannihilating particles with masses $m_i=m_{\rm DM}(1+\Delta_i)$ and internal degrees of freedom $g_i$, the equilibrium weights are Boltzmann suppressed according to the mass splittings. The effective cross section is
\begin{equation}
\langle\sigma_{\rm eff}v\rangle(x)
=\sum_{i,j}\langle\sigma_{ij}v_{ij}\rangle\,r_i(x)r_j(x),
\qquad
r_i(x)=
\frac{g_i(1+\Delta_i)^{3/2}e^{-x\Delta_i}}
{\sum_k g_k(1+\Delta_k)^{3/2}e^{-x\Delta_k}},
\qquad
\Delta_i=\frac{m_i-m_{\rm DM}}{m_{\rm DM}} .
\label{effec.crosec}
\end{equation}
Equation~\eqref{effec.crosec} show why compressed spectra are so important. At freeze-out one typically has $x_F\simeq20$--$30$, so particles within a few to ten percent of the DM mass can still carry a significant Boltzmann weight. Their annihilation channels may then dominate the relic density even if the lightest particle has a weak self-annihilation rate. The usual semi analytic solution gives the relic abundance in terms if the integrated effective annihilation rate \cite{Griest:1990kh}
\begin{equation}
\Omega h^2\simeq
\frac{1.07\times10^9\,\mathrm{GeV}^{-1}}{M_{\rm Pl}\sqrt{g_*(x_F)}}\frac{1}{J(x_F)},\qquad
J(x_F)=\int_{x_F}^{\infty}\frac{\langle\sigma_{\rm eff}v\rangle}{x^2} dx .
\label{eq:dm-freezeout-solution}
\end{equation}
where $J(x_F)$ is the post-freeze-out annihilation integral which accumulates the thermally averaged effective cross section from the freeze-out point $x_F=m_{\rm DM}/T_F$ to late times. In the present model, the model dependent ingredients entering $J(x_F)$ are precisely the masses, mixings and couplings of the $Z_2$-odd sector. The initial states $i,j$ may include $\psi$, $H_i^0$, $A_i^0$ and $\phi^\pm$; their Boltzmann weights depend on the same mass splittings that appear in the scalar spectrum, while their annihilation amplitudes depend on the leptophilic Yukawa couplings, scalar mixing angles, gauge interactions and Higgs-portal couplings. These quantities are not independent of neutrino physics: the one-loop neutrino mass matrix is generated by the same odd fermions and odd scalars, with the same Yukawa structures and loop masses. The Planck relic abundance therefore constrains the dark sector in a way that is correlated with neutrino masses, cLFV, electroweak precision observables and Higgs phenomenology. In the numerical study we impose the Planck $3\sigma$ band
\begin{equation}
  |\Omega h^2-0.1200|<3(0.0012),
\end{equation}
corresponding to the Planck relic density target \cite{Planck:2018vyg}.
\newline
\\
\textbf{Fermionic dark matter:}
When $\psi$ is the lightest odd state, the direct annihilation channel
$\psi\psi\to \ell_\alpha\bar\ell_\beta,\nu_\alpha\bar\nu_\beta$
proceeds through $t$- and $u$-channel exchange of the inert scalars. Its size
is governed by the same new lepton sector Yukawa couplings that enter the
neutrino mass matrix and cLFV amplitudes. Increasing these couplings can
enhance the annihilation rate, but it also enhances processes such as
$\mu\to e\gamma$, $\mu\to3e$, and coherent $\mu-e$ conversion. The fermionic
case therefore does not behave as a single channel WIMP scenario. Instead,
scotogenic fermionic-DM analyses show that the relic density can be obtained
through coannihilations with the inert scalar sector when the fermion and
scalar states are sufficiently degenerate
\cite{Vicente:2014wga,Hagedorn:2018spx,Karan:2023adm}. The same mechanism can
operate here when the inert scalar partners remain close enough to $\psi$ to
participate in freeze-out. The corresponding annihilation and coannihilation
channel decomposition is discussed in Sec.~\ref{SecVI}.
\newline
\\
\textbf{Scalar dark matter:}
When $H_1^0$ is the lightest odd state, the relic-density mechanisms are richer. The scalar candidate has Higgs-portal interactions, inherits gauge interactions through its doublet component, and can coannihilate with nearby CP-even, CP-odd, and charged inert states. The scalar scan therefore probes the Higgs-resonance region $m_{H_1^0}\simeq m_h/2$, gauge and Higgs-mediated annihilation controlled by the doublet-singlet admixture, and coannihilation
with $A_i^0$, $H_2^0$, and $\phi^\pm$ when the odd spectrum is compressed. The accepted points show that the viable scalar regime is best interpreted as a Higgs-funnel/coannihilation region rather than as a pure Higgs-portal limit. The detailed numerical pattern is presented in Sec.~\ref{SecVI}.
%%%%%%%%%%%%%%%%%%%%%%%%%%%%%%%%%%%%%%%%%%%%%%%%%%%%%%%%%%%%%%%%%
\subsection{Direct detection}
%%%%%%%%%%%%%%%%%%%%%%%%%%%%%%%%%%%%%%%%%%%%%%%%%%%%%%%%%%%%%%%%%
Direct detection experiments probe the nonrelativistic elastic scattering of DM on nuclei. In the present analysis the relevant observable is the spin-independent (SI) cross section. The Higgs boson provides the scalar current that matches onto the nucleon matrix element, but the origin of the DM--Higgs vertex depends on the identity of the lightest $Z_2$-odd state. For scalar DM, the vertex $hH_1^0H_1^0$ is present at tree level through the Higgs-portal interactions. For fermionic DM, instead, no independent renormalizable tree-level $h\psi\psi$ interaction exists; the Higgs-mediated description arises only after a loop-induced effective vertex $h\psi\psi$ is generated. At the nucleon level the scalar Higgs current is written as
\begin{equation}
g_{h{\cal N} {\cal N}}=\frac{m_{\cal N}}{v}\left[ \sum_{q=u,d,s} f_{Tq}^{({\cal N})} + \frac{2}{27}f_{TG}^{({\cal N})}
\right], \qquad f_{TG}^{({\cal N})}=1-\sum_{q=u,d,s}f_{Tq}^{({\cal N})} ,
\label{eq:dm-higgs-nucleon}
\end{equation}
where $f_{Tq}^{({\cal N})}$ are the light-quark scalar form factors and $f_{TG}^{({\cal N})}$ accounts for the heavy-quark contribution through the trace anomaly. If $X$ denotes the DM particle and $\mu_{X {\cal N}}=m_X m_{\cal N}/(m_X + m_{\cal N})$ is the reduced mass, Higgs exchange gives the schematic dependence
\begin{equation}
\sigma_{\rm SI}\propto
\frac{\mu_{X {\cal N}}^2}{\pi}
\left|\frac{g_{hXX}g_{h {\cal N} {\cal N}}}{m_h^2}\right|^2 ,
\label{eq:dm-si-schematic}
\end{equation}
with the usual external state normalization depending on whether $X$ is a fermion or a scalar.
\newline
\\
\textbf{Fermionic dark matter:}
The Majorana fermion $\psi$ is a gauge singlet and has no tree-level Higgs portal. Consequently, its elastic scattering on nuclei is naturally suppressed. The leading SI interaction is generated after integrating out the charged inert scalar and charged lepton loop, producing the effective operator
\begin{equation}
{\cal L}_{\rm eff}\supset
-\frac12\,g_{h\psi\psi}^{\rm eff}~h \bar\psi\psi .
\label{eq:dm-hchichi-lagrangian}
\end{equation}
For the CalcHEP implementation used in our numerical scan, the effective vertex inserted for direct detection is
\begin{equation}
g_{h\psi\psi}^{\rm eff}=
-\frac{\kappa_1v}{16\pi^2}
\left[\sum_{\alpha=e,\mu,\tau}(y_{\alpha R}^2+y_{\alpha I}^2)\right]
\frac{M_\psi^2+(m_{\phi^\pm}^2-M_\psi^2)\ln\left(1-M_\psi^2/m_{\phi^\pm}^2\right)}{M_\psi^3}
\label{eq:dm-gchichi}
\end{equation}
for $M_\psi<m_{\phi^\pm}$. The formula makes the main parametric behavior transparent. The DD amplitude grows with $|\kappa_1|$ and with the leptophilic Yukawa norm $\sum_\alpha |y_\alpha|^2$, but it is loop suppressed and decreases when the charged scalar becomes heavy. In the limit $m_{\phi^\pm}^2\gg M_\psi^2$,
\begin{equation}
g_{h\psi\psi}^{\rm eff} \simeq - \frac{\kappa_1v}{32\pi^2}
\left(\sum_\alpha |y_\alpha|^2\right) \frac{M_\psi}{m_{\phi^\pm}^2}.
\label{eq:dm-gchichi-heavy}
\end{equation}
This expression displays the decoupling behavior of the loop-induced Higgs coupling: for fixed $\kappa_1$ and Yukawa norm, the effective vertex decreases as $M_\psi / m_{\phi^\pm}^2$ when the charged scalar becomes heavy. After integrating out the Higgs boson, the leading nucleon coupling is
\begin{equation}
f_{\cal N}^\psi=\frac{g_{h\psi\psi}^{\rm eff}g_{h {\cal N} {\cal N}}}{m_h^2},
\qquad
\sigma_{\rm SI}^{\psi N}=\frac{\mu_{\psi {\cal N}}^2}{\pi}|f_{\cal N}^\psi|^2 .
\label{eq:dm-chi-si}
\end{equation}
The fermionic DD signal is therefore not governed by an isolated Higgs portal. It is enhanced by the charged scalar Higgs coupling $\kappa_1$, by the Yukawa norm $\sum_\alpha|y_\alpha|^2$, and by a charged scalar that is not strongly decoupled. The same ingredients also enter other parts of the phenomenology; $\kappa_1$ affects the charged scalar spectrum and $h \to \gamma \gamma$, the Yukawa combinations contribute to cLFV and to the one-loop neutrino mass matrix, and the odd sector masses control coannihilation, electroweak precision observables and thermal freeze-out.
\newline
\\
\textbf{Scalar dark matter:}
For the scalar DM assignment, the relevant field is the CP-even neutral state $H_1^0$. Unlike the fermionic case, its coupling to the Higgs is already present at the renormalizable level. The diagonal trilinear interaction controlling elastic Higgs-mediated scattering is
\begin{equation}
g_{h H_1^0 H_1^0} = -\sqrt2~c_Hs_H\mu_\phi - \upsilon~c_H^2(\kappa_1 + \kappa_2 + \kappa_3) - \upsilon~s_H^2\kappa_4 .
\label{dm-hh1h1}
\end{equation}
The scalar potential also contains the vertices $h H_2^0 H_2^0$ and $h H_2^0 H_2^0$. They are relevant for scalar sector interactions and may enter annihilation or coannihilation processes involving $H_2^0$. The $h H_2^0 H_2^0$ vertex contains two $H_2^0$ fields and would be the diagonal portal for an $H_2^0$ DM assignment. The mixed
vertex $h H_1^0 H_2^0$ changes the identity of the neutral scalar and corresponds, in DD, to an inelastic transition $H_1^0 {\cal N} \to H_2^0 {\cal N}$ rather than elastic scattering. Thus the standard elastic SI cross section for $H_1^0$ DM is governed by $g_{h H_1^0 H_1^0}$.
The dependence of Eq. \ref{dm-hh1h1} shows that the scalar DD rate depends on the physical portal combination involving $\kappa_1 + \kappa_2 + \kappa_3$, $\kappa_4$, the CP-even mixing angle, and the dimensionful parameter $\mu_\phi$. A suppressed diagonal portal reduces the elastic Higgs-exchange rate, while the same scalar potential parameters also affect Higgs-mediated annihilation, scalar mass splittings, coannihilation channels and Higgs diphoton phenomenology. For an interaction written as $ -1/2~g_{hH_1H_1}~h (H_1^0)^2$, Higgs exchange gives the parametric dependence
\begin{equation}
\sigma_{\rm SI}^{H_1^0 {\cal N}} \propto \frac{\mu_{H_1^0 {\cal N}}^2}{m_{H_1^0}^2}
\left| \frac{g_{h H_1^0 H_1^0} g_{h {\cal N}{\cal N}}}{m_h^2} \right|^2,
\label{dm-scalar-si}
\end{equation}
with the usual scalar-DM external state normalization. Since this interaction is present at tree level for $H_1^0$, the scalar SI rate is parametrically less loop suppressed than in the fermionic case. This is the central contrast with the fermionic case: scalar DD is a tree-level Higgs-portal problem, whereas fermionic direct detection is induced only after the loop-level effective $h\psi\psi$ interaction is generated.
%%%%%%%%%%%%%%%%%%%%%%%%%%%%%%%%%%%%%%%%%%%%%%%%%%%%%%%%%%%%%%%%%
\section{Results}
\label{SecVI}
%%%%%%%%%%%%%%%%%%%%%%%%%%%%%%%%%%%%%%%%%%%%%%%%%%%%%%%%%%%%%%%%%
\subsection{Statistical framework}
\label{sec_stat}
%%%%%%%%%%%%%%%%%%%%%%%%%%%%%%%%%%%%%%%%%%%%%%%%%%%%%%%%%%%%%%%%%
The statistical analysis was performed by scanning the parameter space of the model and comparing each point with the relevant experimental and theoretical constraints. For each parameter point $\theta$, the model predictions are obtained from the full chain of derived masses, mixing matrices, loop-induced neutrino observables, Higgs observables, electroweak precision quantities, cLFV observables, and DM observables. The fit is based on a total negative log-likelihood,
\begin{equation}
  \mathcal{L}_{\rm tot}(\theta) = \sum_i \mathcal{L}_i(\theta),
\end{equation}
where each term $\mathcal{L}_i$ represents one experimental measurement, bound, or theoretical consistency requirement. The best-fit point is defined as the point that minimizes $Q(\theta)=\mathcal{L}_{\rm tot}(\theta)$, namely $\theta_{\rm bf} = \arg\min_{\theta} Q(\theta)$. For Gaussian observables, the likelihood contribution is written as
\begin{equation}
  \mathcal{L}_{O}
  =
  \frac{1}{2}
  \left(
    \frac{O_{\rm th}(\theta)-O_{\rm exp}}{\sigma_O}
  \right)^2,
\end{equation}
while correlated observables are treated using their covariance matrix. In particular, the electroweak oblique parameters are included through
\begin{equation}
  \mathcal{L}_{ST} = \Delta X^T C^{-1} \Delta X,  \qquad \Delta X =
  \begin{pmatrix}
  S-S_0 \\
  T-T_0
  \end{pmatrix},
\end{equation}
with $(S_0,T_0)=(-0.05,0)$ and a correlation coefficient of $93\%$. The remaining theoretical and experimental bounds are imposed as hard constraints: a point is retained only if it satisfies all such requirements.

The neutrino oscillation likelihood is constructed from the one-dimensional likelihood profiles of the \texttt{NuFIT 6.0} global analysis~\cite{Esteban:2024eli}. The scan includes the three leptonic mixing angles, the Dirac CP phase, and the two independent neutrino mass-squared splittings. Normal and inverted neutrino mass orderings are treated separately, using the corresponding likelihood profiles. The total likelihood used in the fit can be schematically written as
\begin{equation}
  \mathcal{L}_{\rm tot}
  =
  \mathcal{L}_{\nu}
  + \mathcal{L}_{h\gamma\gamma}
  + \mathcal{L}_{ST}
  + \mathcal{L}_{\Omega h^2}
  + \mathcal{L}_{\rm cuts}.
\end{equation}
Here $\mathcal{L}_{\nu}$ denotes the neutrino oscillation likelihood, including
$\theta_{12}$, $\theta_{13}$, $\theta_{23}$, $\delta_{\rm CP}$,
$\Delta m^2_{21}$, and $\Delta m^2_{3\ell}$. The Higgs diphoton signal strength is treated as a Gaussian observable, while the relic abundance is constrained around the observed DM density. The term $\mathcal{L}_{\rm cuts}$ collects the remaining hard constraints, including the cosmological bound on the sum of neutrino masses, scalar sector consistency, perturbative unitarity, cLFV limits, and DD bounds. The scan is performed over the free parameters summarized in Table~\ref{tab:scan_parameters}. Mass parameters and positive couplings are sampled logarithmically where appropriate, while signed couplings are sampled directly as real variables. The ranges shown in the table give the broad domain explored in the numerical analysis. Focused runs used smaller intervals inside the domain only after viable regions had been identified.
\begin{table}[H]
\centering
\begin{tabular}{lll}
\hline
\textbf{Parameters} & \textbf{Range} & \textbf{Prior} \\
\hline
$m_{\rm DM}$ & $[50,\,1500]~{\rm GeV}$ & log \\
$\Delta_X\equiv m_X-m_{\rm DM}$  & $[1,\,350]~{\rm GeV}$ & linear \\
$M_N$ & $[100,\,5000]~{\rm GeV}$ & log \\
$s_H$ & $[10^{-3},\,0.997]$ & linear \\
$\lambda_2,\lambda_3$ & $[10^{-4},\,6]$ for $\psi$ DM; $[10^{-4},\,0.8]$ for $H_1^0$ DM & log \\
$\kappa_1$ & $[-8,\,8]$ for $\psi$ DM; $[-0.5,\,0.5]$ for $H_1^0$ DM & linear \\
$\lambda_2,\lambda_3$ & $[10^{-4},\,6]$ & log \\
$\kappa_1$ & $[-8,\,8]$ & linear \\
$\kappa_4,\kappa_5$ & $[-0.5,\,0.5]$ & linear \\
$\mathrm{Re}\,y'_{11},\mathrm{Im}\,y'_{11}$ & $[-3,\,3]$ & signed log \\
$\mathrm{Re}\,Y_\alpha,\mathrm{Im}\,Y_\alpha$ & $[-1,\,1]$ & signed log \\
$\mathrm{Re}\,y_\alpha,\mathrm{Im}\,y_\alpha$ & $[-3.2,\,3.2]$ & signed log \\
$\mathrm{Re}\,y_\alpha,\mathrm{Im}\,y_\alpha$ & $[-3.2,\,3.2]$ & signed log \\
\hline
\end{tabular}
\caption{Free parameters used in the four scans, together with their ranges and sampling priors. Here $\alpha=e,\mu,\tau$. In each dark-matter assignment, the physical spectrum is reconstructed from $m_{\rm DM}$ and the corresponding mass gaps to the other $Z_2$-odd states $X$.}
\label{tab:scan_parameters}
\end{table}
The experimental inputs and bounds included in the likelihood are summarized in Table~\ref{tab:constraints}. For the neutrino sector, the full likelihood profiles are used rather than Gaussian approximations.
\begin{table}[H]
\centering
\begin{tabular}{c c c}
\hline
Observable / constraint & Value or bound & Treatment \\
\hline
$\theta_{12},\theta_{13},\theta_{23}$ & NuFIT profiles & likelihood  \cite{Esteban:2024eli} \\
$\Delta m^2_{21},\Delta m^2_{3\ell}$ & NuFIT profiles & likelihood  \cite{Esteban:2024eli}\\
$\delta_{\rm CP}$ & NuFIT profile & likelihood \cite{Esteban:2024eli} \\
$\sum_i m_i$ & $<0.12~{\rm eV}$ & hard cut \\
$R_{\gamma\gamma}$ & $1.0883 \pm 0.092374$ & Gaussian \cite{ATLAS:2022vkf} \\
$S$ & $-0.05$ & correlated Gaussian \\
$T$ & $0$ & correlated Gaussian \\
$\Omega h^2$ & $0.120$ & Gaussian \\
$l_\alpha \to l_\beta \gamma$ & Table \ref{tab:clfv} & hard cut \\
$l_\alpha \to 3 l_\beta$ & Table \ref{tab:clfv} & hard cut \\
$CR(\mu {\cal N} \to e {\cal N})$ & Table \ref{tab:clfv} & hard cut \\
Direct detection & current bounds & hard cut \\
Scalar potential & satisfied & hard cut \\
Perturbative unitarity & satisfied & hard cut \\
\hline
\end{tabular}
\caption{Main constraints included in the fit.}
\label{tab:constraints}
\end{table}
Since the viable region occupies only a small fraction of the full parameter space, the search for the best-fit point is performed in stages. A broad initial scan is first used to explore the full parameter volume and locate promising regions. Points with good likelihood values are then used to identify distinct basins of viable solutions. Each basin is subsequently refined with a focused differential-evolution minimization. The final best-fit point is selected as the valid point with the smallest value of the total negative log-likelihood after all constraints have been applied.
%%%%%%%%%%%%%%%%%%%%%%%%%%%%%%%%%%%%%%%%%%%%%%%%%%%%%%%%%%%%%%%%%
\subsection{Results of the fits}
%%%%%%%%%%%%%%%%%%%%%%%%%%%%%%%%%%%%%%%%%%%%%%%%%%%%%%%%%%%%%%%%%
The results of the global fit for the viable DM scenarios are presented in this section. We performed four broad scans with the same statistics, $10^6$ proposed points in each case: fermionic DM with NO, fermionic DM with IO, scalar DM with NO, and scalar DM with IO. After the likelihood requirements and hard cuts are applied, the accepted samples contain 157, 156, 1641 and 1636 points, respectively. The figures are split by observable class. For each case we show the dark sector and electroweak observables, the neutrino observables, and the cLFV observables. The color scale gives the likelihood ratio relative to the best-fit point, so brighter points lie closer to the global minimum.

\begin{figure}[tb]
\centering
\includegraphics[width=0.32\textwidth]{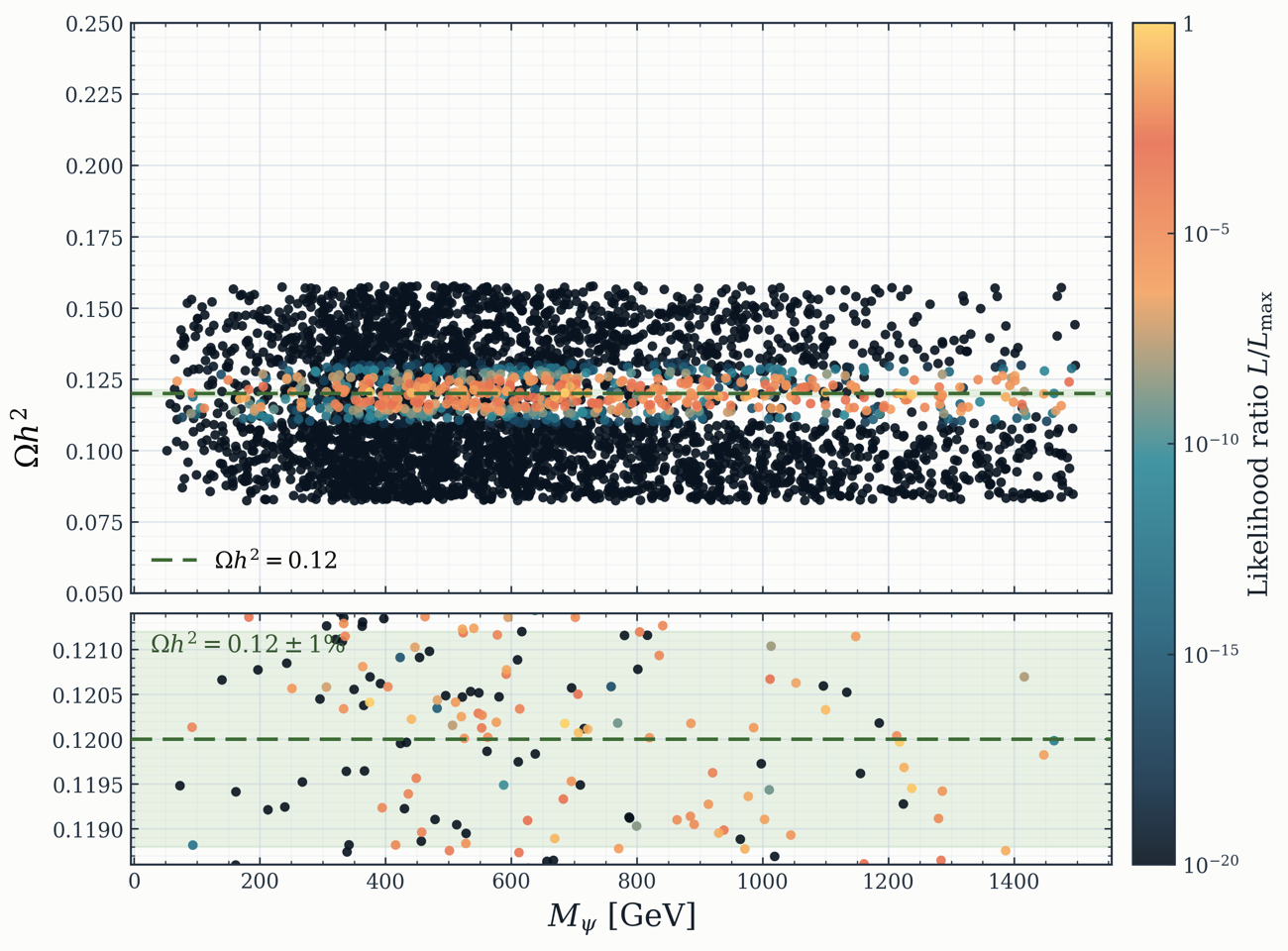}
\includegraphics[width=0.32\linewidth]{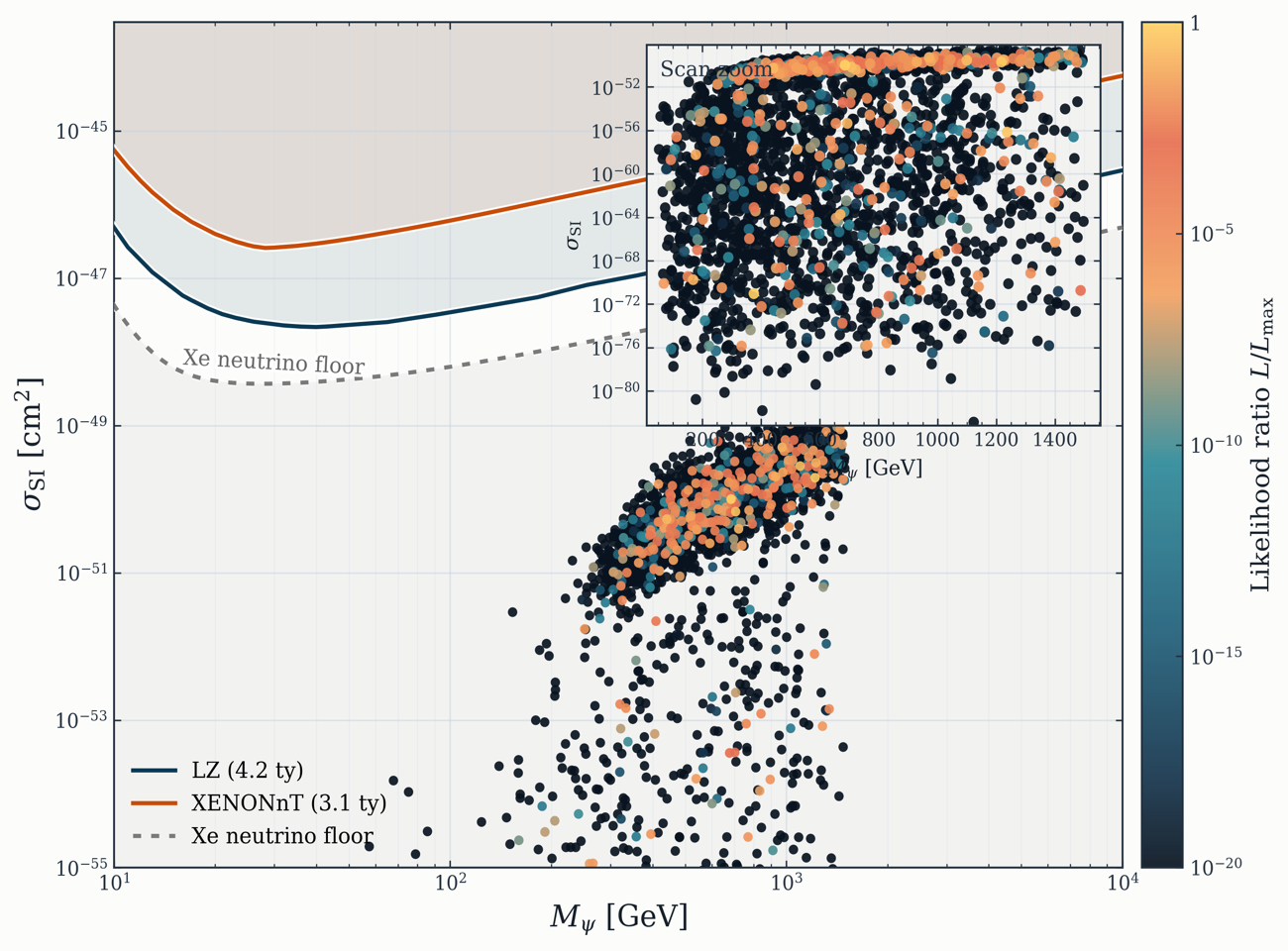}
\includegraphics[width=0.32\linewidth]{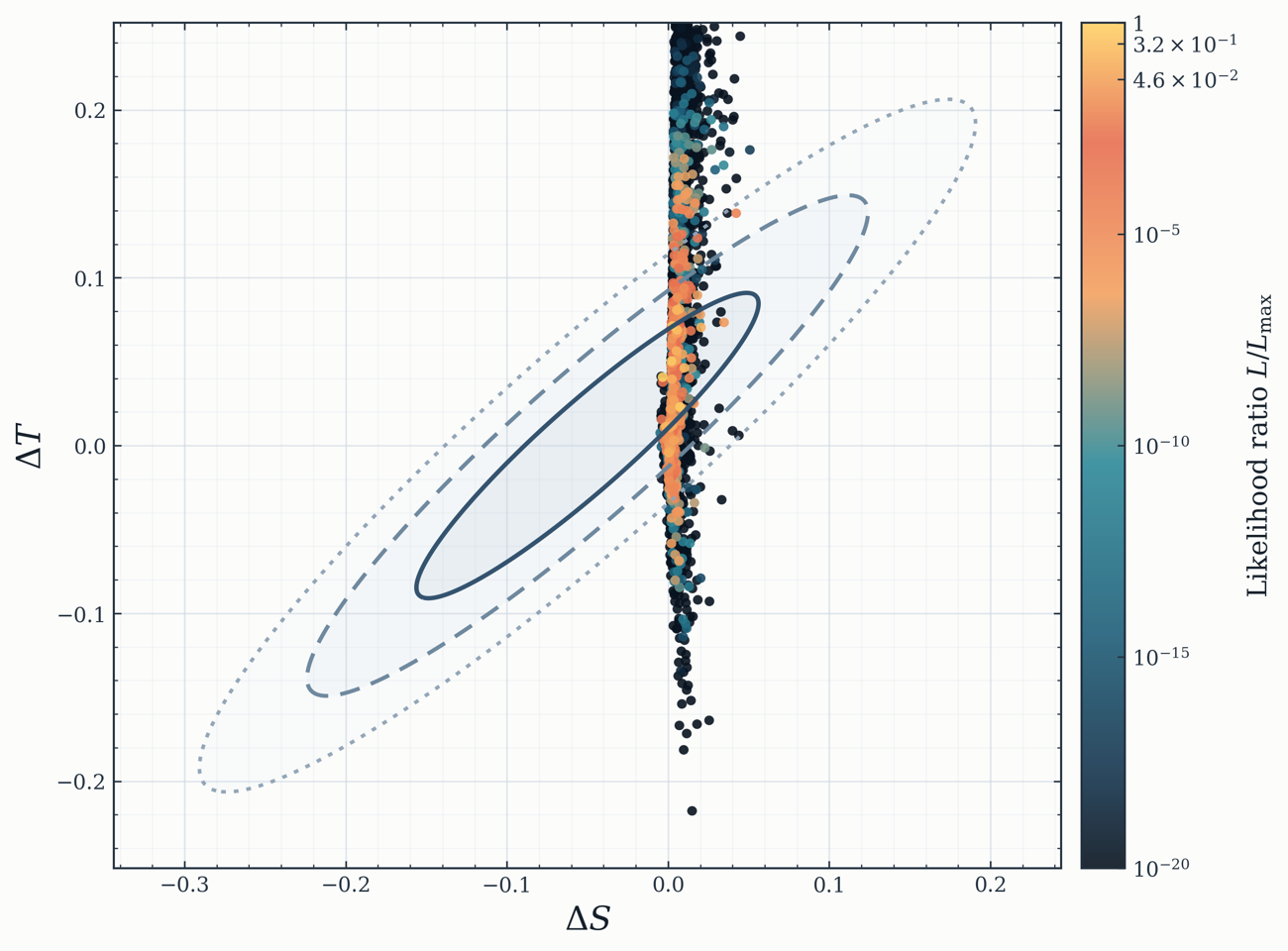}
\caption{Results for fermionic DM with NO. Left panel: relic abundance as a function of $M_\psi$. Middle panel: spin-independent DD cross section as a function of $M_\psi$. Right panel: allowed points in the $(\Delta S,\Delta T)$ plane. Current LZ and XENONnT bounds, the xenon neutrino-floor reference and the electroweak confidence ellipses are shown where relevant.}
\label{f2}
\end{figure}

\begin{figure}[H]
\centering
\includegraphics[width=0.32\linewidth]{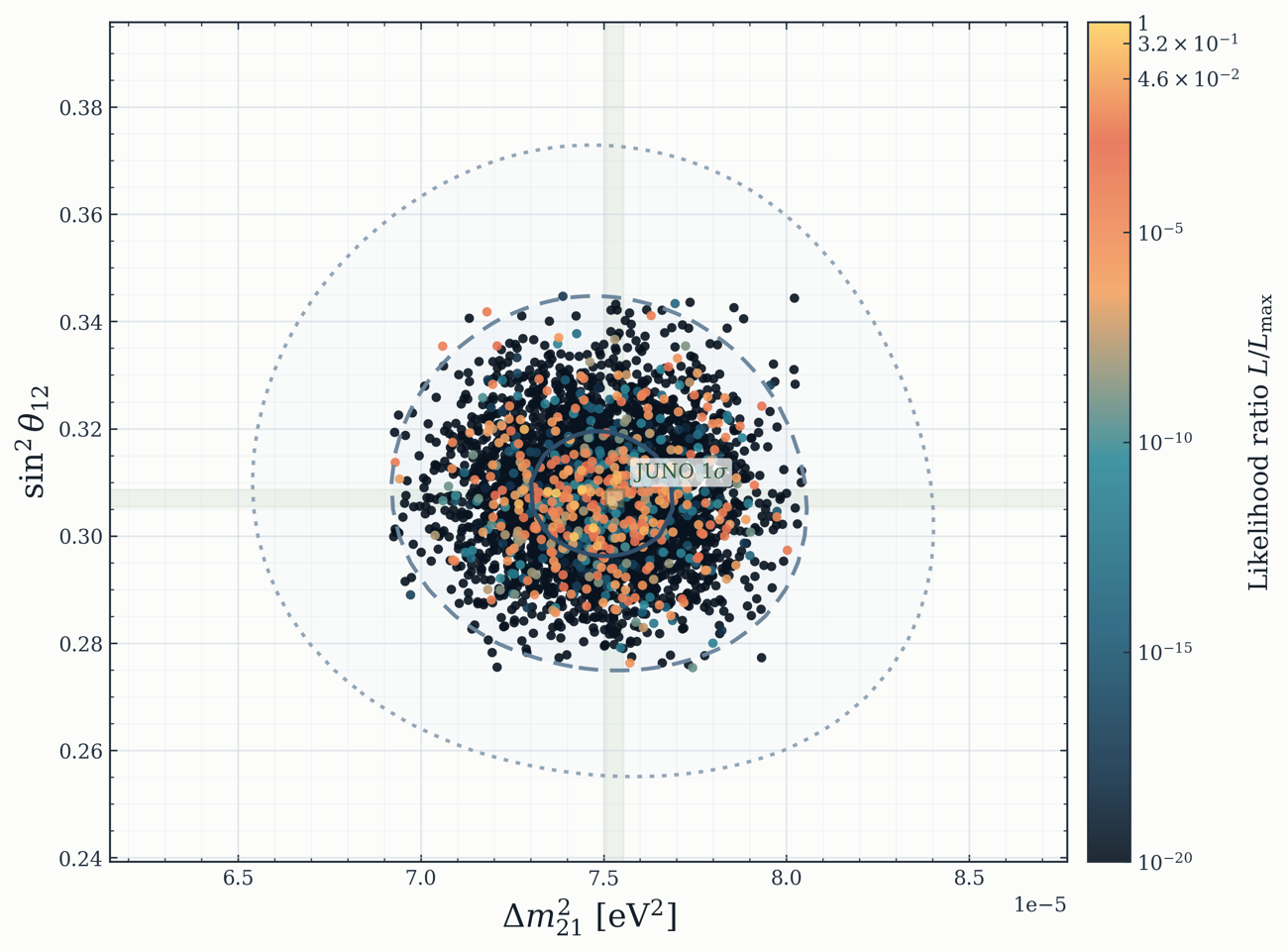}
\includegraphics[width=0.32\linewidth]{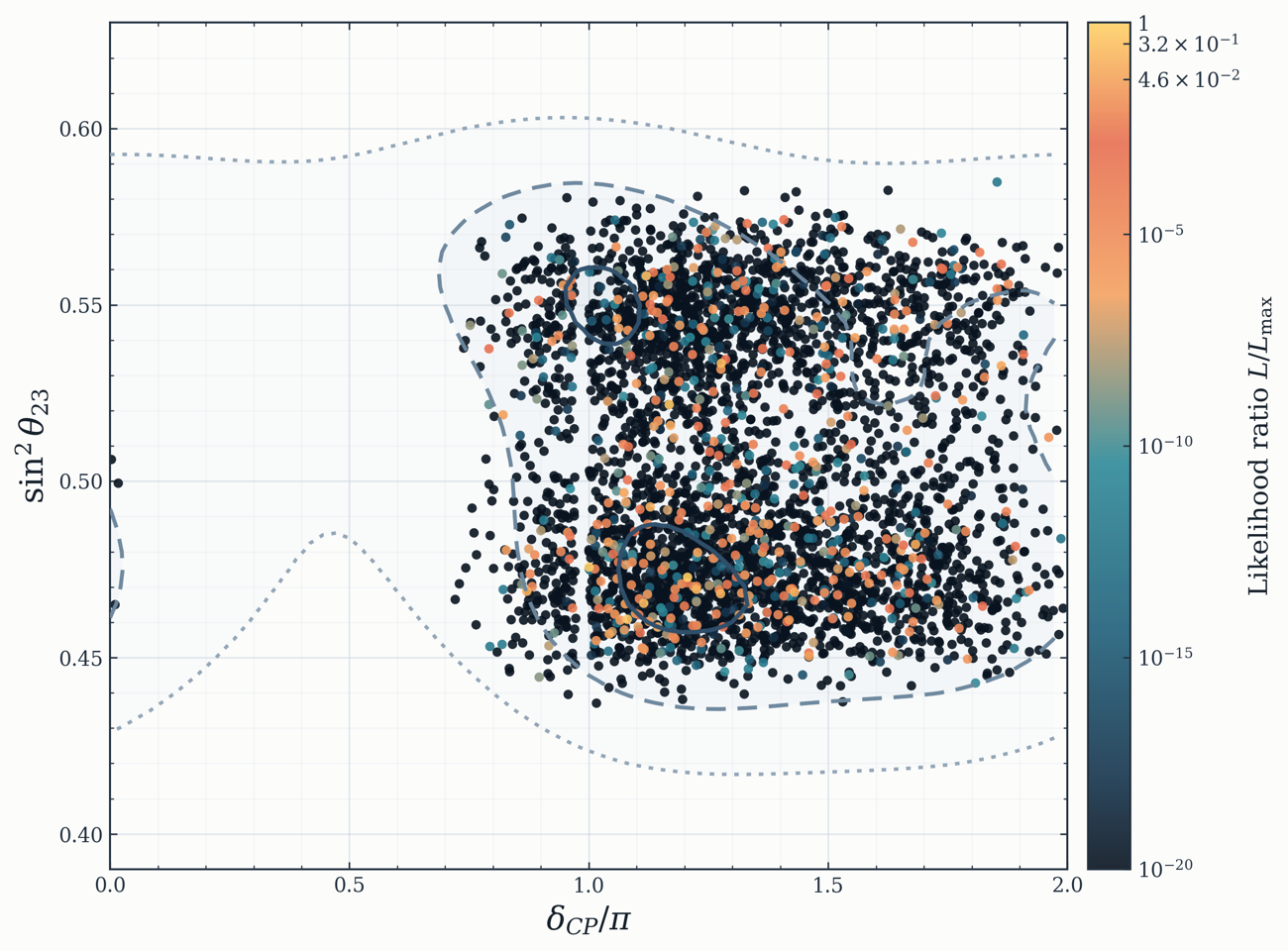}
\includegraphics[width=0.32\linewidth]{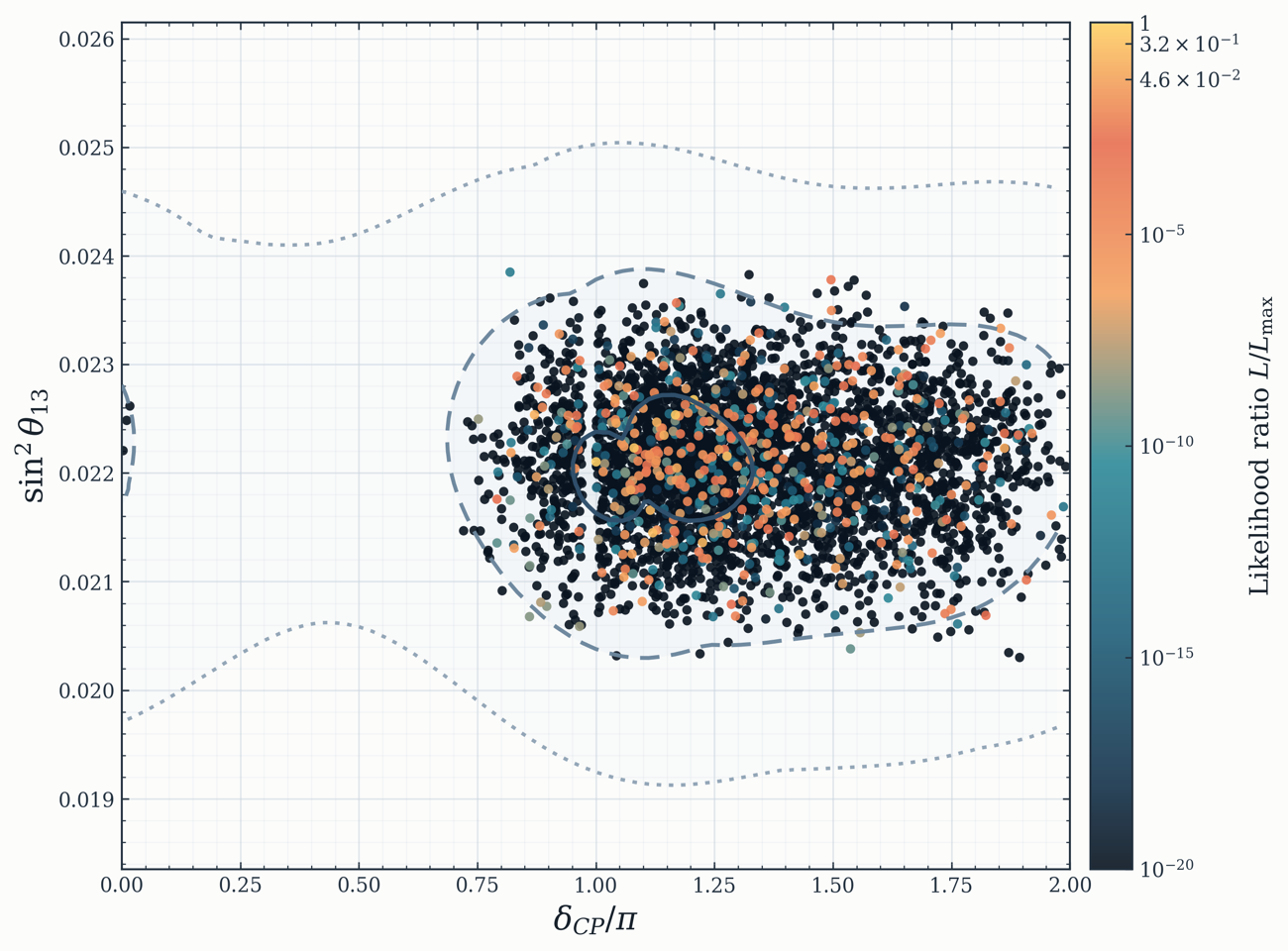}
\includegraphics[width=0.32\linewidth]{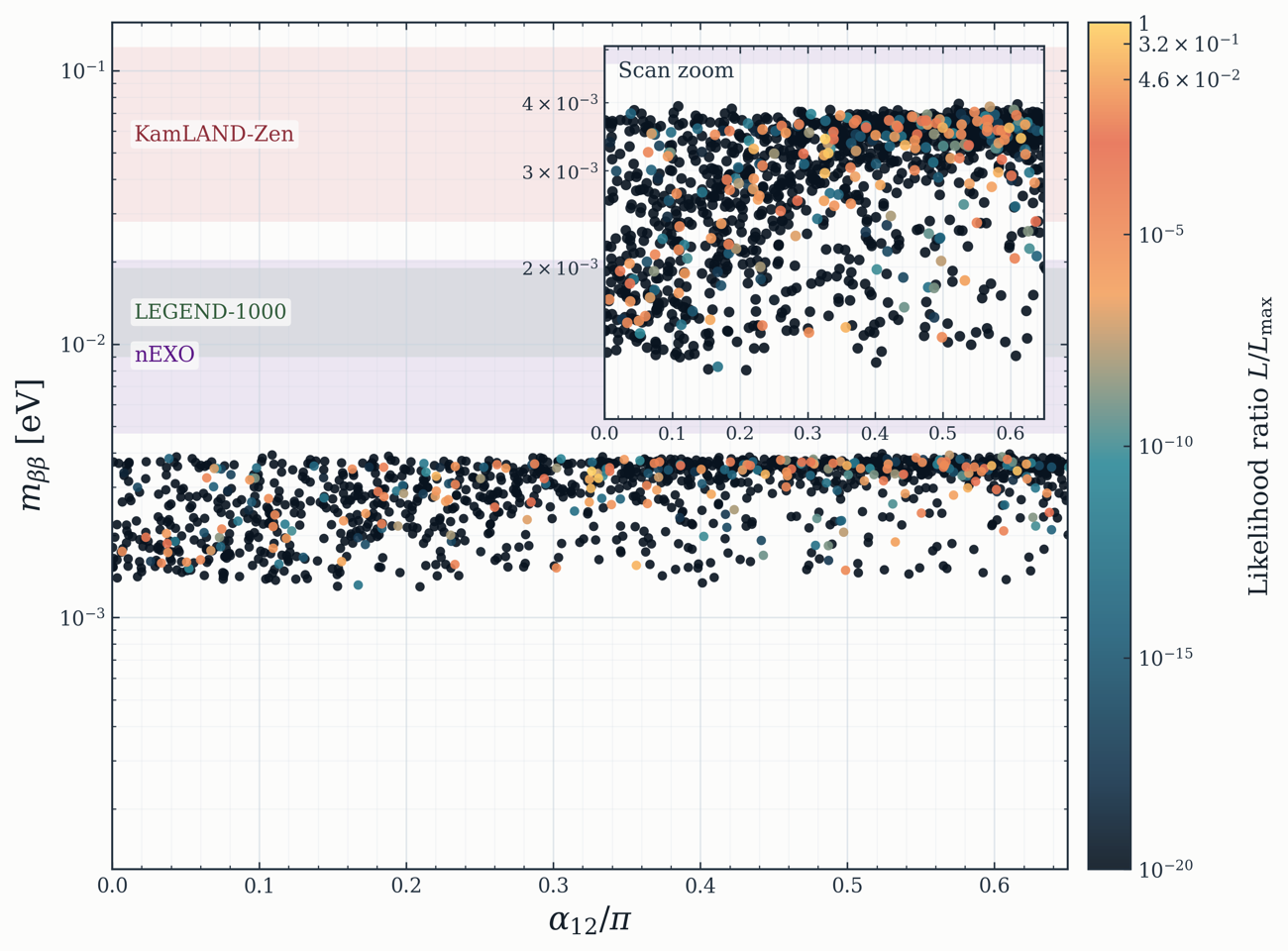}
\includegraphics[width=0.32\linewidth]{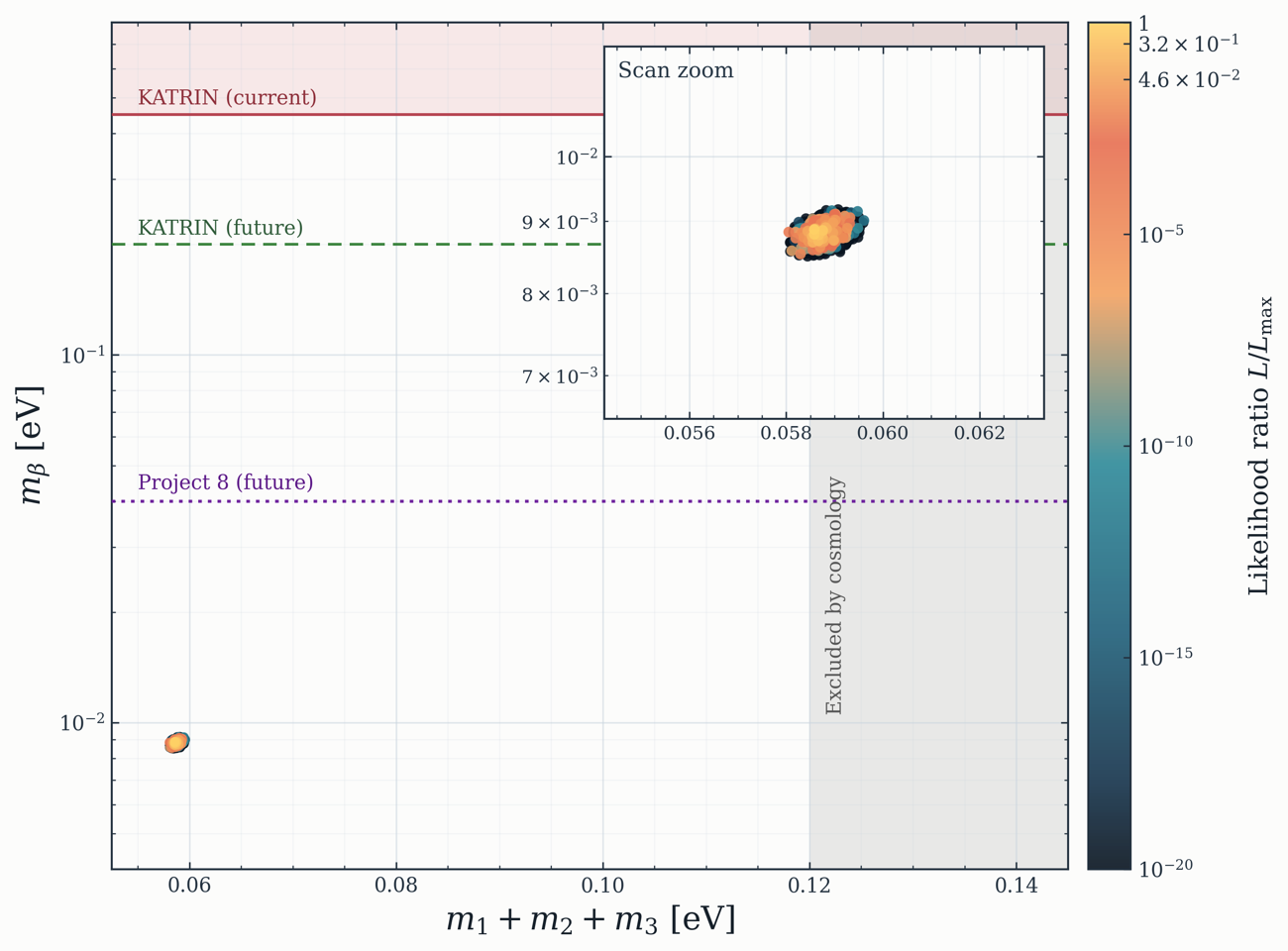}
\caption{Neutrino-sector predictions for fermionic DM with NO. Top-left panel: solar projection in the $(\Delta m^2_{21},\sin^2\theta_{12})$ plane. Top-middle and top-right panels: $\delta_{\rm CP}$ projections with $\sin^2\theta_{23}$ and $\sin^2\theta_{13}$, respectively. Bottom-left panel: $m_{\beta\beta}$ as a function of the Majorana phase $\alpha_{12}$. Bottom-right panel: $m_\beta$ as a function of $\sum_i m_i$. The color scale denotes the likelihood ratio relative to the best-fit point. Current experimental limits and projected future sensitivities are shown where applicable.}
\label{f3}
\end{figure}

\begin{figure}[H]
\centering
\includegraphics[width=0.32\textwidth]{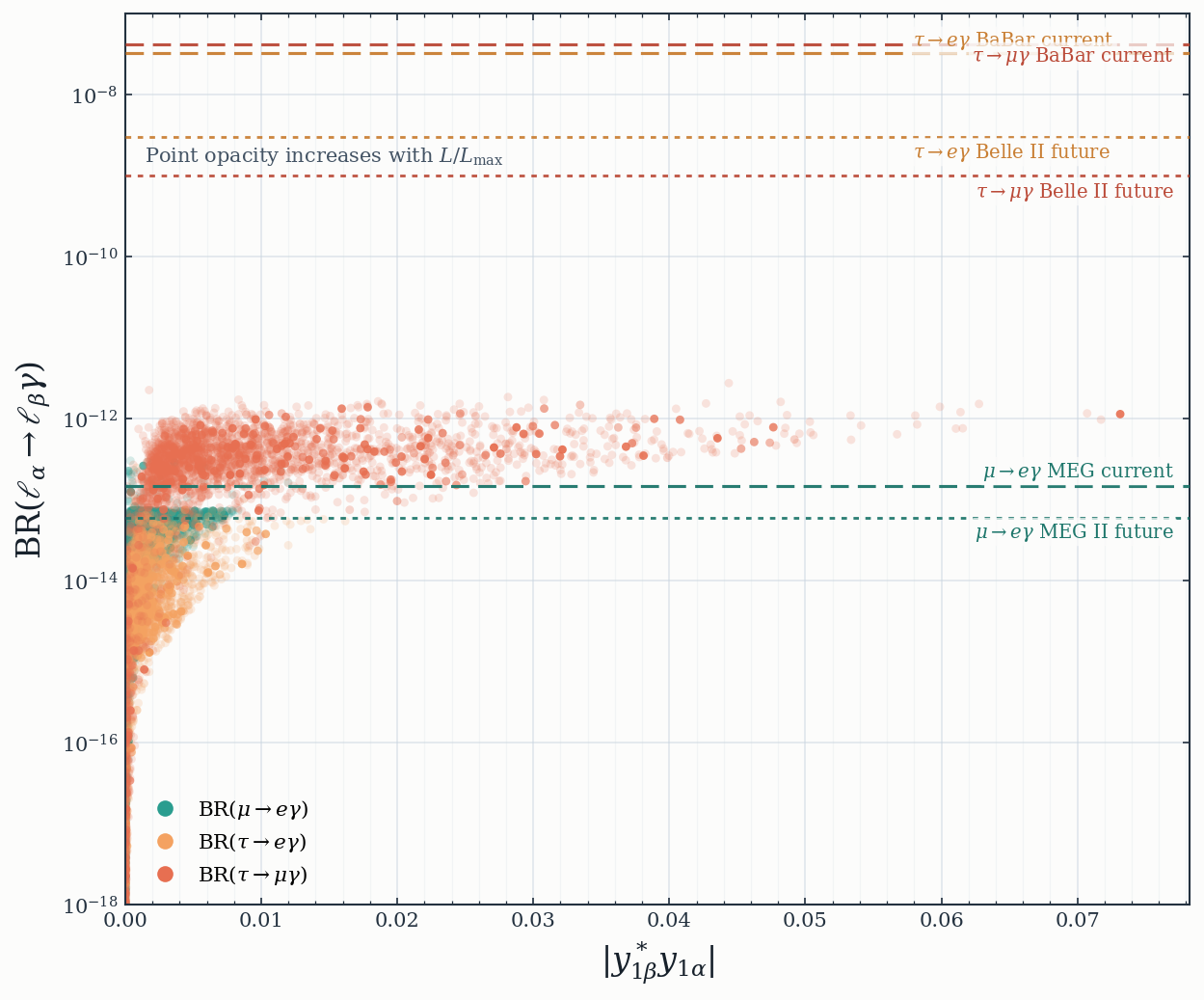}
\includegraphics[width=0.32\linewidth]{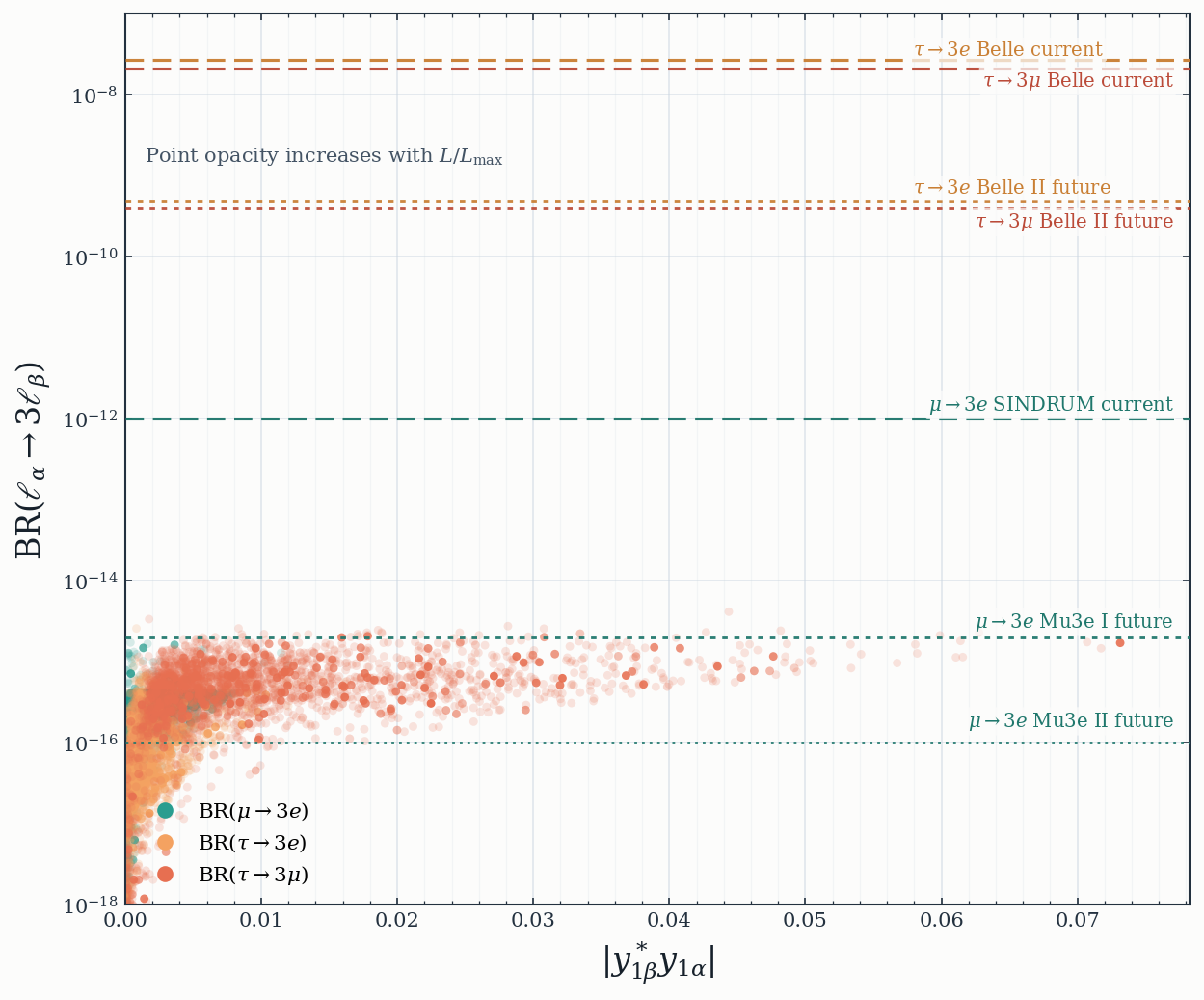}
\includegraphics[width=0.32\linewidth]{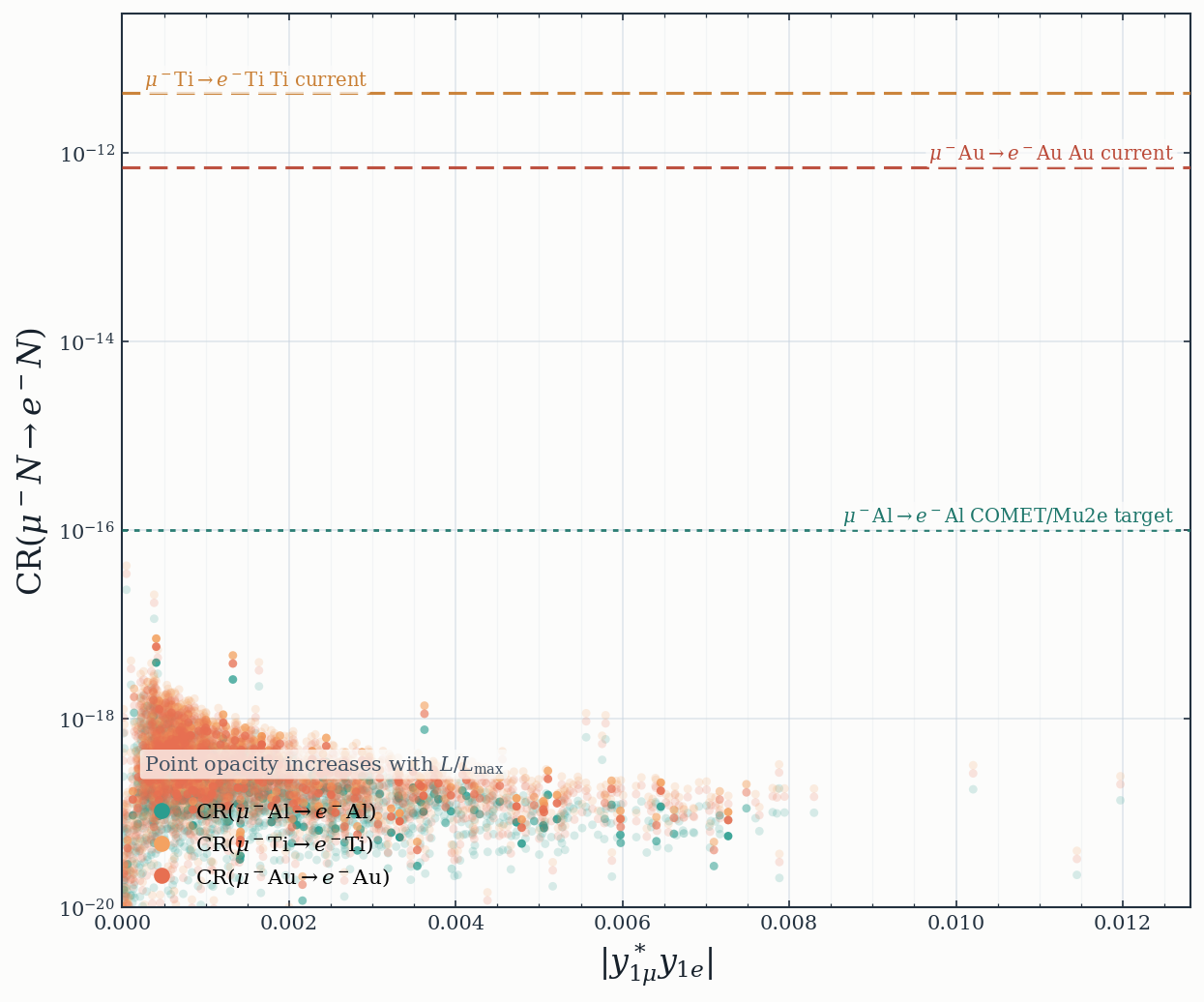}
\caption{cLFV predictions for fermionic DM with NO. Left panel: radiative decays $\ell_\alpha\to\ell_\beta\gamma$. Middle panel: three-body decays $\ell_\alpha\to3\ell_\beta$. Right panel: coherent $\mu-e$ conversion in nuclei. Present limits and projected sensitivities are those listed in Table~\ref{tab:clfv}.}
\label{f4}
\end{figure}

\begin{figure}[H]
\centering
\includegraphics[width=0.32\textwidth]{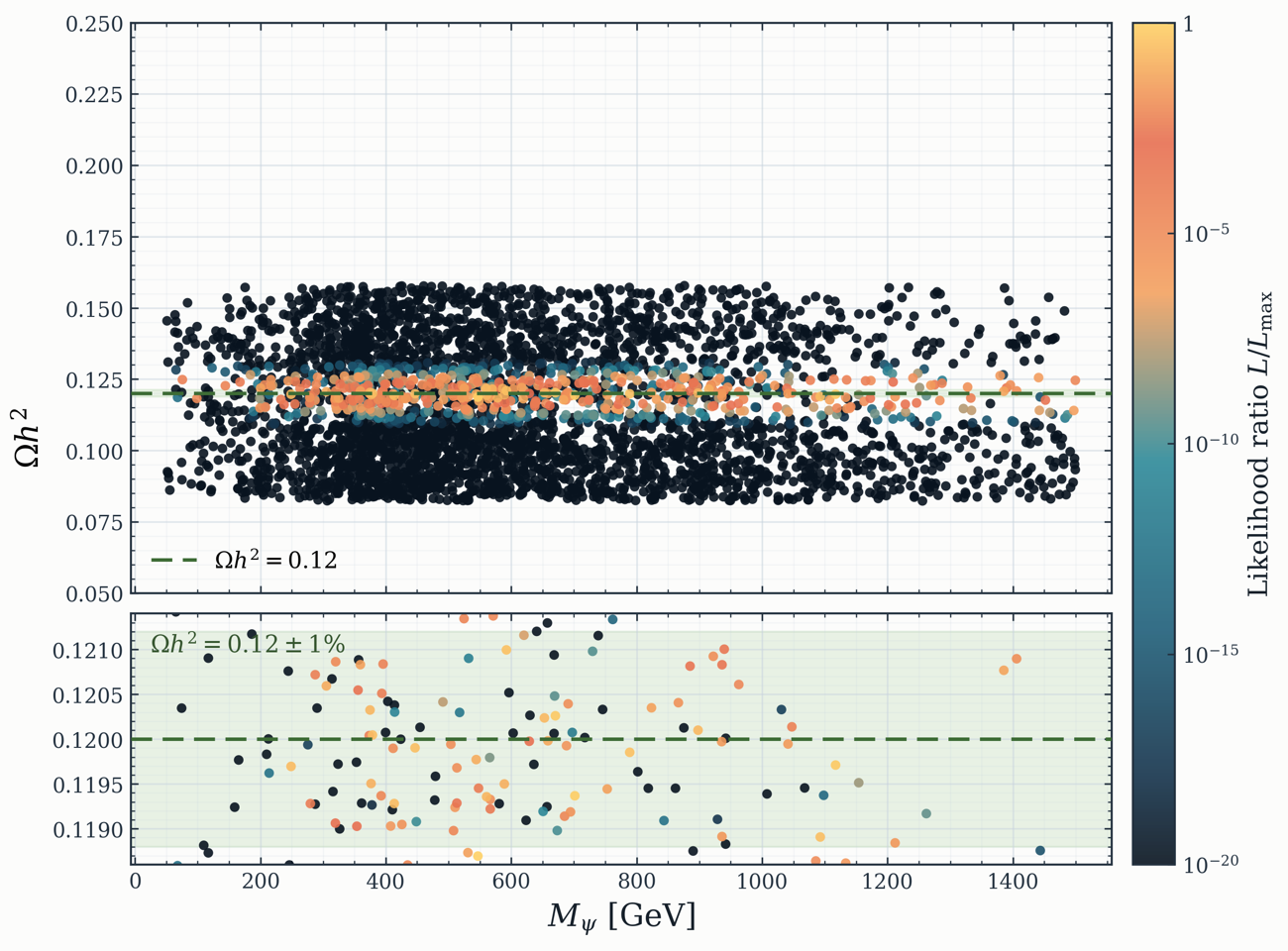}
\includegraphics[width=0.32\linewidth]{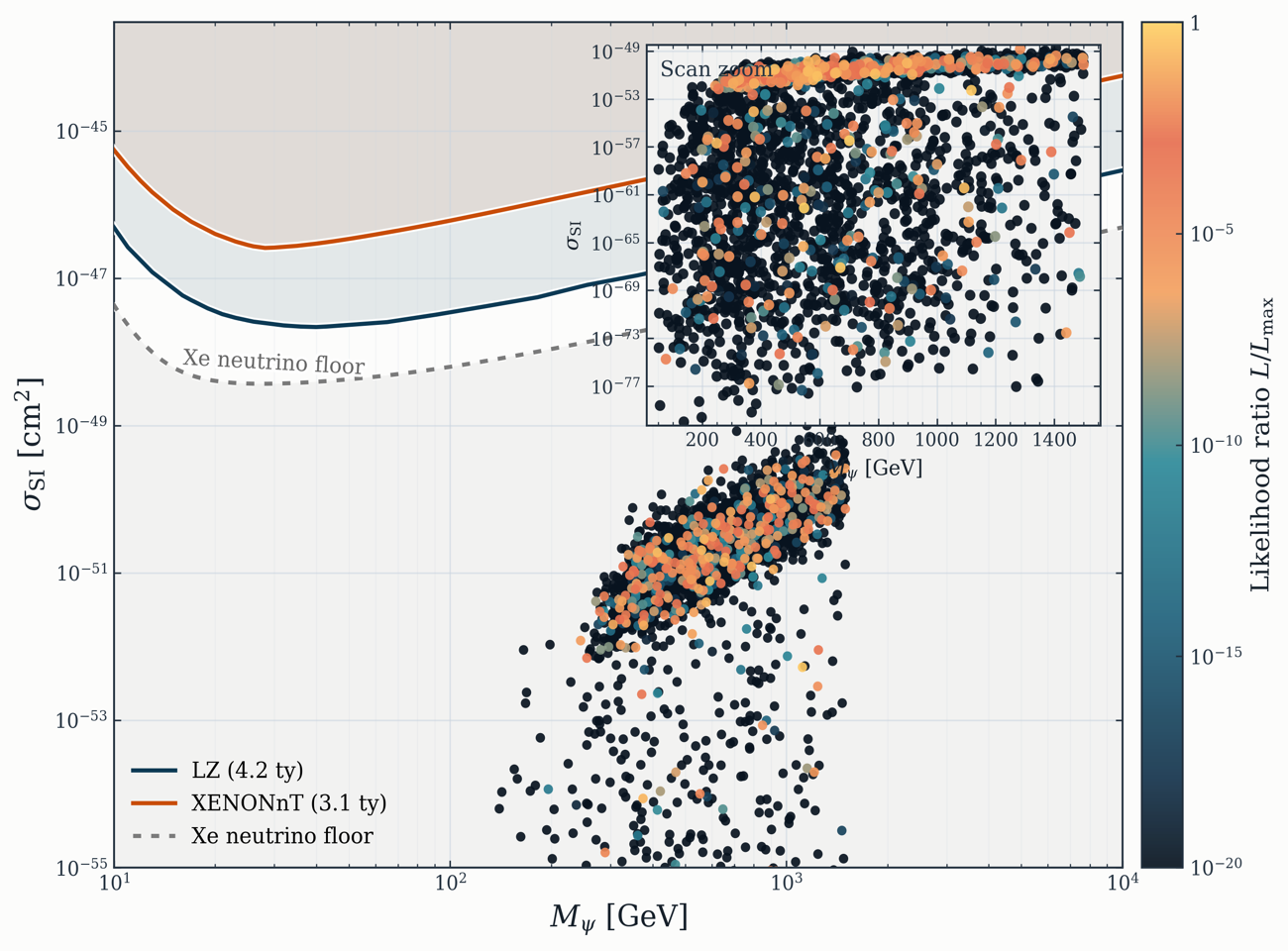}
\includegraphics[width=0.32\linewidth]{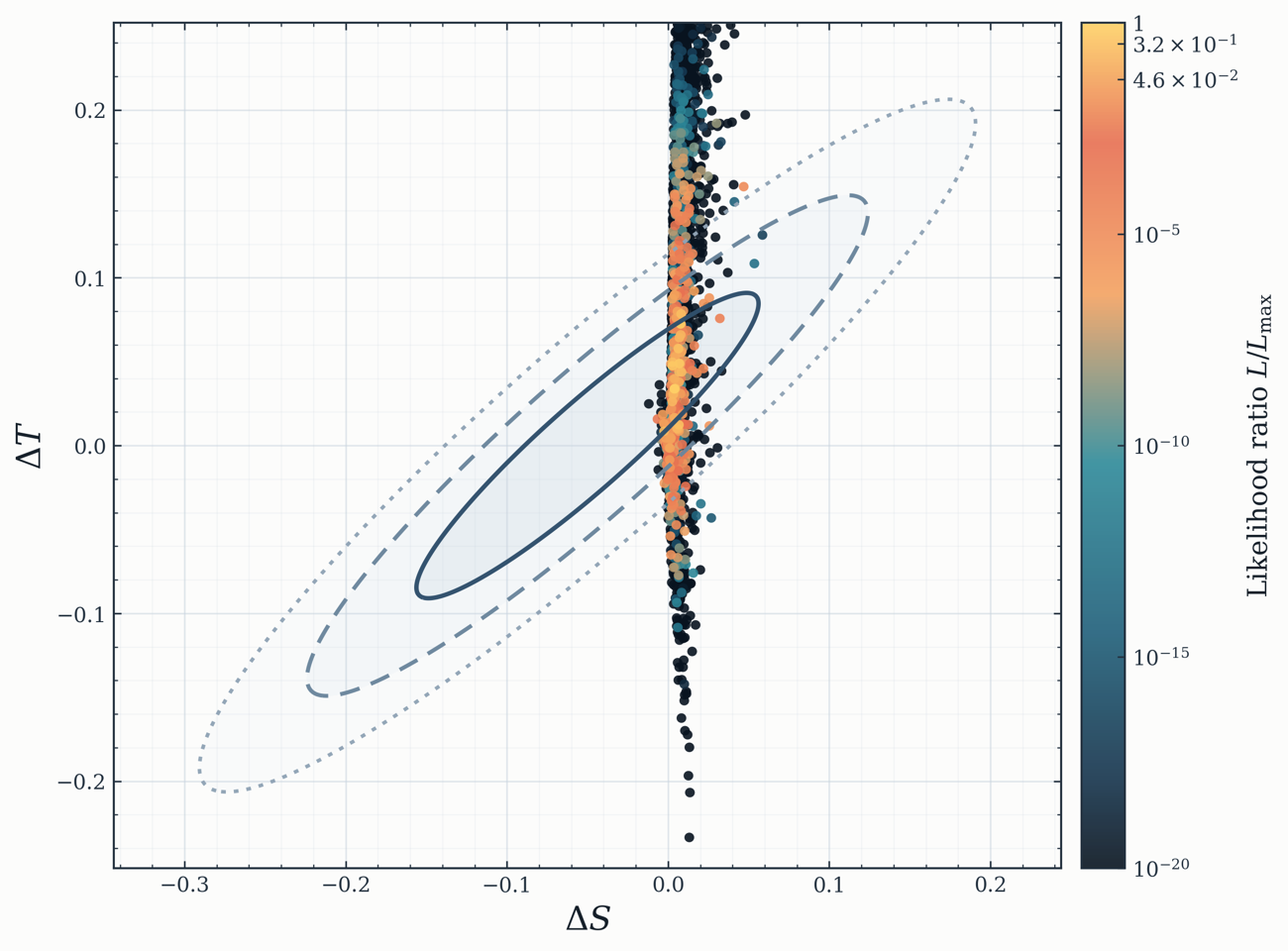}
\caption{Same as Fig. \ref{f2} but for IO.}
\label{f5}
\end{figure}

\begin{figure}[H]
\centering
\includegraphics[width=0.32\linewidth]{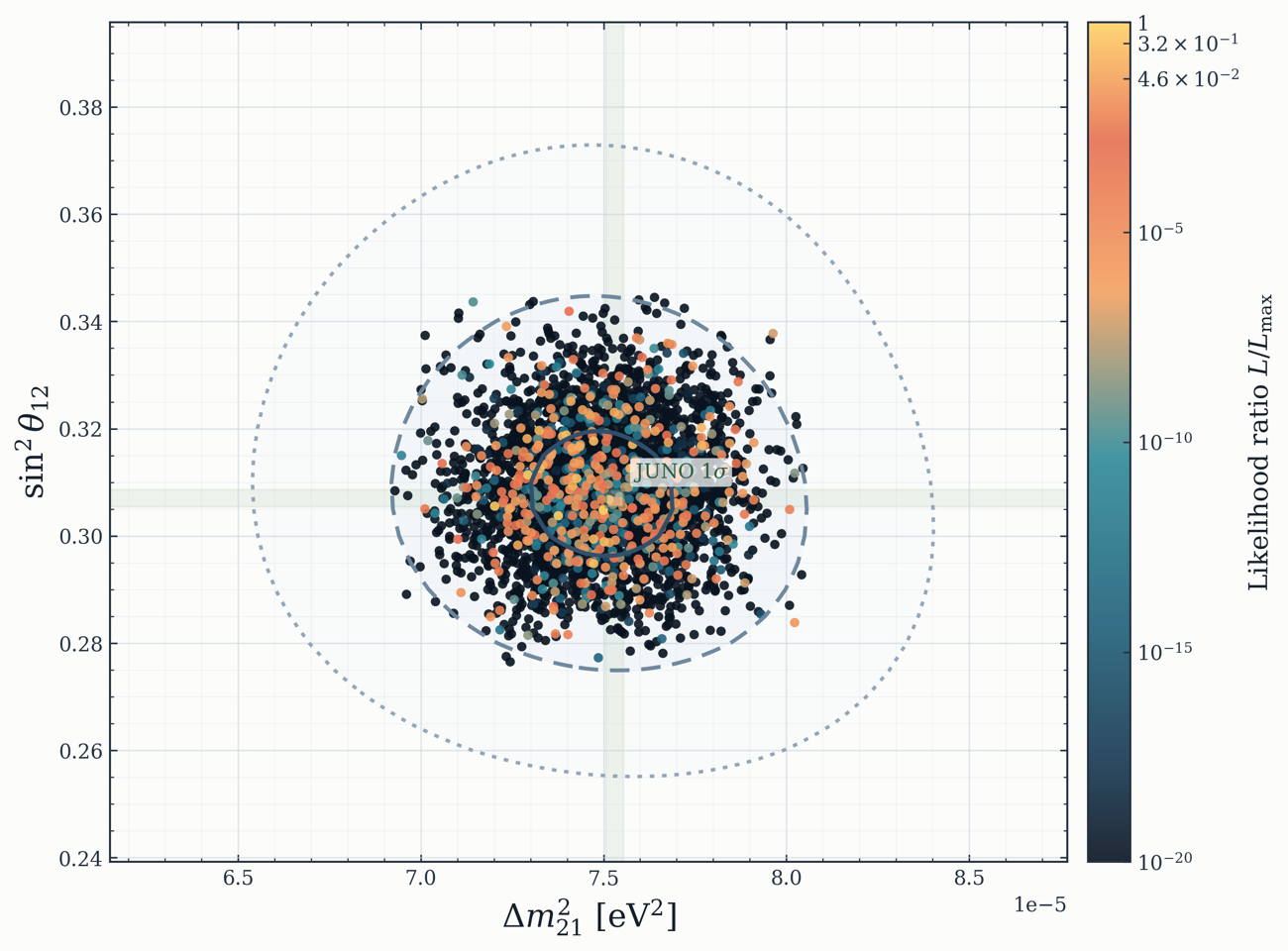}
\includegraphics[width=0.32\linewidth]{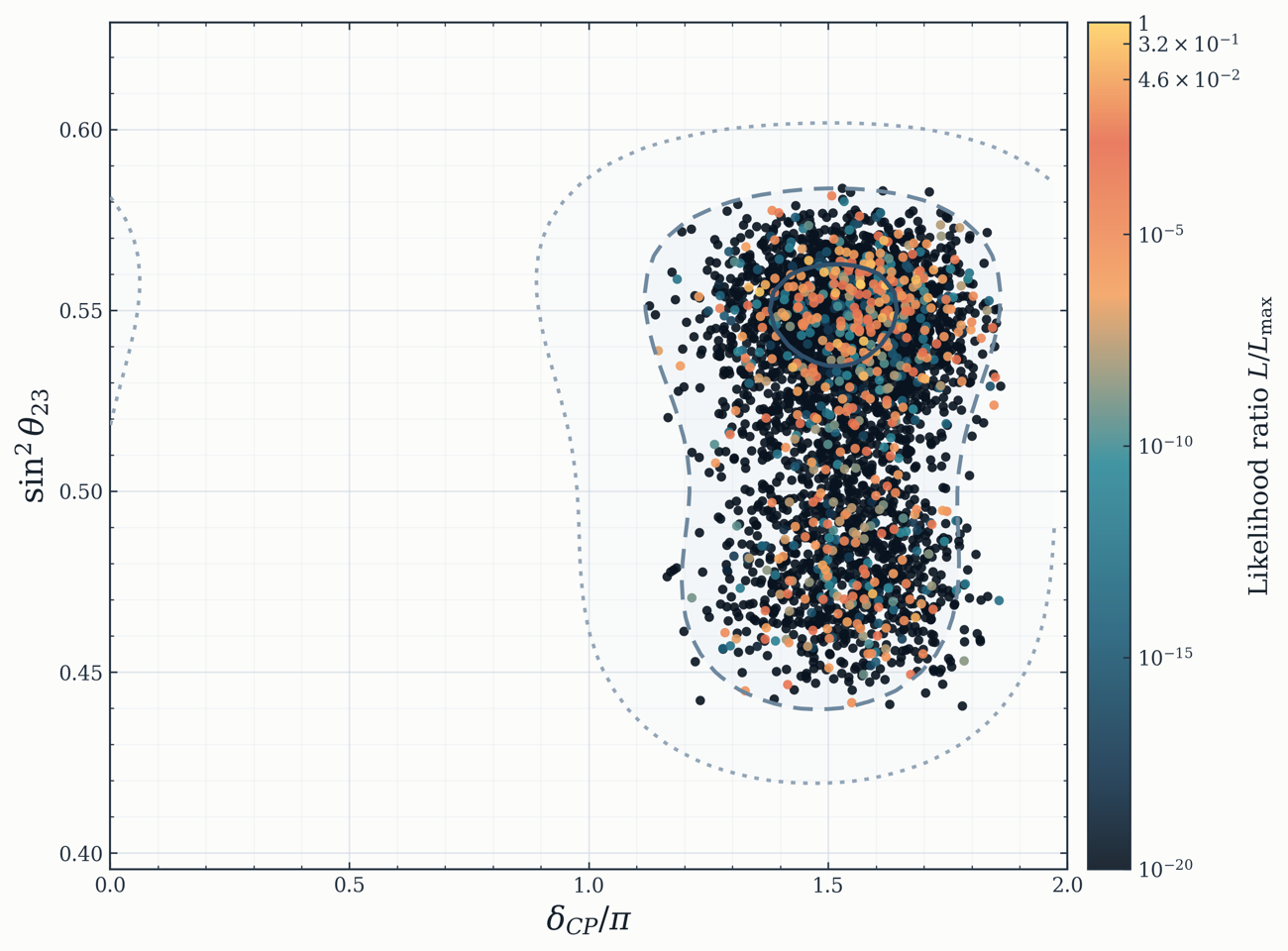}
\includegraphics[width=0.32\linewidth]{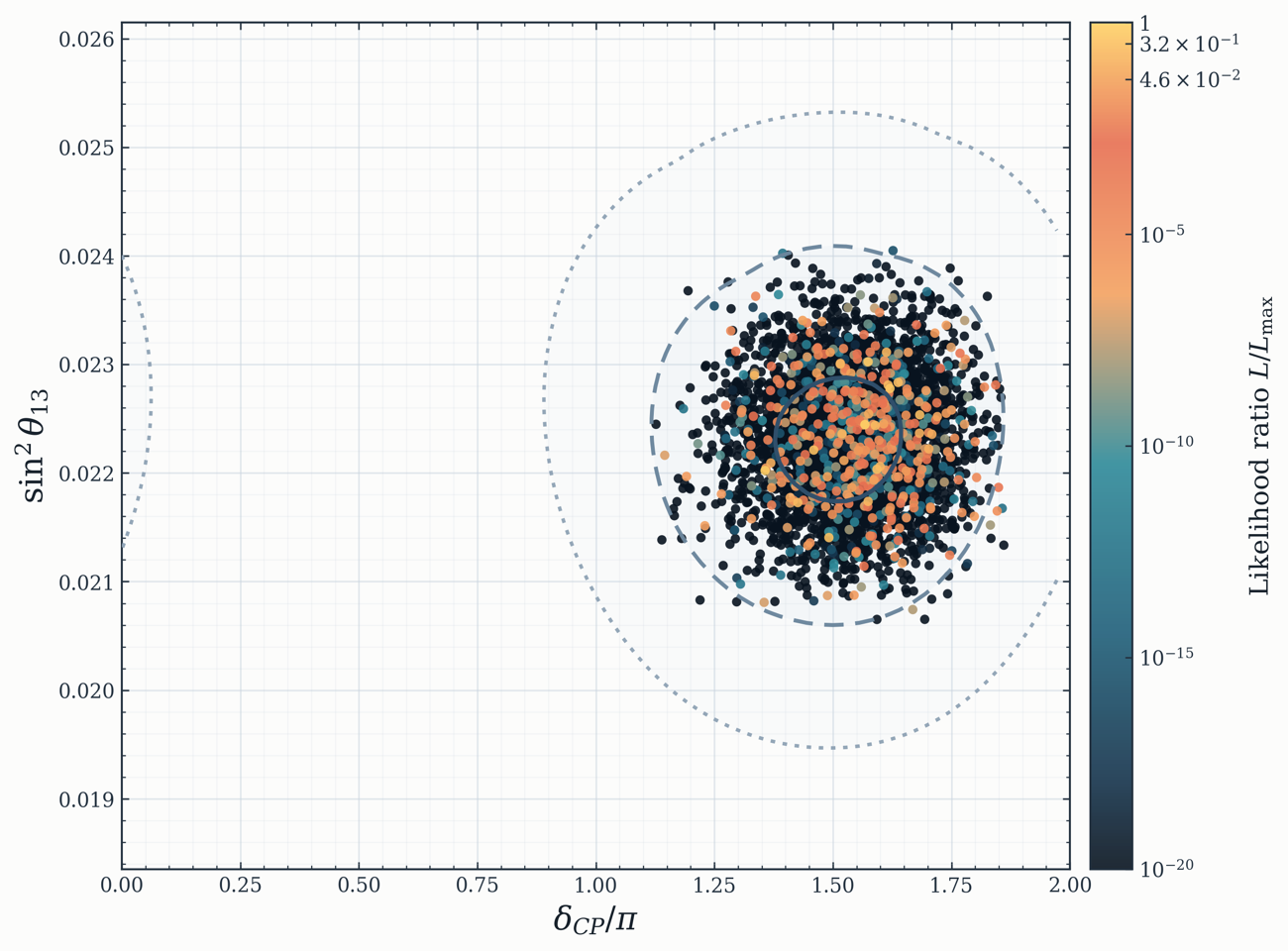}
\includegraphics[width=0.32\linewidth]{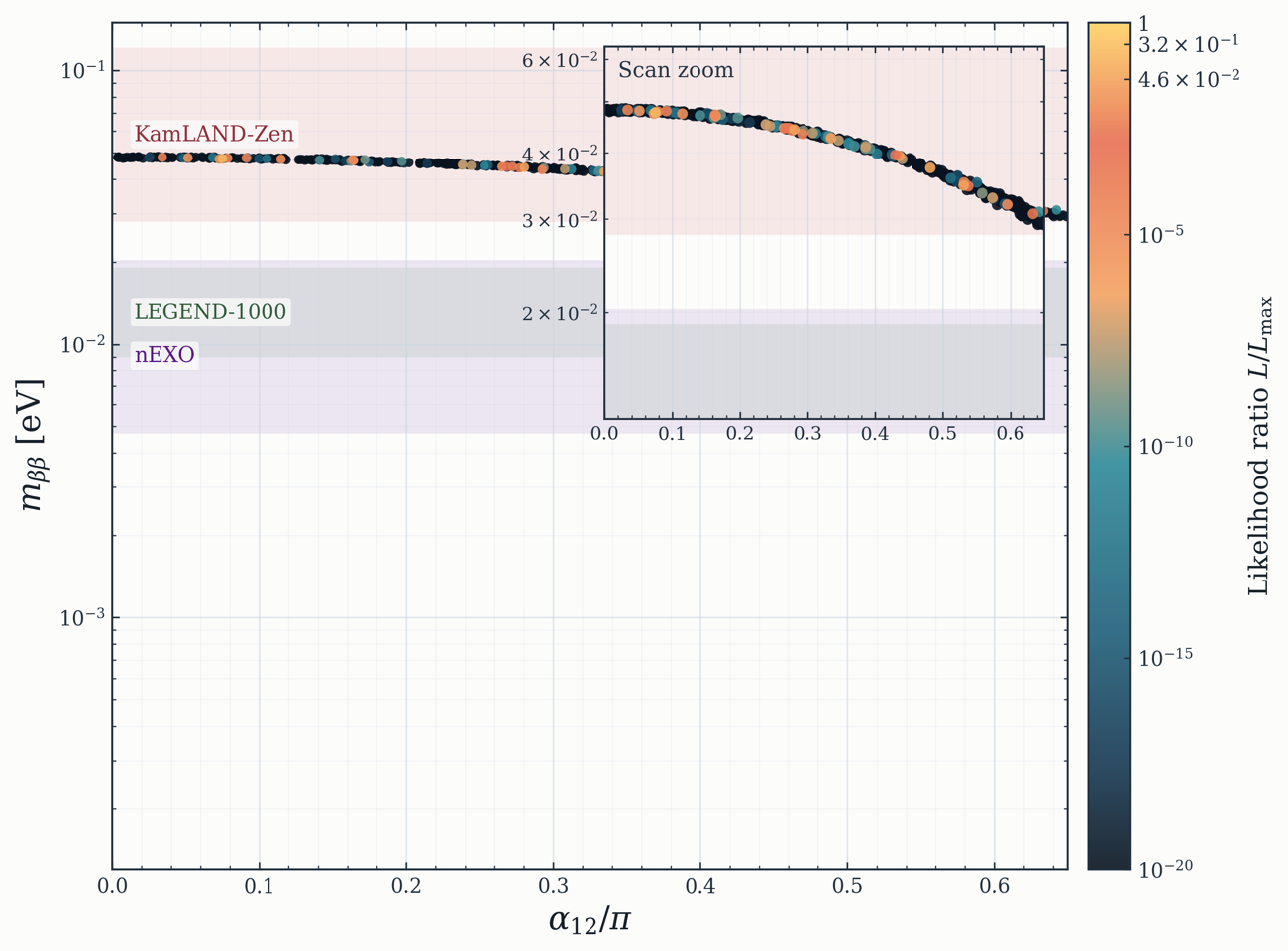}
\includegraphics[width=0.32\linewidth]{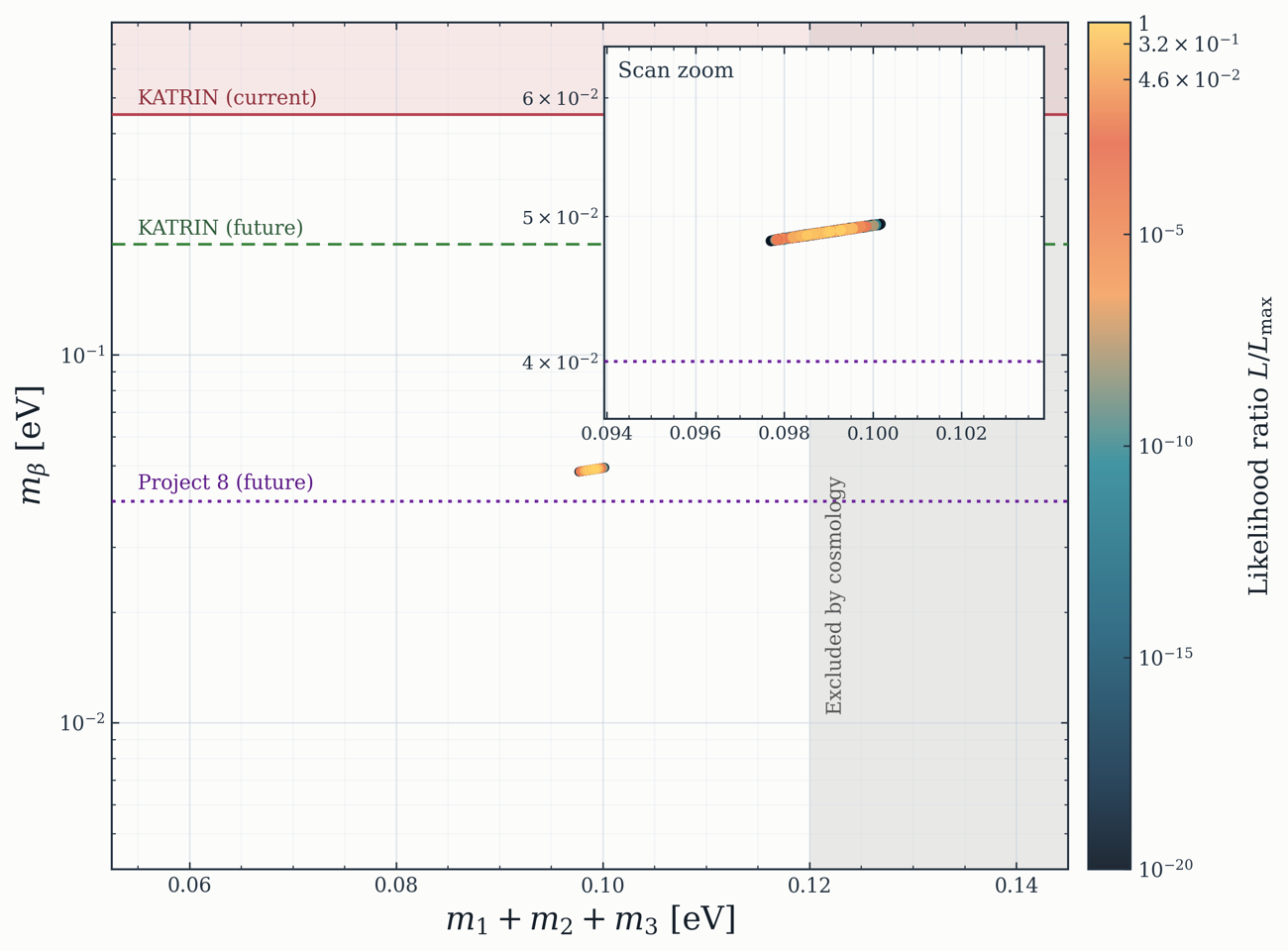}
\caption{Same as Fig. \ref{f3} but for IO.}
\label{f6}
\end{figure}

\begin{figure}[H]
\centering
\includegraphics[width=0.32\textwidth]{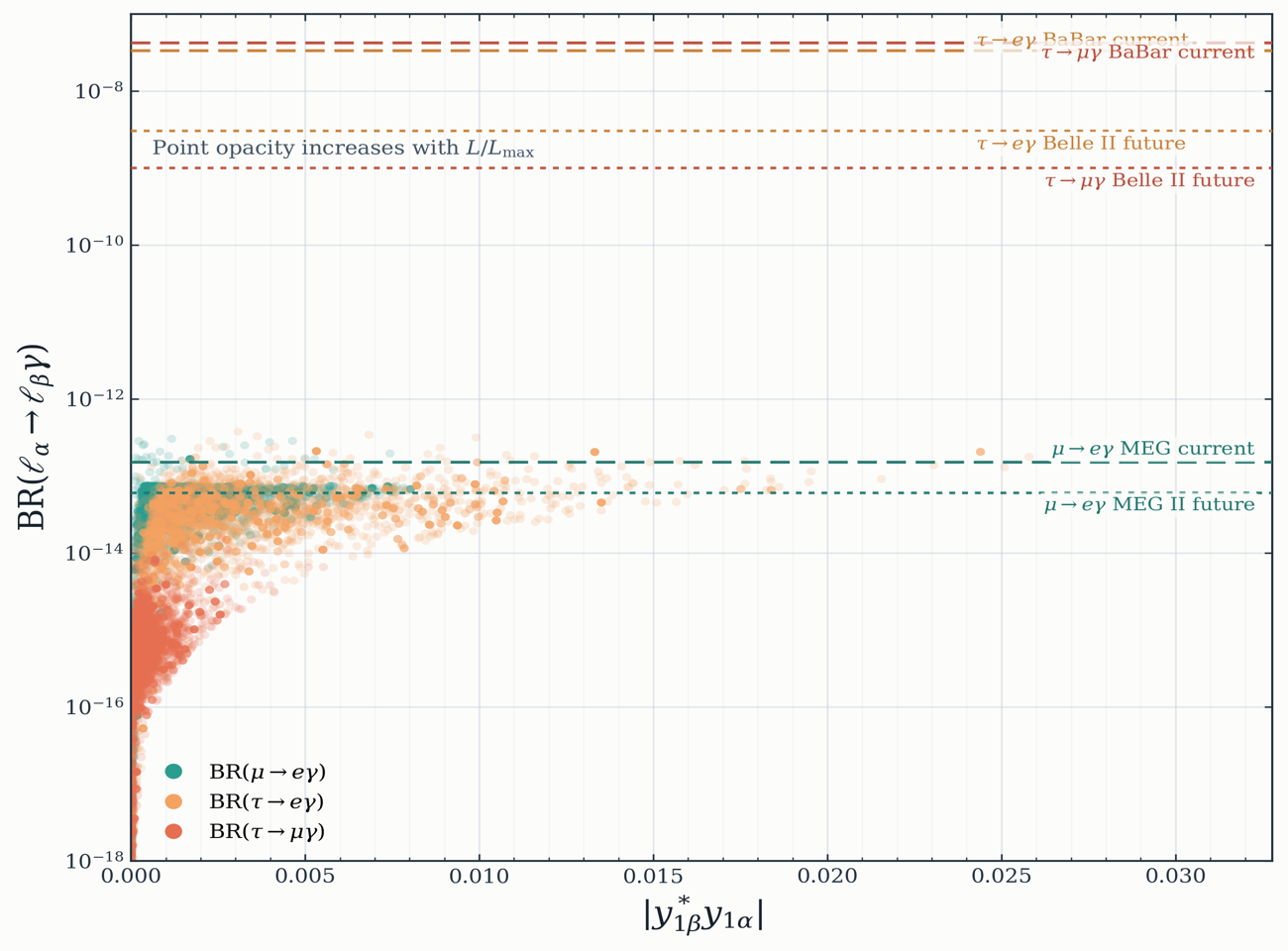}
\includegraphics[width=0.32\linewidth]{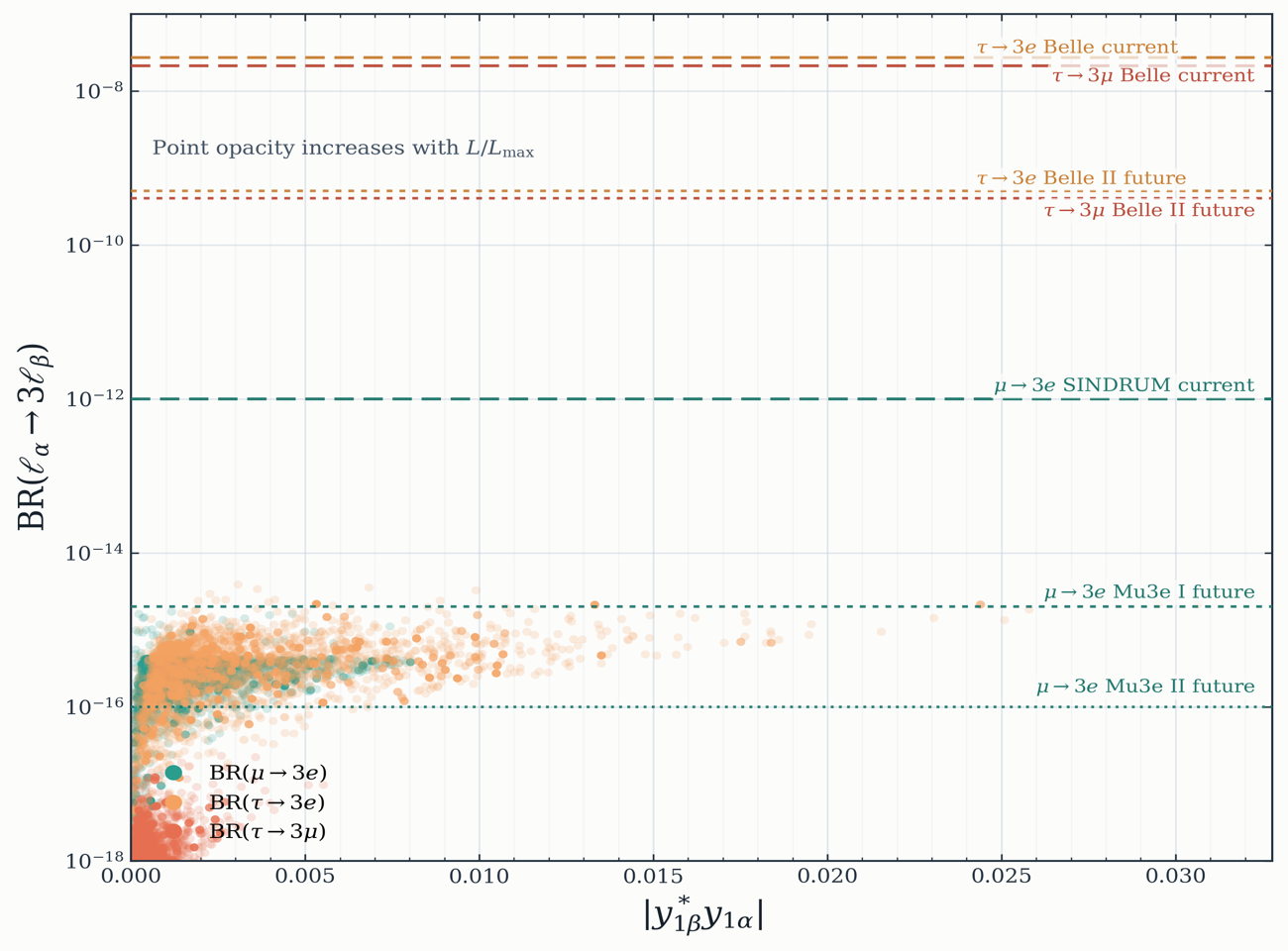}
\includegraphics[width=0.32\linewidth]{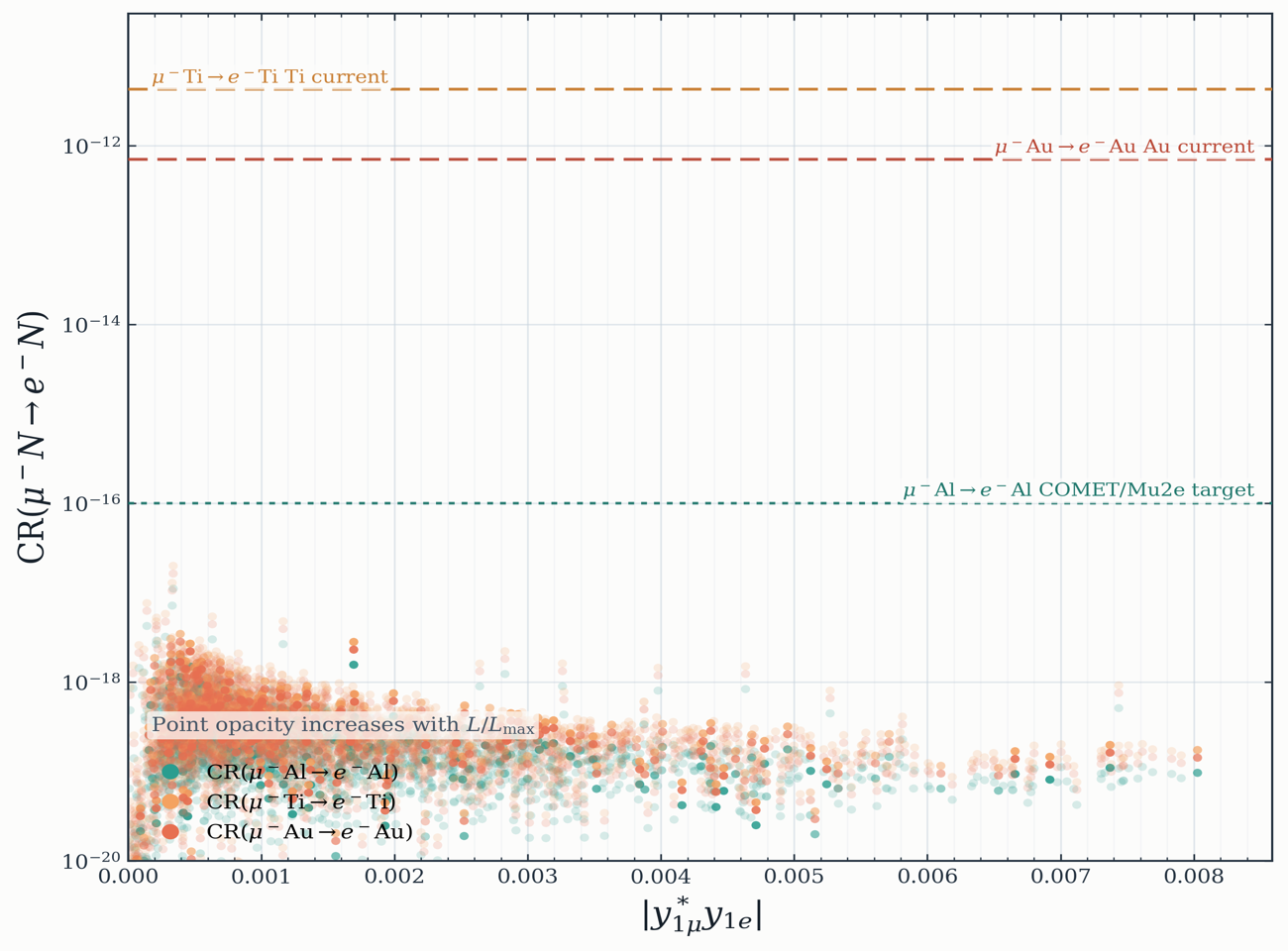}
\caption{Same as Fig. \ref{f4} but for IO.}
\label{f7}
\end{figure}

\begin{figure}[H]
\centering
\includegraphics[width=0.32\textwidth]{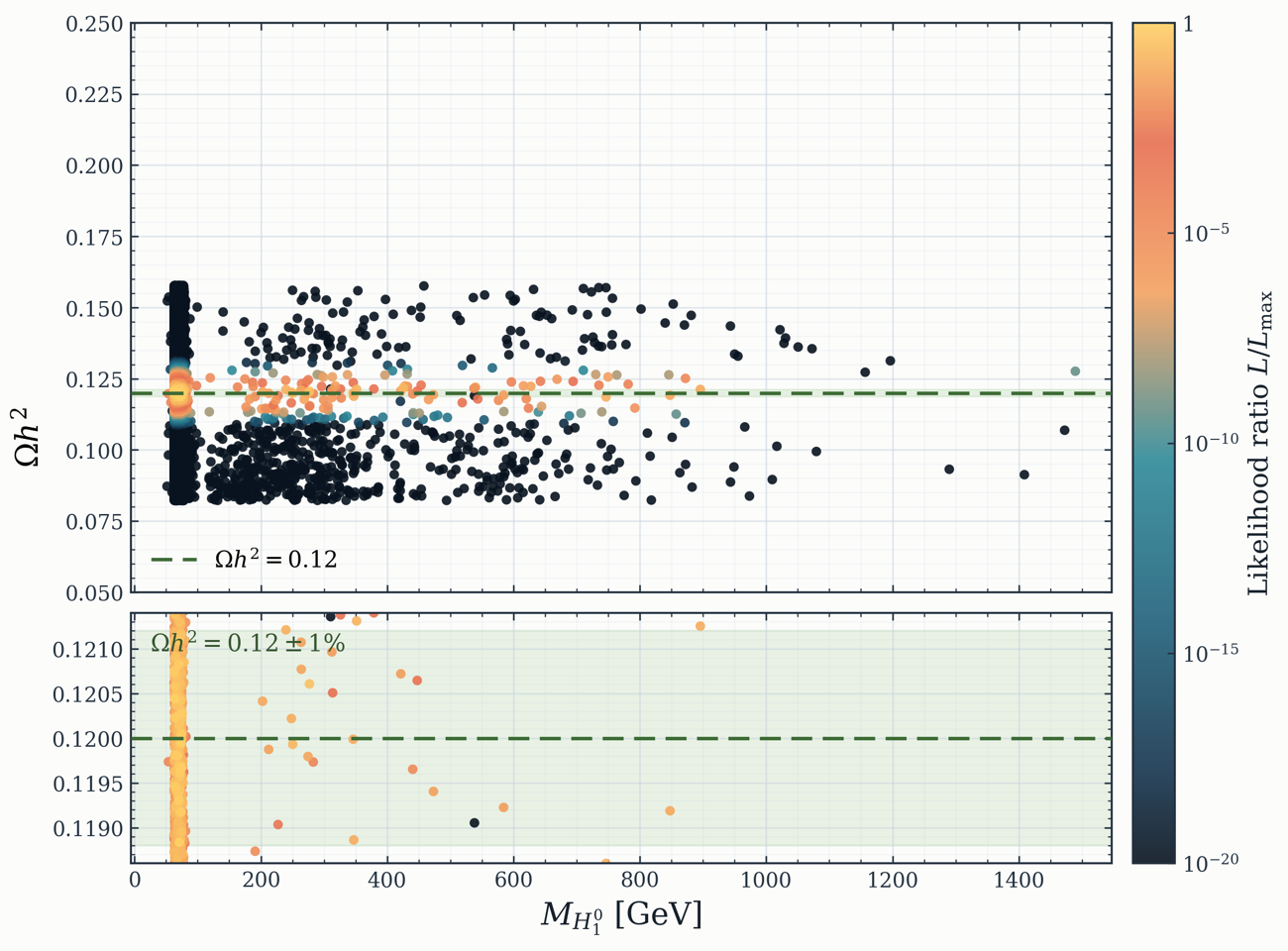}
\includegraphics[width=0.32\linewidth]{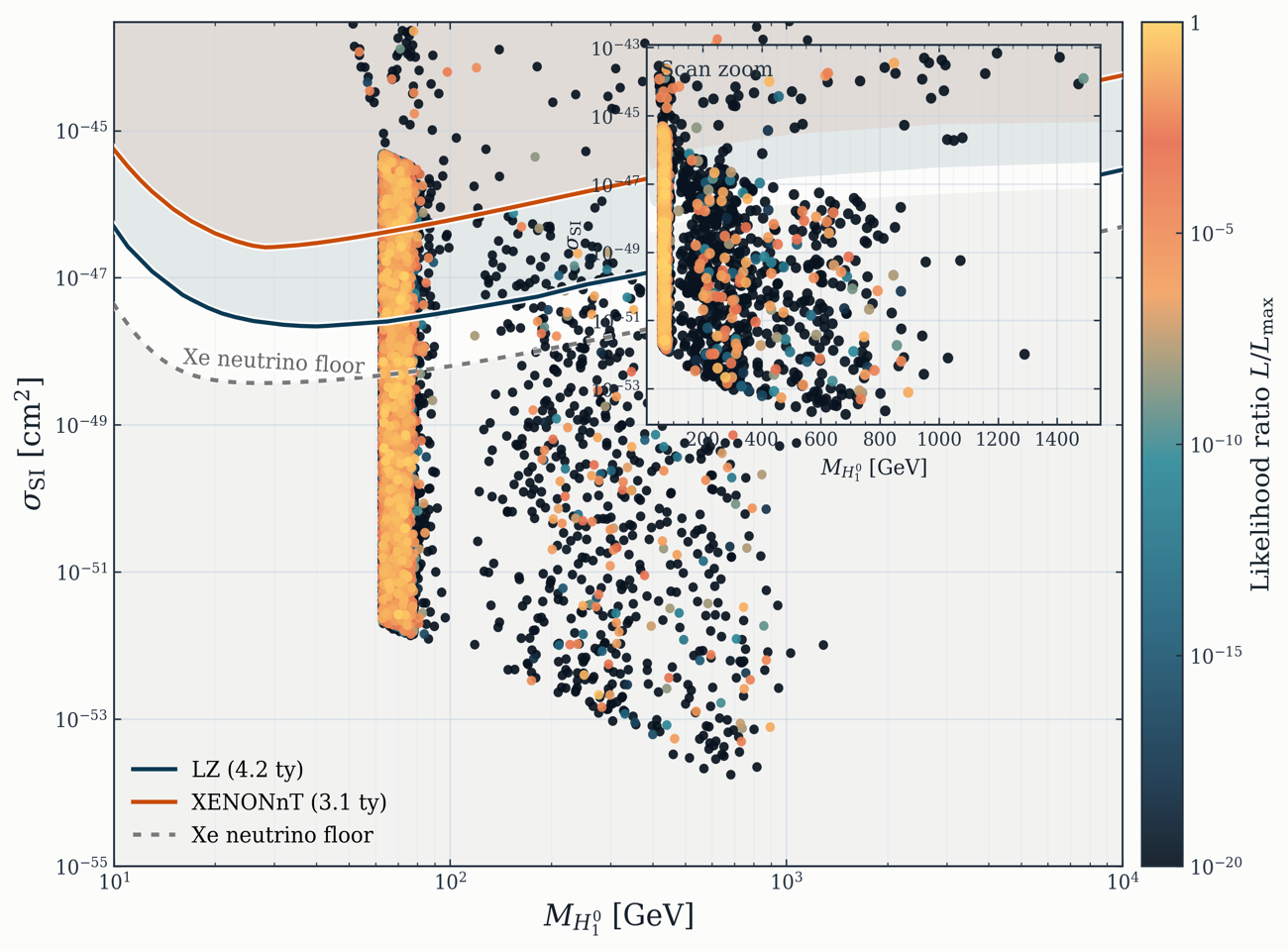}
\includegraphics[width=0.32\linewidth]{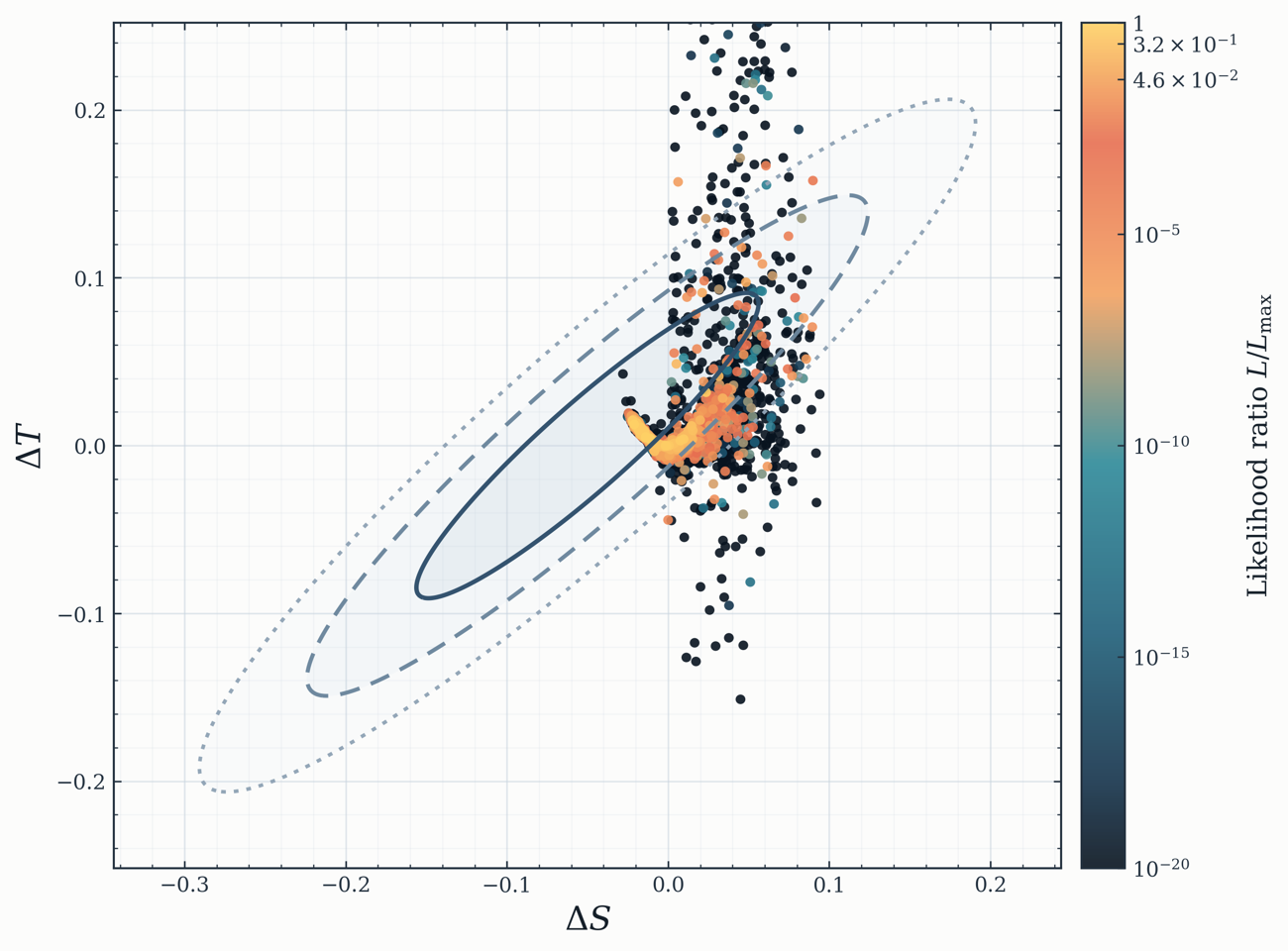}
\caption{Scalar DM with NO. Left panel: relic density as a function of $m_{H_1^0}$. Middle panel: spin-independent DD cross section as a function of $m_{H_1^0}$. Right panel: allowed points in the $(\Delta S,\Delta T)$ plane. The color scale and experimental references follow the conventions of Fig.~\ref{f2}.}
\label{f8}
\end{figure}

\begin{figure}[H]
\centering
\includegraphics[width=0.32\linewidth]{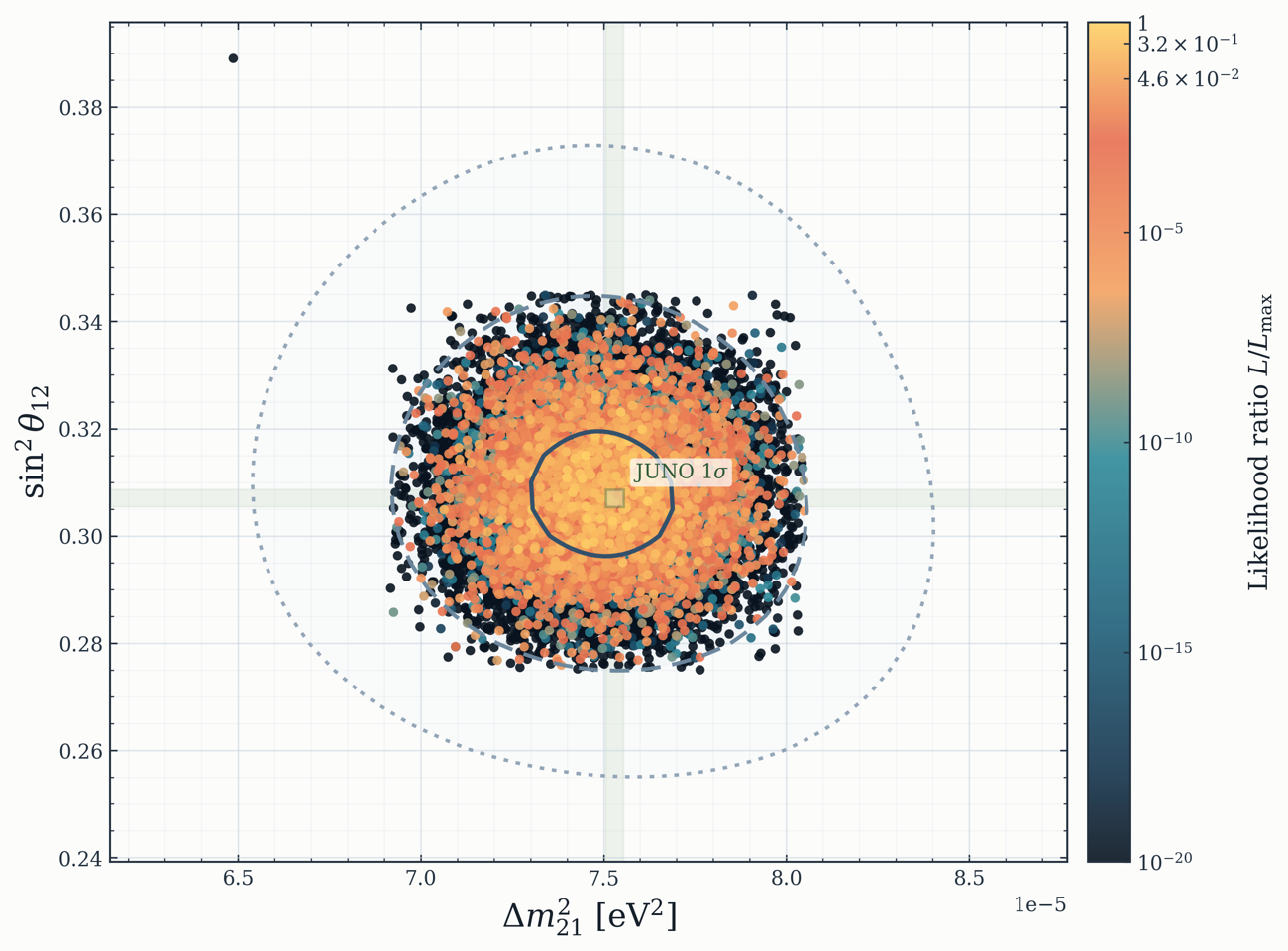}
\includegraphics[width=0.32\linewidth]{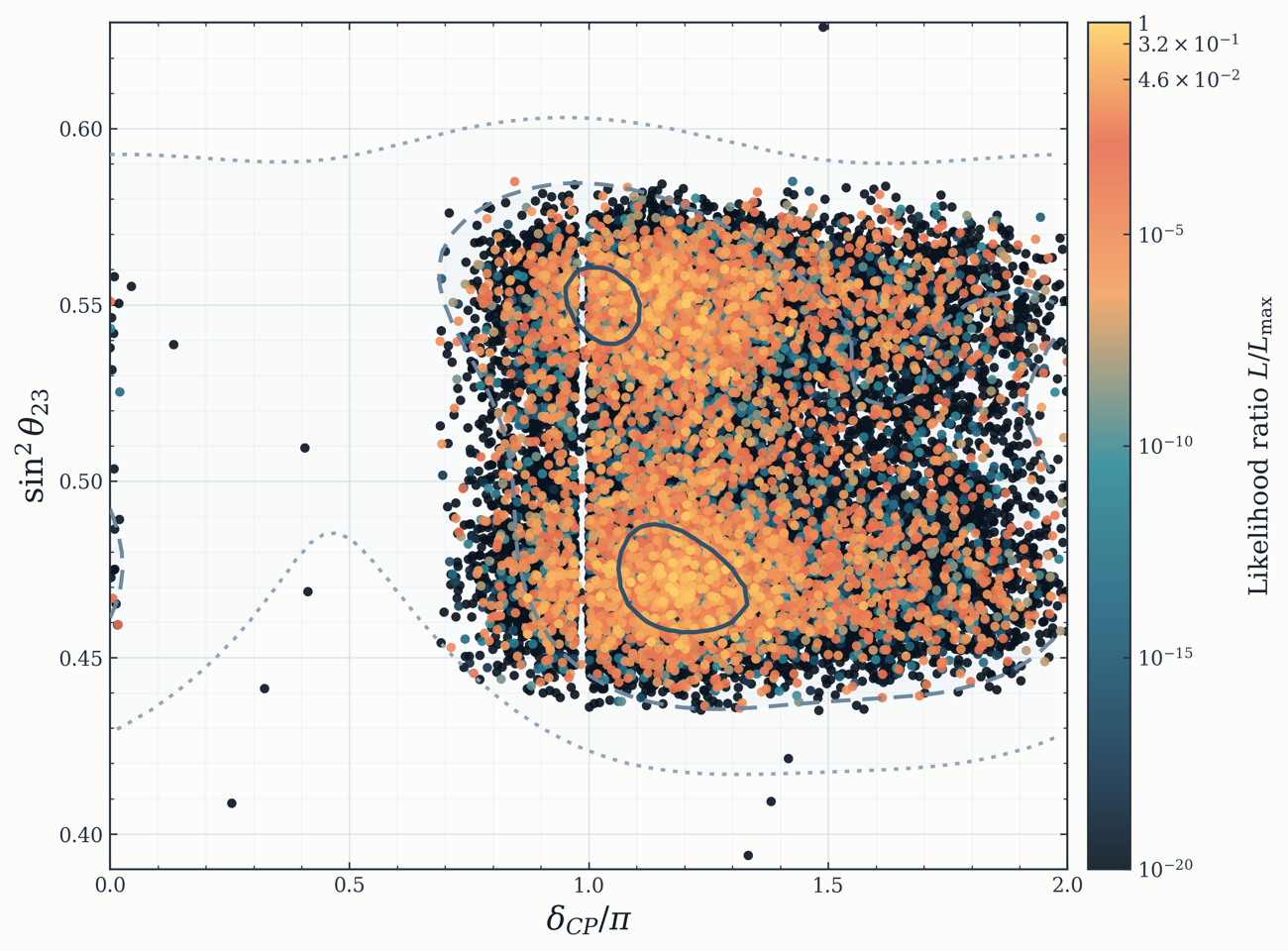}
\includegraphics[width=0.32\linewidth]{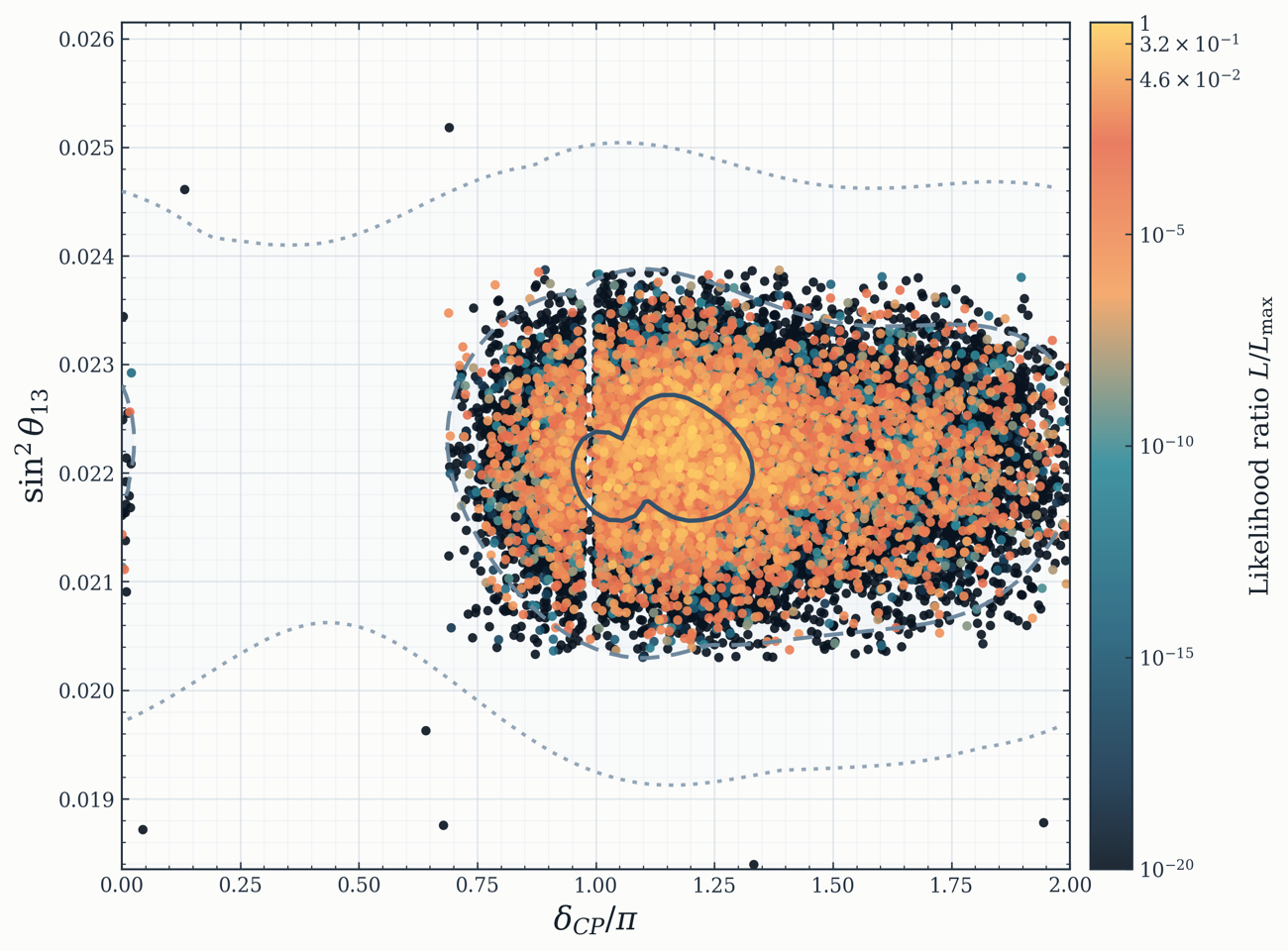}
\includegraphics[width=0.32\linewidth]{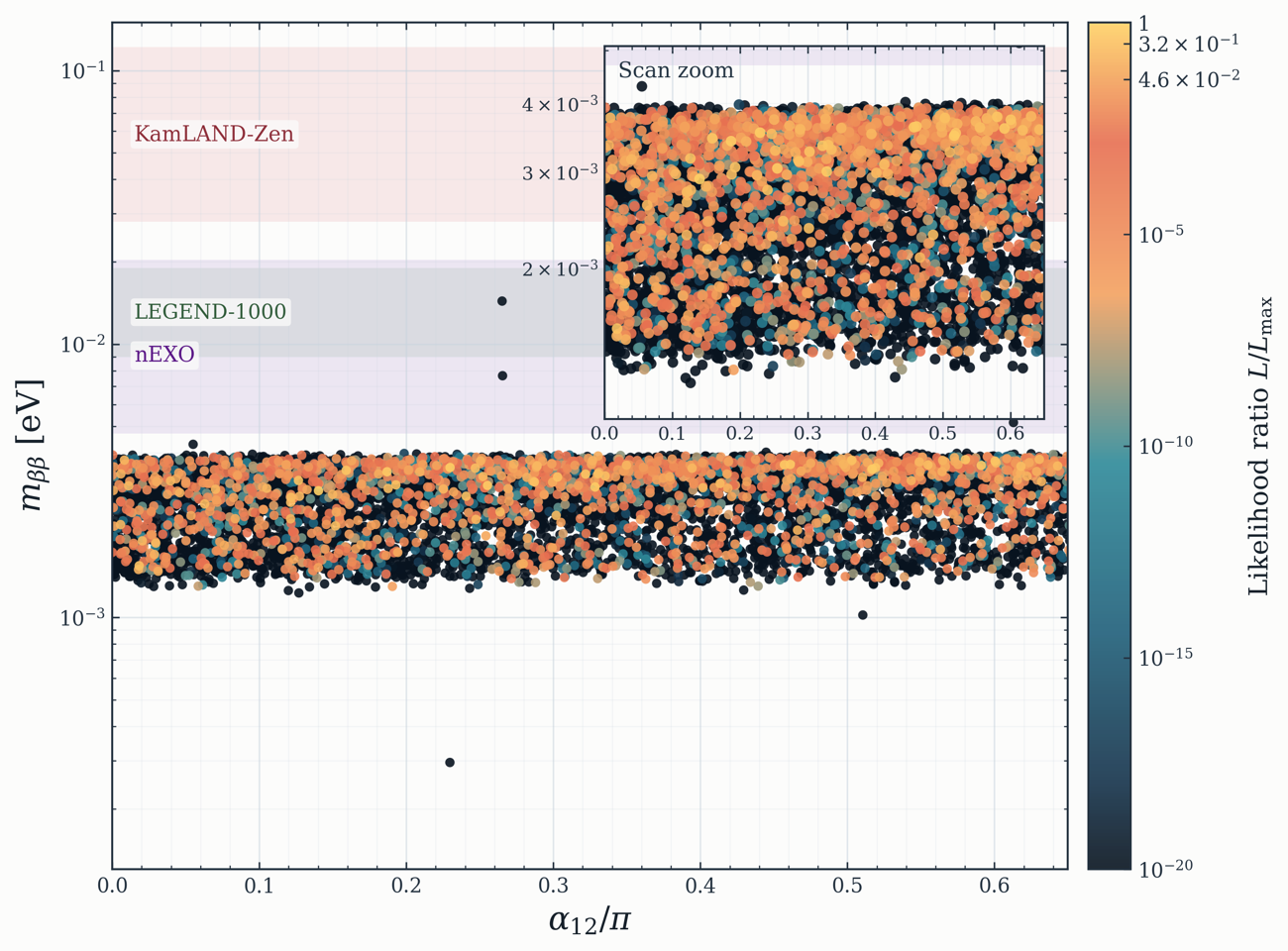}
\includegraphics[width=0.32\linewidth]{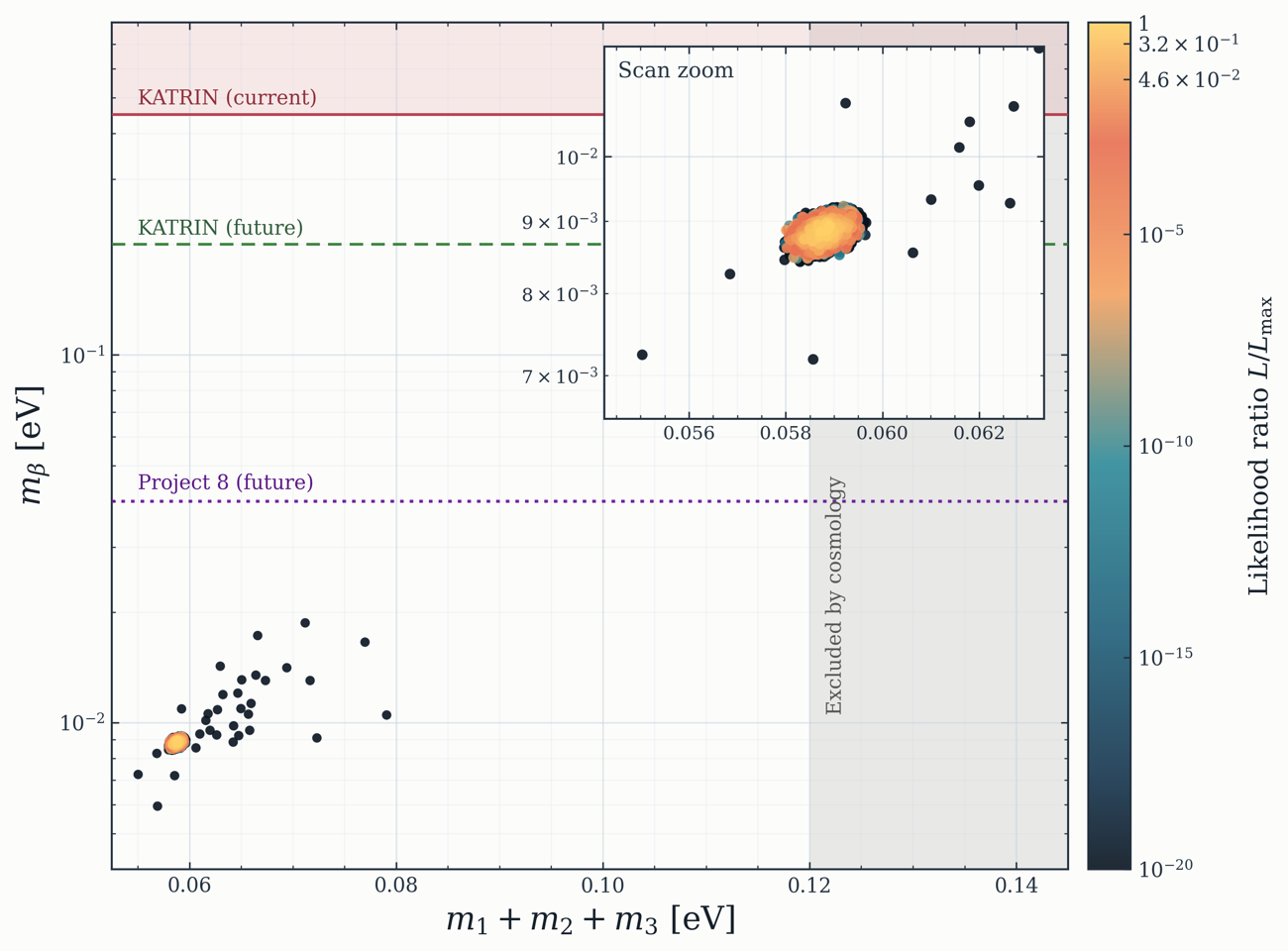}
\caption{Neutrino sector predictions for scalar DM with NO. The panel order, color code and external references are the same as in Fig. \ref{f3}.}
\label{f9}
\end{figure}

\begin{figure}[H]
\centering
\includegraphics[width=0.32\textwidth]{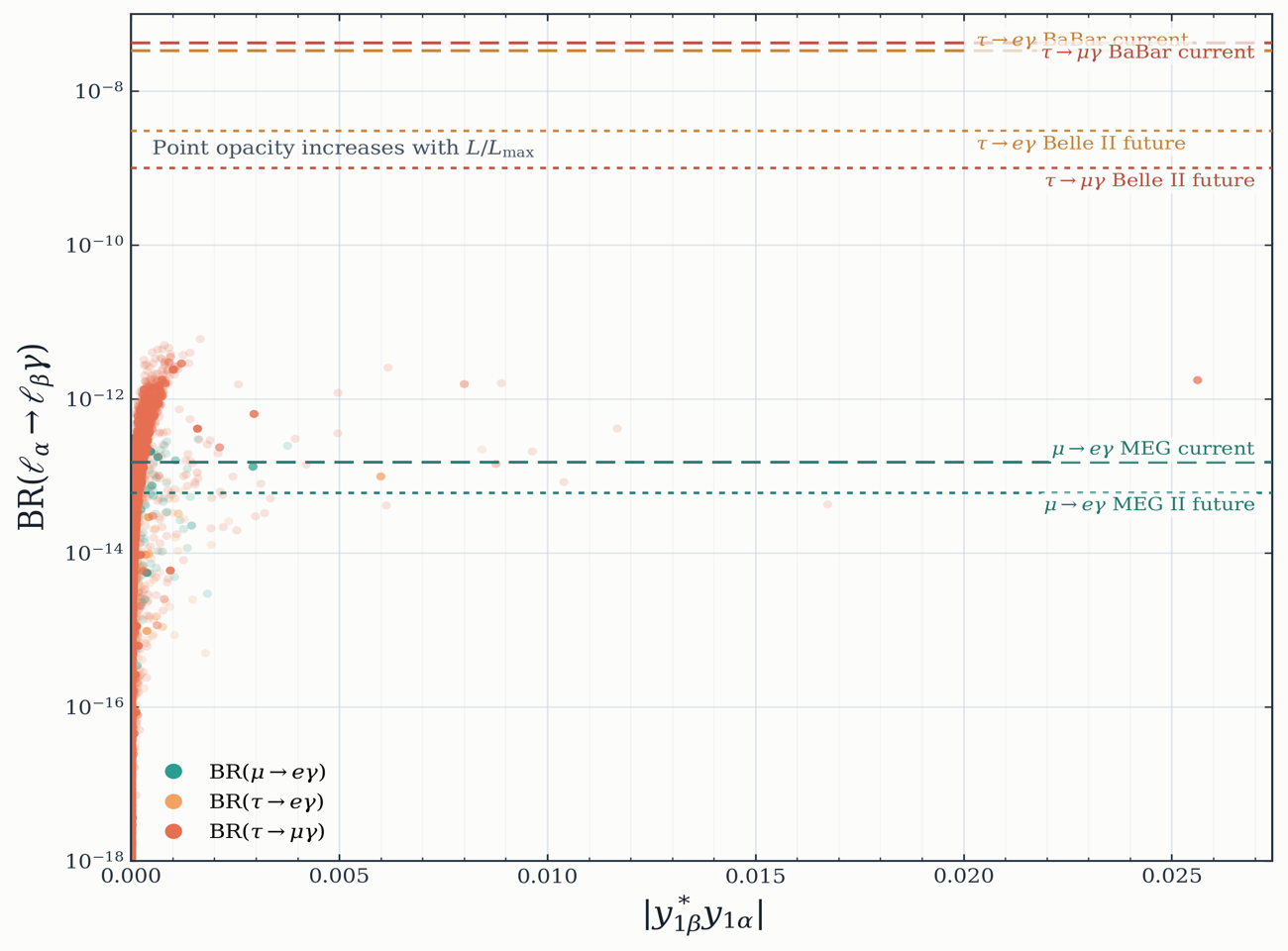}
\includegraphics[width=0.32\linewidth]{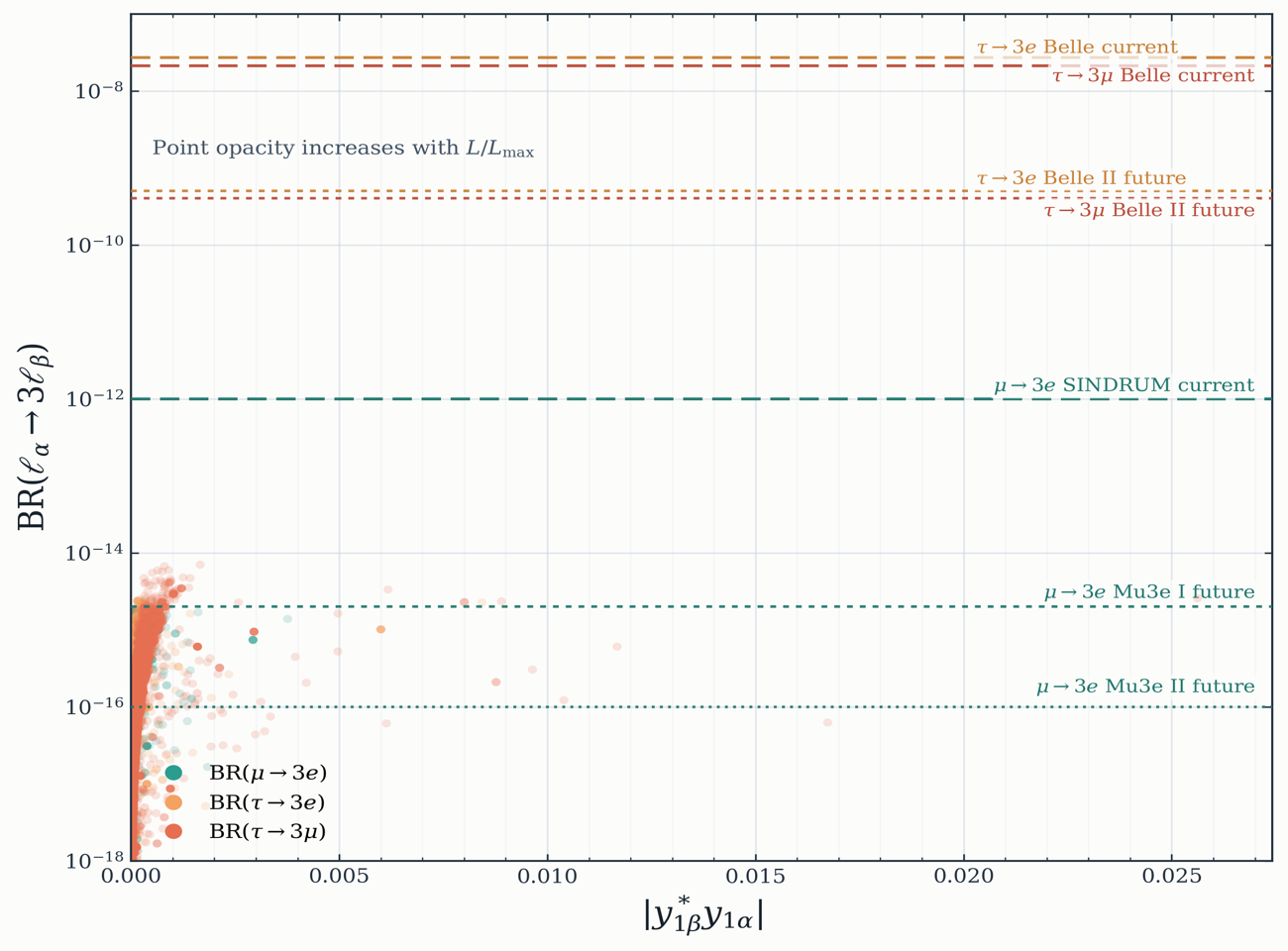}
\includegraphics[width=0.32\linewidth]{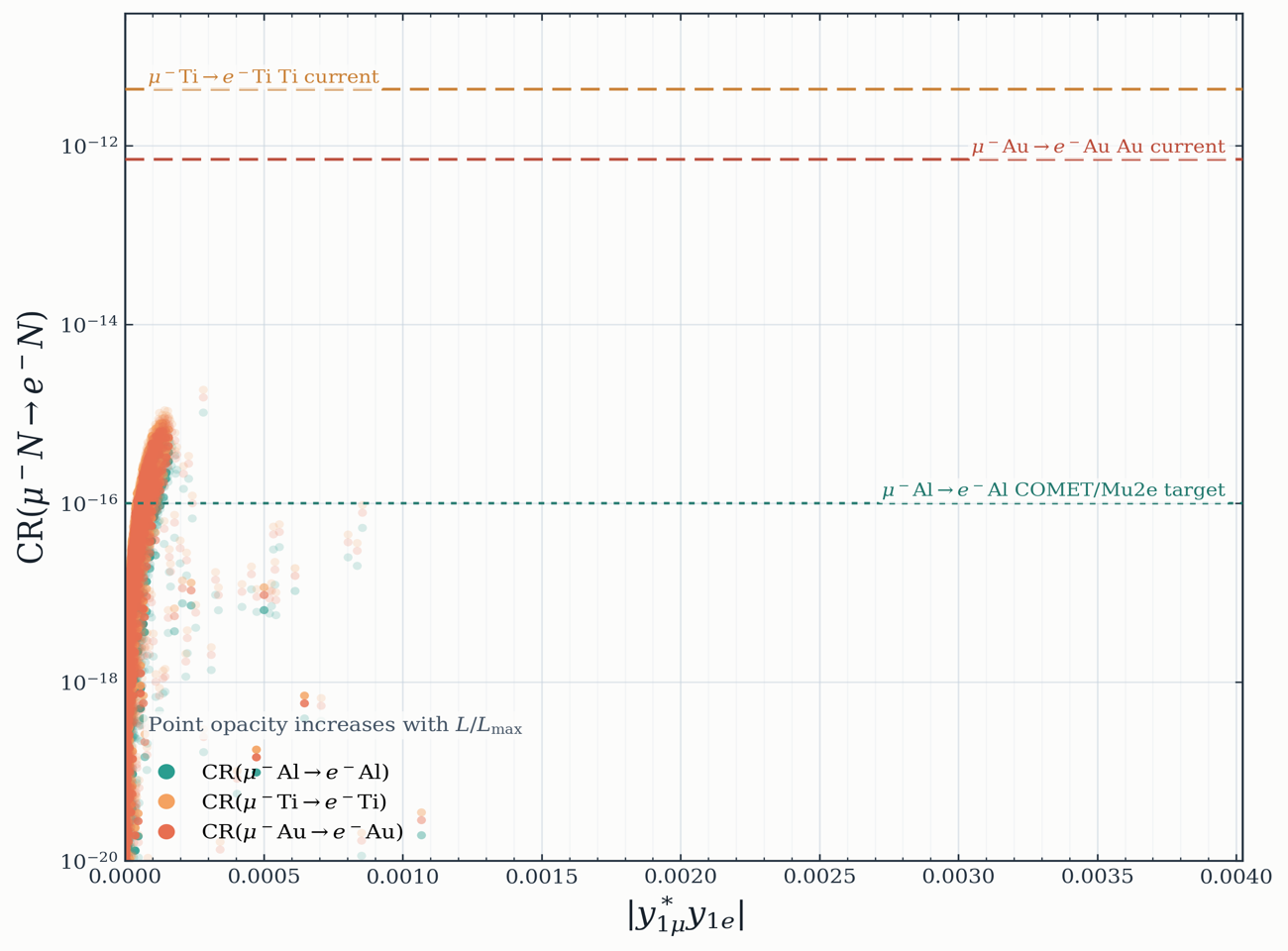}
\caption{cLFV predictions for scalar DM with NO. The panel order, color convention and experimental lines are the same as in Fig.~\ref{f4}.}
\label{f10}
\end{figure}

\begin{figure}[H]
\centering
\includegraphics[width=0.32\textwidth]{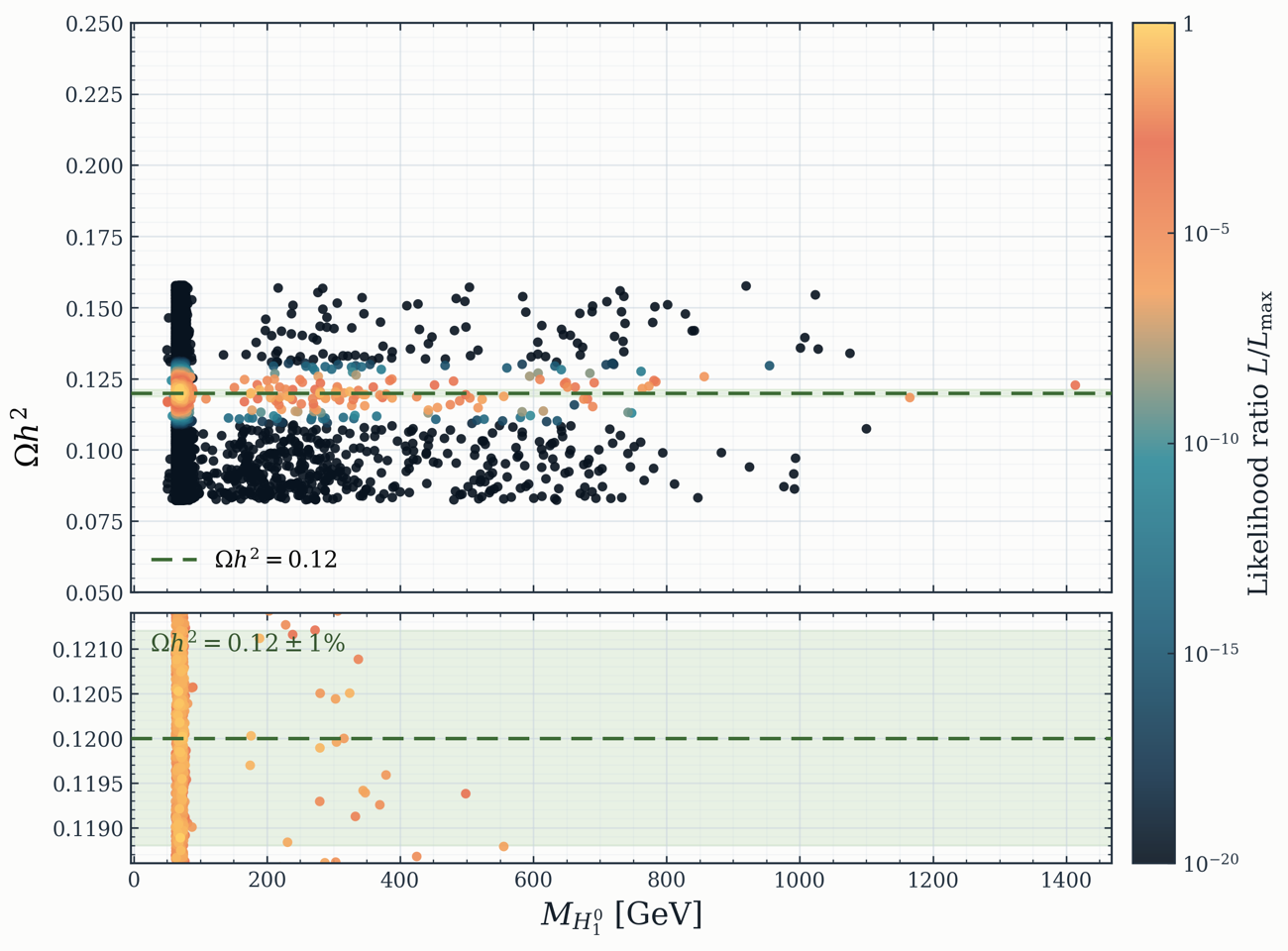}
\includegraphics[width=0.32\linewidth]{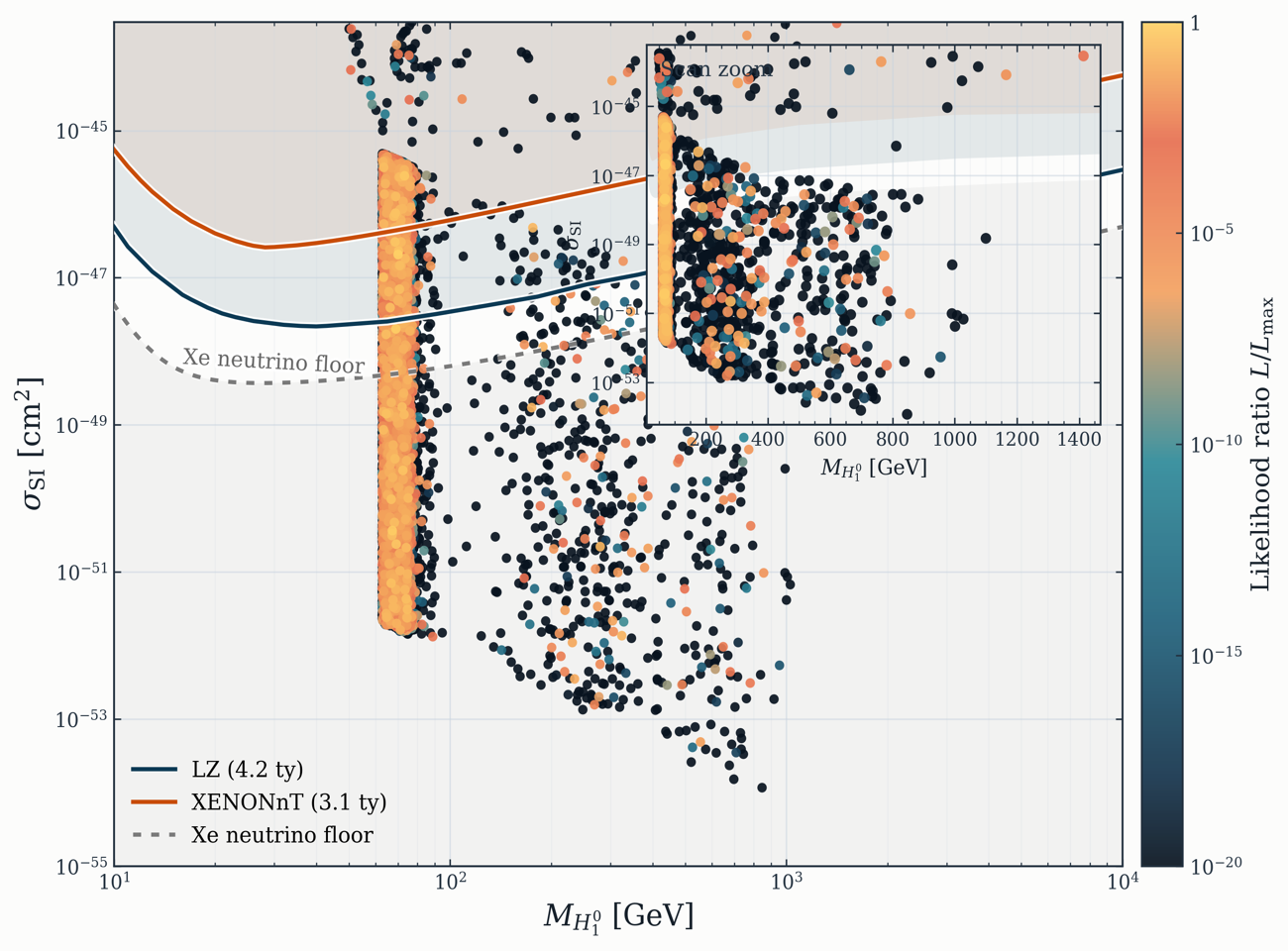}
\includegraphics[width=0.32\linewidth]{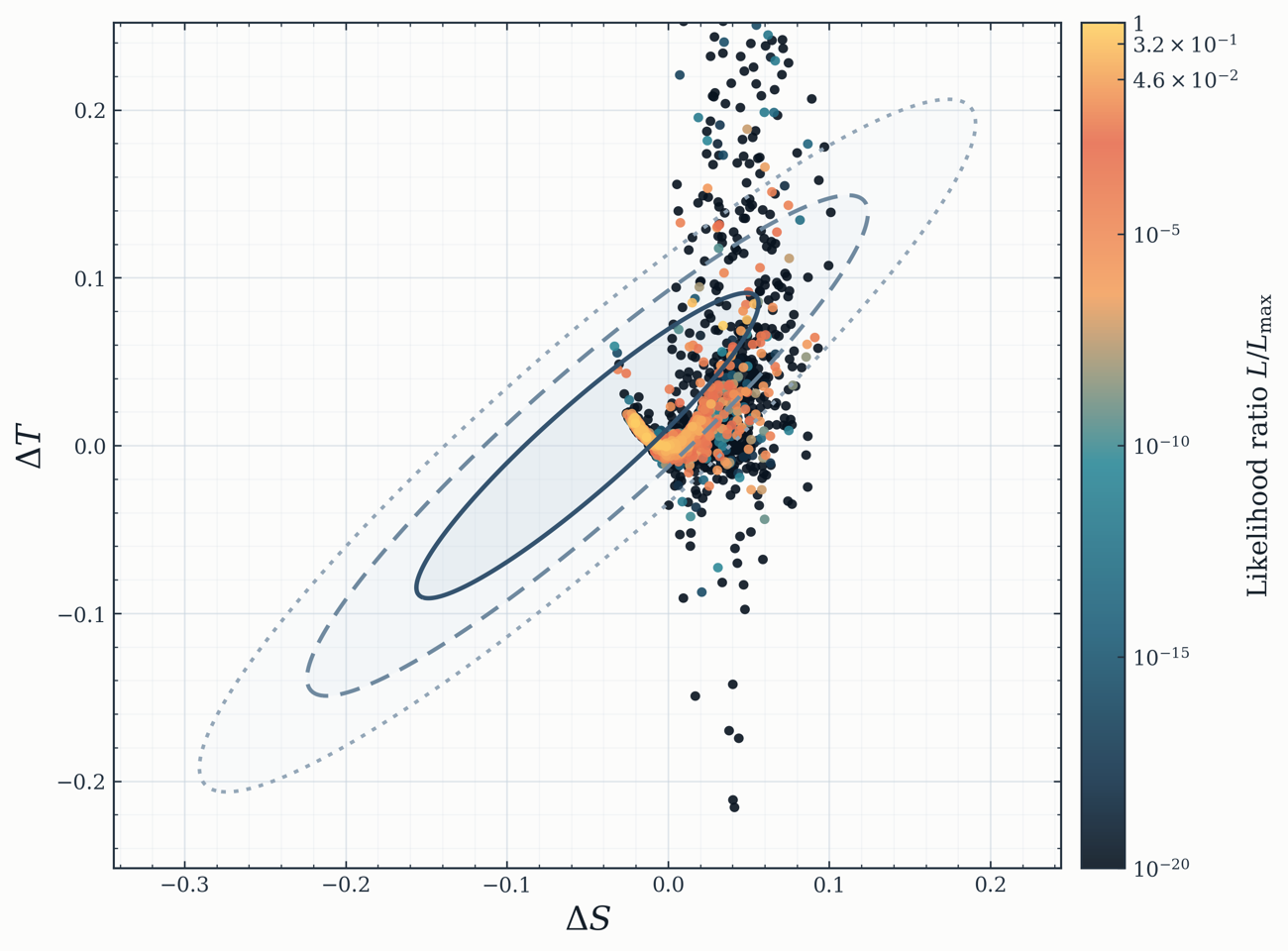}
\caption{Same as Fig. \ref{f8} but for IO.}
\label{f11}
\end{figure}

\begin{figure}[H]
\centering
\includegraphics[width=0.32\linewidth]{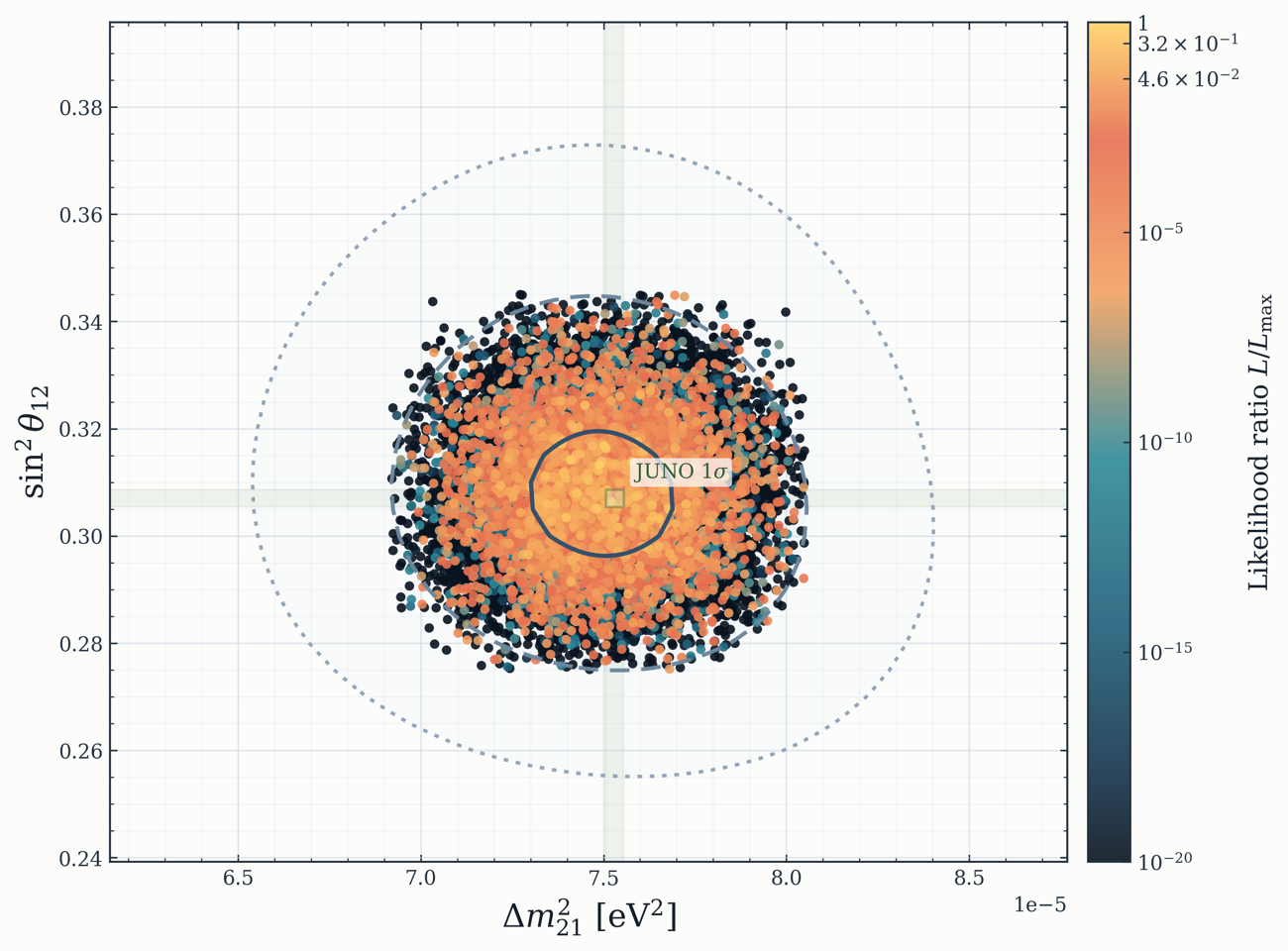}
\includegraphics[width=0.32\linewidth]{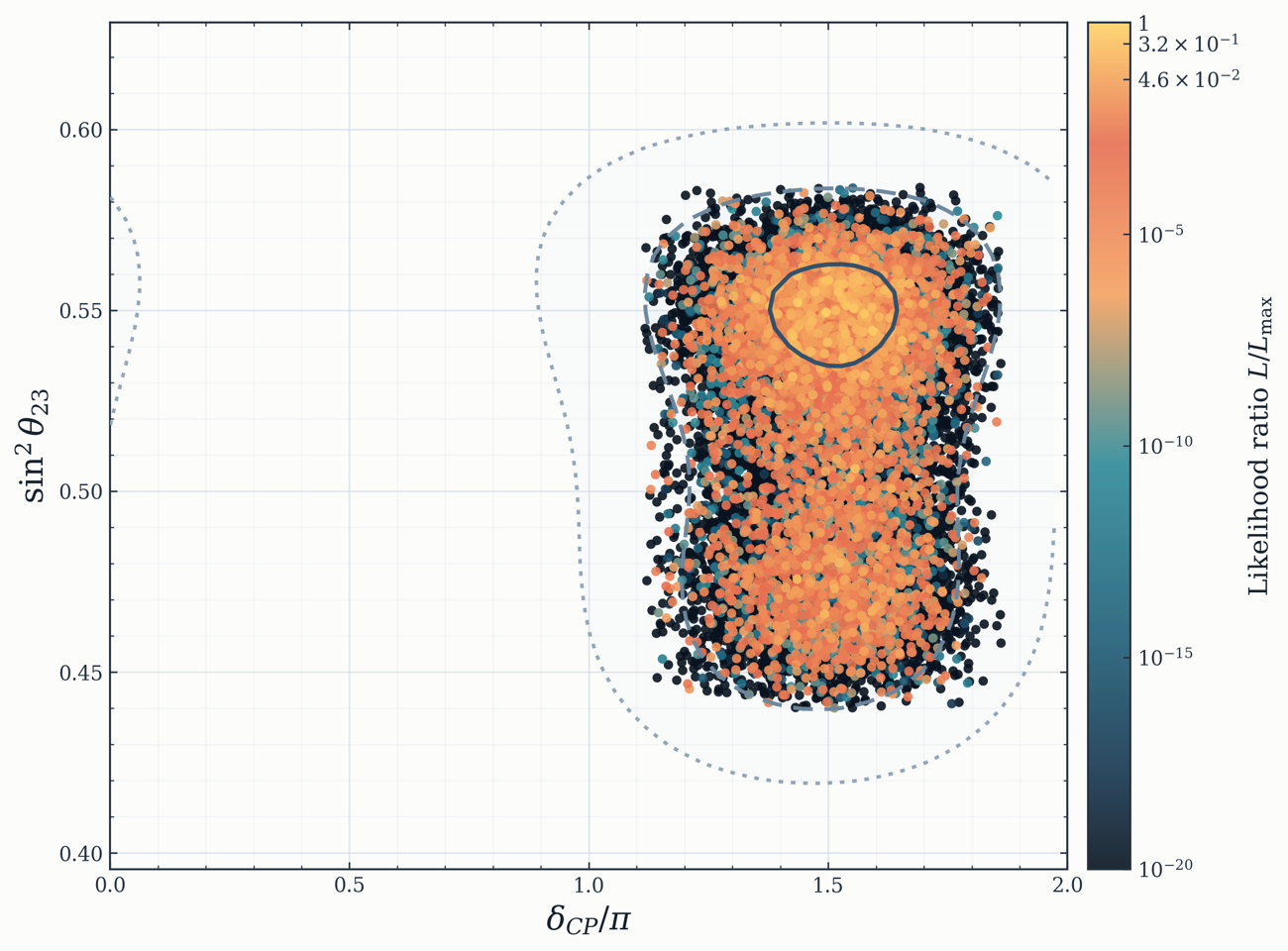}
\includegraphics[width=0.32\linewidth]{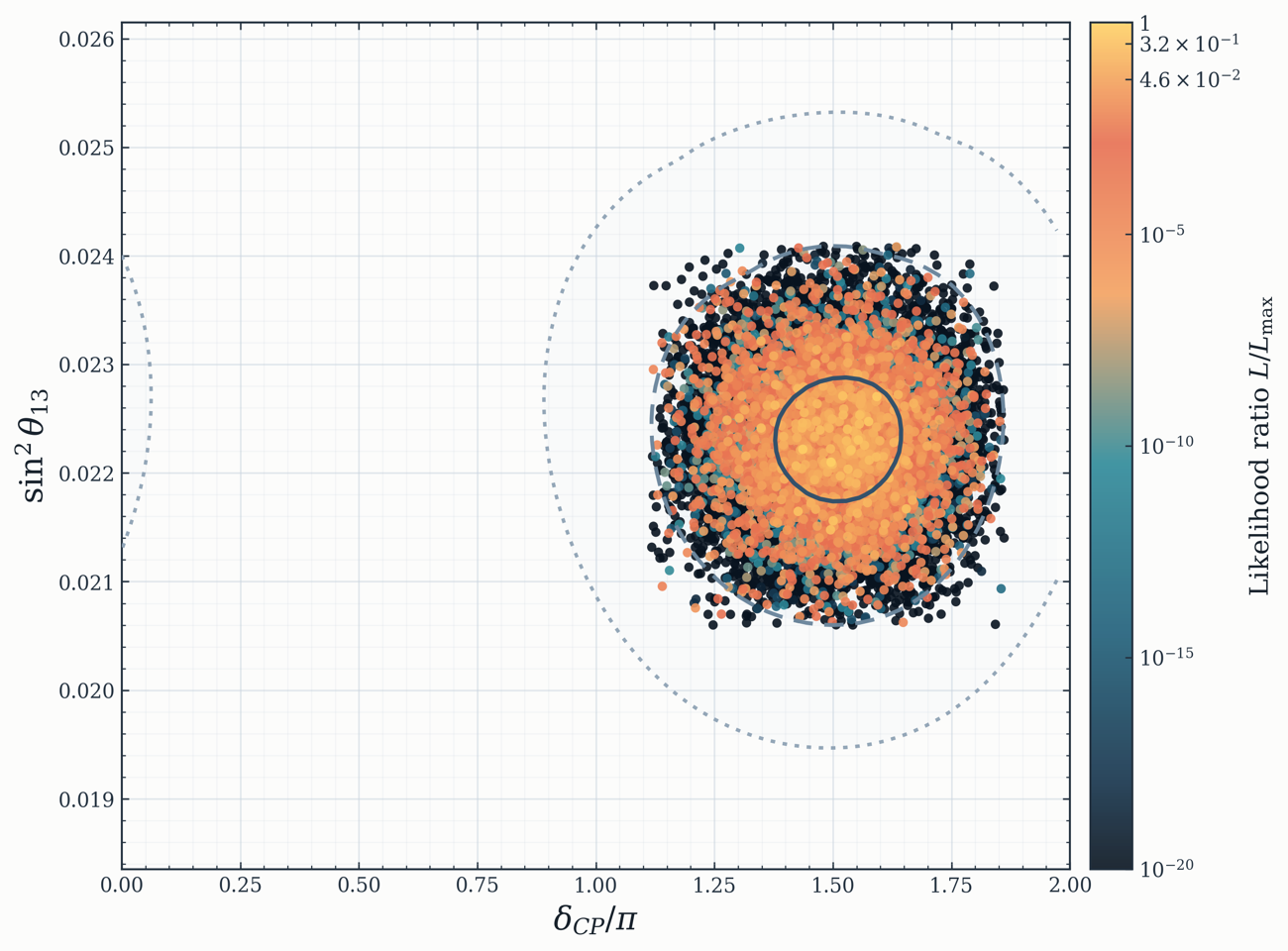}
\includegraphics[width=0.32\linewidth]{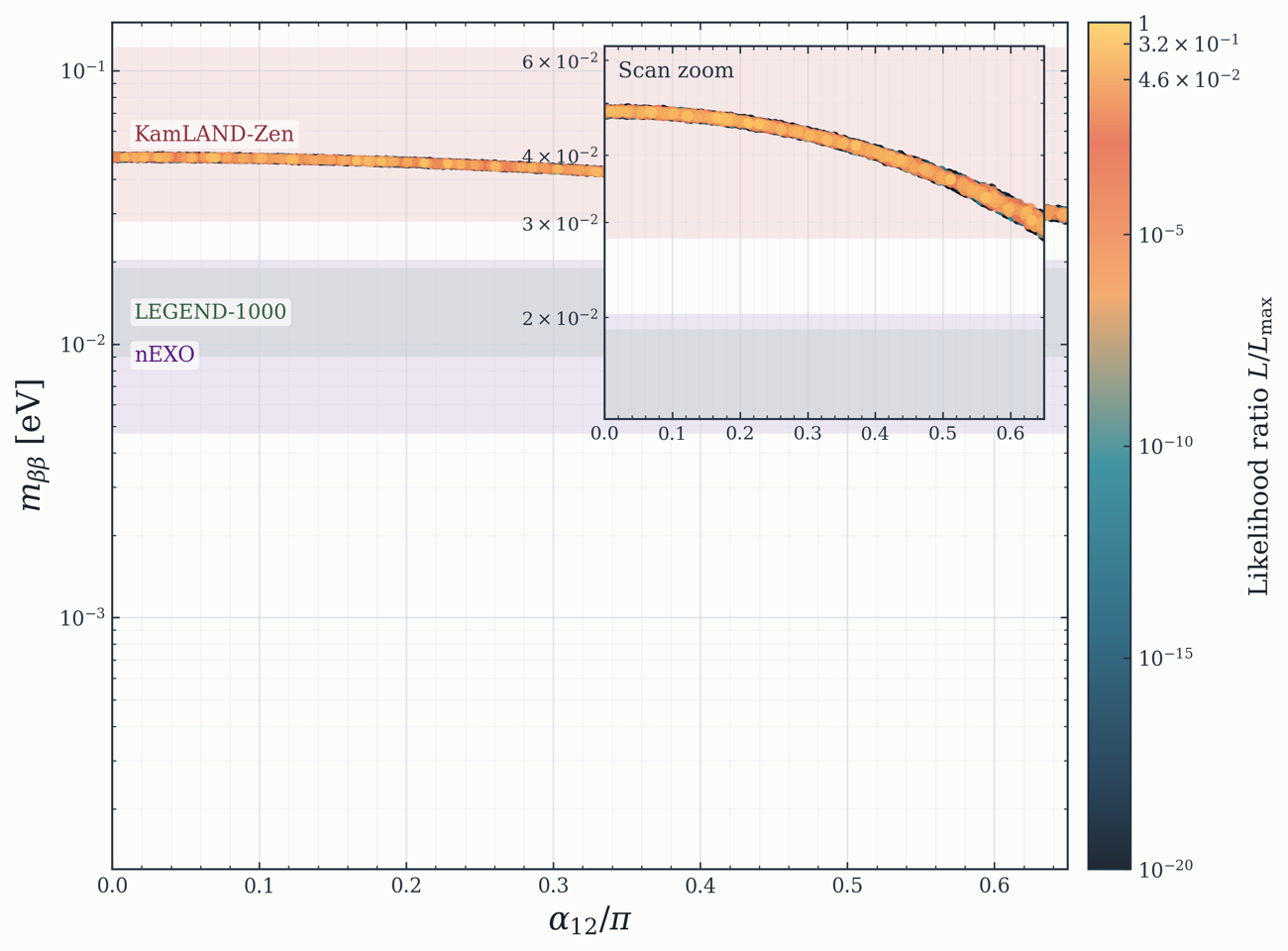}
\includegraphics[width=0.32\linewidth]{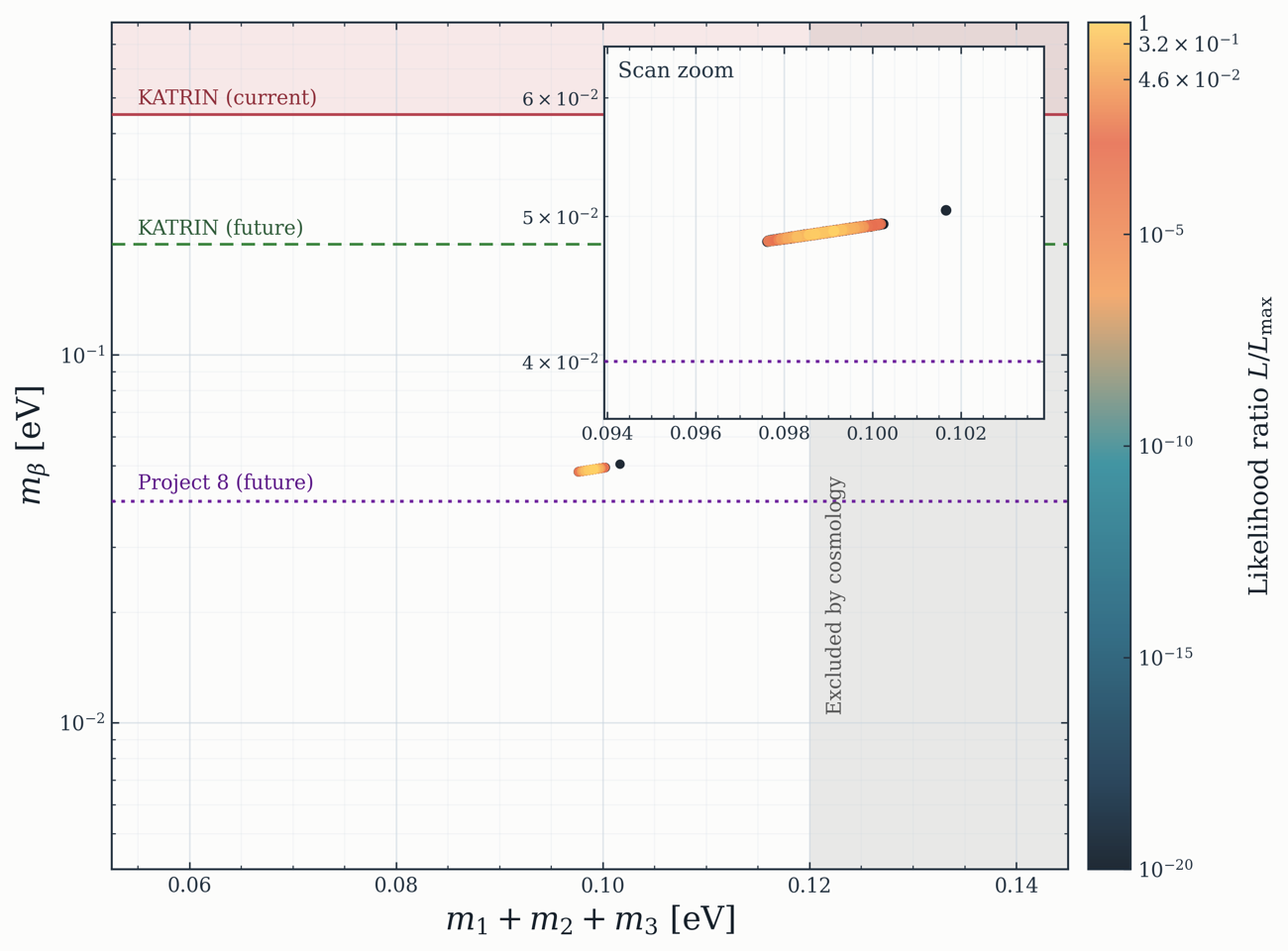}
\caption{Same as Fig. \ref{f9} but for IO.}
\label{f12}
\end{figure}

\begin{figure}[H]
\centering
\includegraphics[width=0.32\textwidth]{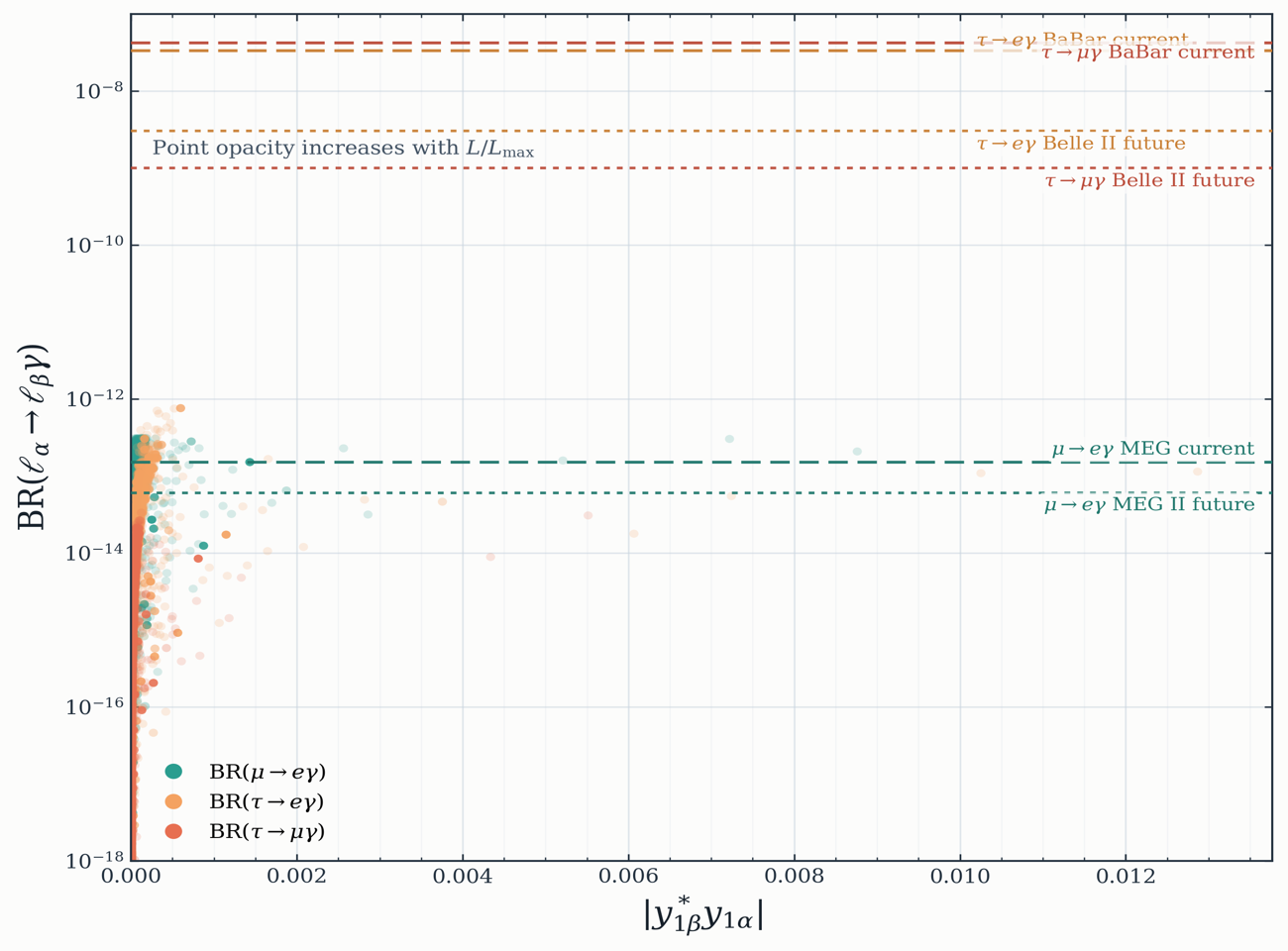}
\includegraphics[width=0.32\linewidth]{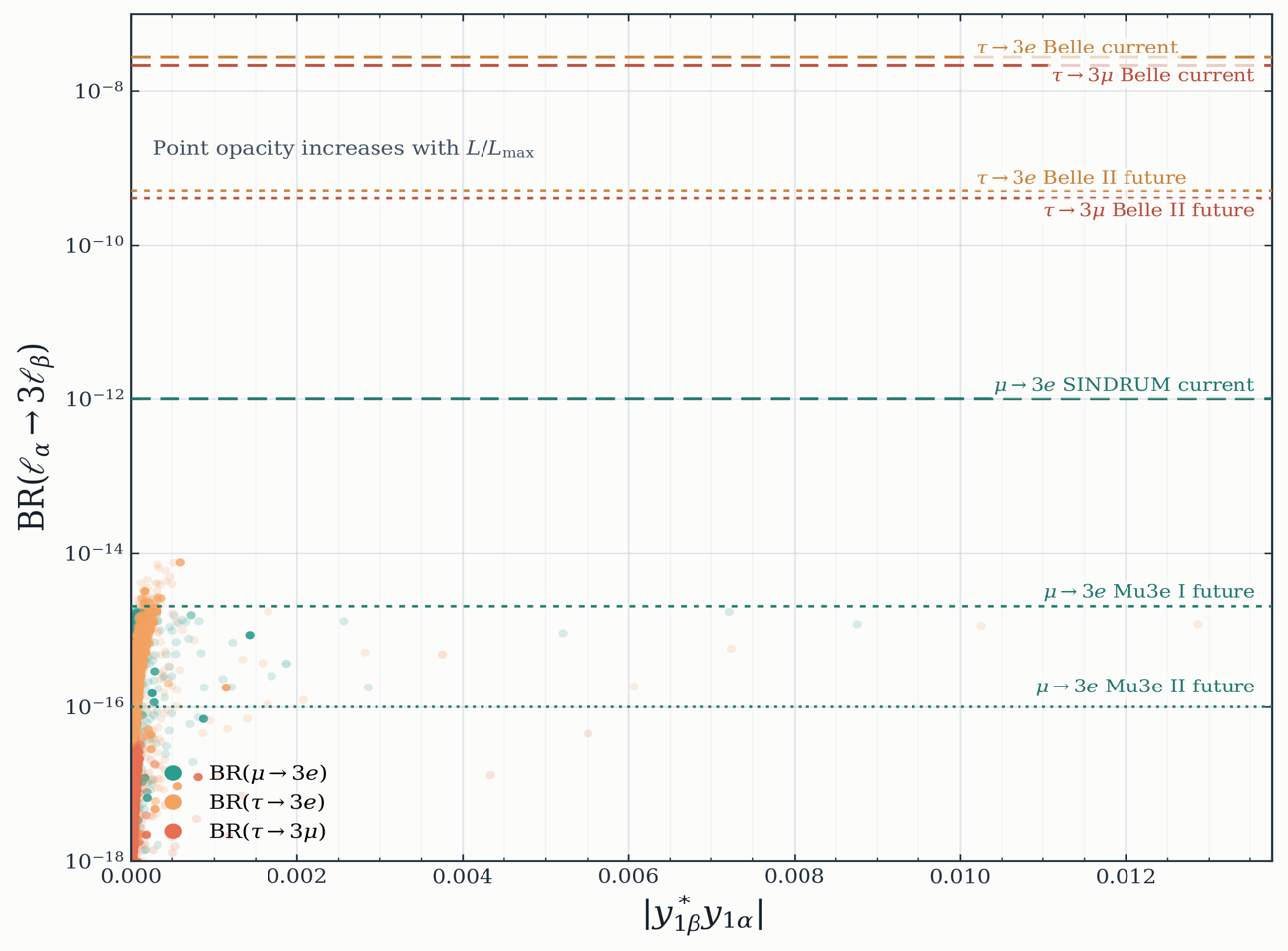}
\includegraphics[width=0.32\linewidth]{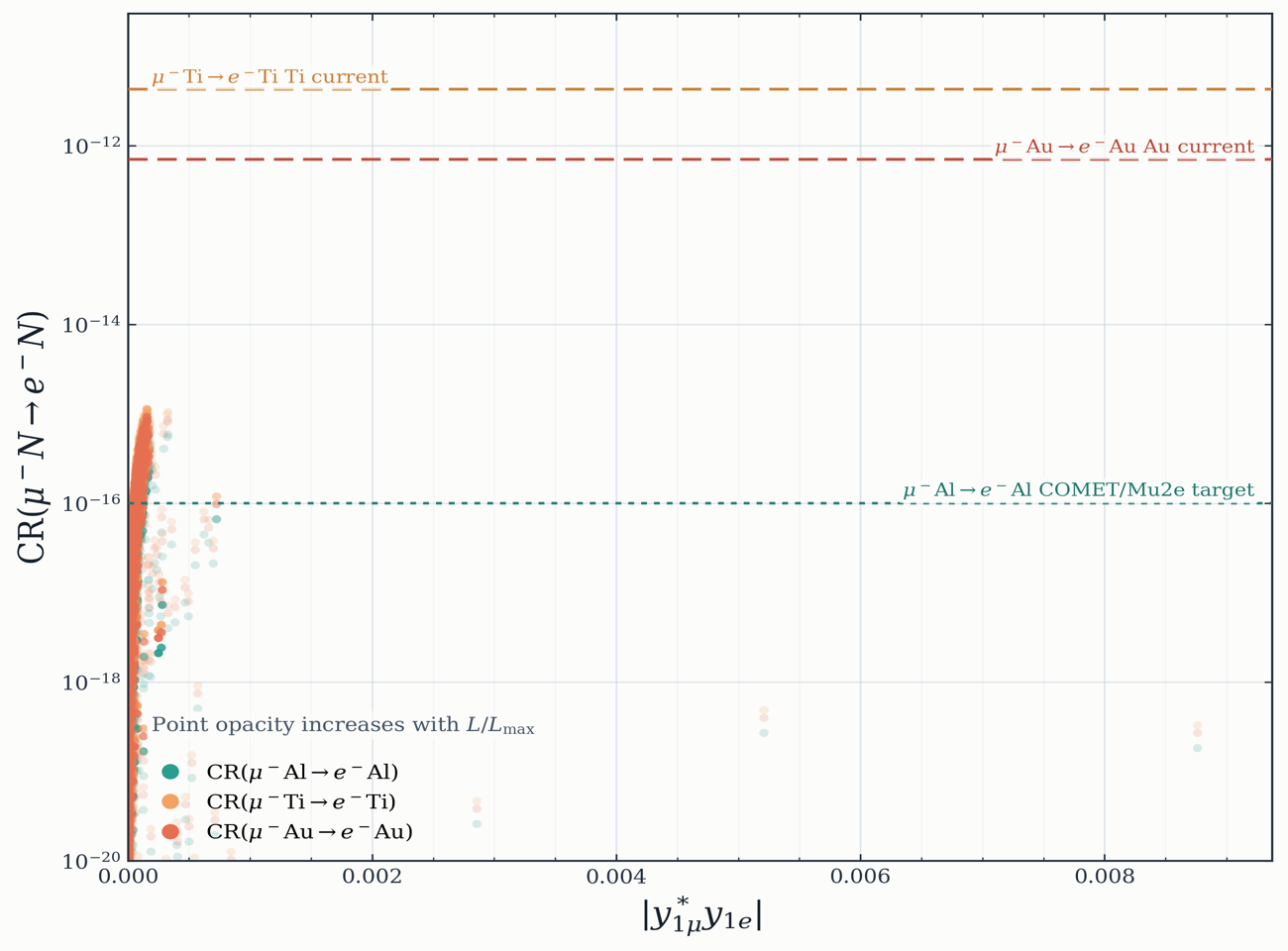}
\caption{Same as Fig. \ref{f10} but for IO.}
\label{f13}
\end{figure}

A common feature of the allowed parameter space is the nontrivial interplay between the relic density requirement and the flavor structure responsible for neutrino mass generation. The relic abundance selects regions in which the annihilation rate is compatible with $\Omega h^2\simeq 0.12$, while neutrino oscillation data restrict the allowed Yukawa structure through the measured mixing angles and mass-squared splittings. At the same time, cLFV processes impose stringent constraints on the size and flavor alignment of the new Yukawa couplings. The viable points therefore emerge only from regions in which the dark, scalar, and flavor sectors are simultaneously compatible with present data.
\newline
\\
\textbf{Fermionic dark matter:}
The results for the fermionic DM candidate are shown in Figs.~\ref{f2}--\ref{f7}. The accepted points cover $M_\psi \simeq (92-1447)~\rm GeV$ for NO, and $M_\psi \simeq (144-1452)~\rm GeV$ for IO. The relic density panels show the expected narrow band around the Planck value, but the freeze-out channel decomposition shows that the accepted fermionic points are not pure $\psi\psi$ annihilation channel. The NO best-fit has $x_F=27.05$ and is dominated by neutral inert scalar annihilations, mainly $A_1^0A_1^0\to W^+W^-$ $(57\%)$, $H_1^0H_1^0\to W^+W^-$ $(19\%)$, and $A_1^0A_1^0\to ZZ$ $(6\%)$, with subleading $H_1^0H_2^0$ and $\phi^\pm A_1^0$ coannihilations. The IO best fit has $x_F=26.90$ and is instead driven by charged and neutral scalar coannihilation, especially $\phi^+\phi^-\to W^+W^-$ $(51\%)$, $\phi^\pm A_2^0\to ZW^\pm$ $(8\%)$, $A_2^0A_2^0\to ZZ$ $(6\%)$, and $A_2^0A_2^0\to hh$ $(5\%)$. Thus, as in coannihilation-dominated scotogenic fermion-DM scenarios~\cite{Vicente:2014wga,Hagedorn:2018spx,Karan:2023adm}, the relic density is obtained because nearby inert scalars remain thermally populated during freeze-out.
\newline
The SI cross section is strongly suppressed for the fermionic DM\footnote{A larger DD rate would require a larger charged-scalar Higgs coupling, larger relevant Yukawa combinations, and a charged scalar that is not strongly decoupled. These changes are constrained because the same parameters also affect $h\to\gamma\gamma$, cLFV, and electroweak precision observables.}. For both hierarchies the accepted points span $\sigma_{\rm SI} <1.1\times10^{-49}~{\rm cm}^2$. All accepted fermionic points lie below the present LZ and XENONnT exclusion curves~\cite{LZ:2024zvo,XENON:2023cxc}, and also below the xenon neutrino-floor reference shown in the figures \cite{OHare:2021utq}. This is a physical consequence of the model where the $h\psi\psi$ interaction is absent at tree level and is generated only through the loop-induced coupling in Eq.~\eqref{eq:dm-gchichi}. The oblique parameters remain inside the correlated electroweak ellipse  the experimentally allowed ellipse.
The neutrino sector is well reproduced in both mass orderings. The solar panels show the $(\Delta m^2_{21},\sin^2\theta_{12})$ projection and include the \texttt{NuFIT 6.0} preferred regions together with the recent JUNO band, which directly probes this pair of solar oscillation observables~\cite{JUNO:2025gmd}. The other oscillation panels show that $\sin^2\theta_{13}$, $\sin^2\theta_{23}$ and $\delta_{\rm CP}$ remain within the allowed regions. The rank-two mass matrix fixes the absolute-mass pattern very sharply: NO gives $m_1\simeq0$ and $\sum_i m_i\simeq0.058$--$0.059~{\rm eV}$, while IO gives $m_3\simeq0$ and $\sum_i m_i\simeq0.098$--$0.100~{\rm eV}$, safely below the cosmological bound. For beta-decay endpoint searches, the NO prediction, $m_\beta\simeq(8.5$--$9.1)\times10^{-3}~{\rm eV}$, is far below the current KATRIN reach and below the Project~8 benchmark shown in the figures. The IO prediction, $m_\beta\simeq0.048$--$0.049~{\rm eV}$, is also well below present KATRIN limits, but it lies close to the Project~8 target sensitivity~\cite{KATRIN:2024mass,Project8:2022bla}. The neutrinoless-double-beta observable shows a stronger ordering dependence. In NO, $m_{\beta\beta}\simeq(1.4$--$3.8)\times10^{-3}~{\rm eV}$ lies below the projected reach of the ton-scale searches displayed in the figures. In IO, $m_{\beta\beta}\simeq0.016$--$0.048~{\rm eV}$ falls in the range targeted by LEGEND-1000, nEXO and CUPID, and the upper part of the interval approaches the present KamLAND-Zen sensitivity, subject to nuclear-matrix-element uncertainties~\cite{KamLAND-Zen:2016pfg,GERDA:2020xhi,CUORE:2022fgx,LEGEND:2021bnm,nEXO:2021ujk,CUPID:2022olt}. Thus neutrinoless double beta decay provides a much stronger test of the IO realization than of the NO realization.
The cLFV projections provide a complementary test of the Yukawa structure. The accepted fermionic points satisfy the current limits on $\mu\to e\gamma$, $\mu\to3e$, $\tau$ radiative decays, $\tau\to3\ell$, and coherent $\mu-e$ conversion. In the accepted samples, ${\rm BR}(\mu\to e\gamma)$ can reach $1.3\times10^{-13}$ for NO and $1.6\times10^{-13}$ for IO, close to the present MEG II sensitivity~\cite{MEGII:2025muegamma}. The corresponding $\mu\to3e$ rates remain below $10^{-15}$, but are relevant for the Mu3e program~\cite{Blondel:2021fji}. The coherent conversion rates in Al, Ti, and Au are below current limits, with the largest Al rates of order $10^{-18}$ in the fermionic samples, making Mu2e and COMET important future probes~\cite{Mu2e:2014fns,Kurup:2011zza}.
\newline
\\
\textbf{Scalar dark matter:}
The scalar DM results are shown in Figs.~\ref{f8}--\ref{f13}. In this part of the analysis, $H_1^0$ is chosen as the lightest odd scalar. The accepted samples are larger than in the fermionic case, with 1641 valid NO points and 1636 valid IO points. The scalar mass range starts near the Higgs pole, $m_{H_1^0}\simeq62.8~{\rm GeV}$, and extends to about $896~{\rm GeV}$ for NO and $773~{\rm GeV}$ for IO. The median accepted mass lies near $70~{\rm GeV}$ in both orderings, reflecting the importance of the Higgs-funnel region, but the freeze-out channel decomposition confirms that the accepted $H_1^0$-DM samples are not described by a pure Higgs-portal limit. The NO best fit has $x_F=23.95$ and is dominated by $H_1^0A_1^0$ coannihilation, with approximately $15\%$ each into $d\bar d$, $s\bar s$ and $b\bar b$, about $12\%$ each into $u\bar u$ and $c\bar c$, and further neutrino and charged lepton final states. The IO best fit has $x_F=24.10$ and is dominated by $H_1^0A_2^0$ coannihilation, with about $10\%$ each into $d\bar d$, $s\bar s$ and $b\bar b$, about $7\%$ each into $u\bar u$ and $c\bar c$, plus smaller $H_1^0H_2^0$, $A_2^0A_2^0$, $H_1^0A_1^0$ and $H_2^0A_2^0$ channels. These channels show that the relic density is controlled either by the Higgs-pole region or by coannihilation with nearby inert states.
The DD behavior of the $H_1^0$ candidate differs sharply from the fermionic case. Since $hH_1^0H_1^0$ is present at tree level, $\sigma_{\rm SI}$ can be much larger. In Figs.~\ref{f8} and~\ref{f11}, points above the LZ curve are excluded by present liquid-xenon data, while the region below that curve remains compatible with current DD searches. The neutrino-floor curve is shown as a sensitivity benchmark. The accepted $H_1^0$-DM samples include points between the LZ curve and the neutrino-floor reference, reaching $\sigma_{\rm SI}\simeq5.6\times10^{-48}~{\rm cm}^2$ in NO and $\sigma_{\rm SI}\simeq8.3\times10^{-48}~{\rm cm}^2$ in IO. This makes the scalar realization more accessible to next-generation xenon experiments such as DARWIN and XLZD~\cite{DARWIN:2016hyl,XLZD:2024nsu}.
The relic density panels show why the scalar points are not distributed uniformly in mass. The visible accumulation near $m_{H_1^0}\simeq m_h/2$ corresponds to the Higgs-resonance region, where efficient annihilation can occur even for a small Higgs-portal coupling. This also helps keep the DD rate below the LZ bound. At larger masses, the viable points form a thinner tail. In this region the relic abundance is mainly supported by coannihilation with nearby neutral or charged inert states, as indicated by the channel decomposition above.
The neutrino panels for the $H_1^0$-DM scans reproduce the same rank-two pattern found in the fermionic scans: $m_1\simeq0$ in NO and $m_3\simeq0$ in IO. Consequently, the conclusions for $m_\beta$, $m_{\beta\beta}$ and $\sum_i m_i$ are essentially fixed by the mass ordering rather than by the identity of the DM particle. The IO sample again gives the larger neutrinoless-double-beta range, $m_{\beta\beta}\simeq0.015$--$0.049~{\rm eV}$, while the NO sample remains in the meV region. 
The cLFV rates satisfy the present bounds, although the accepted $H_1^0$-DM samples can reach ${\rm BR}(\mu\to e\gamma)\simeq3\times10^{-13}$ and $CR(\mu{\rm Al}\to e{\rm Al})\simeq6\times10^{-16}$. These rates are below current limits but fall in the range targeted by the next generation of muon experiments.
\begin{figure}[H]
\centering
\includegraphics[width=0.33\textwidth]{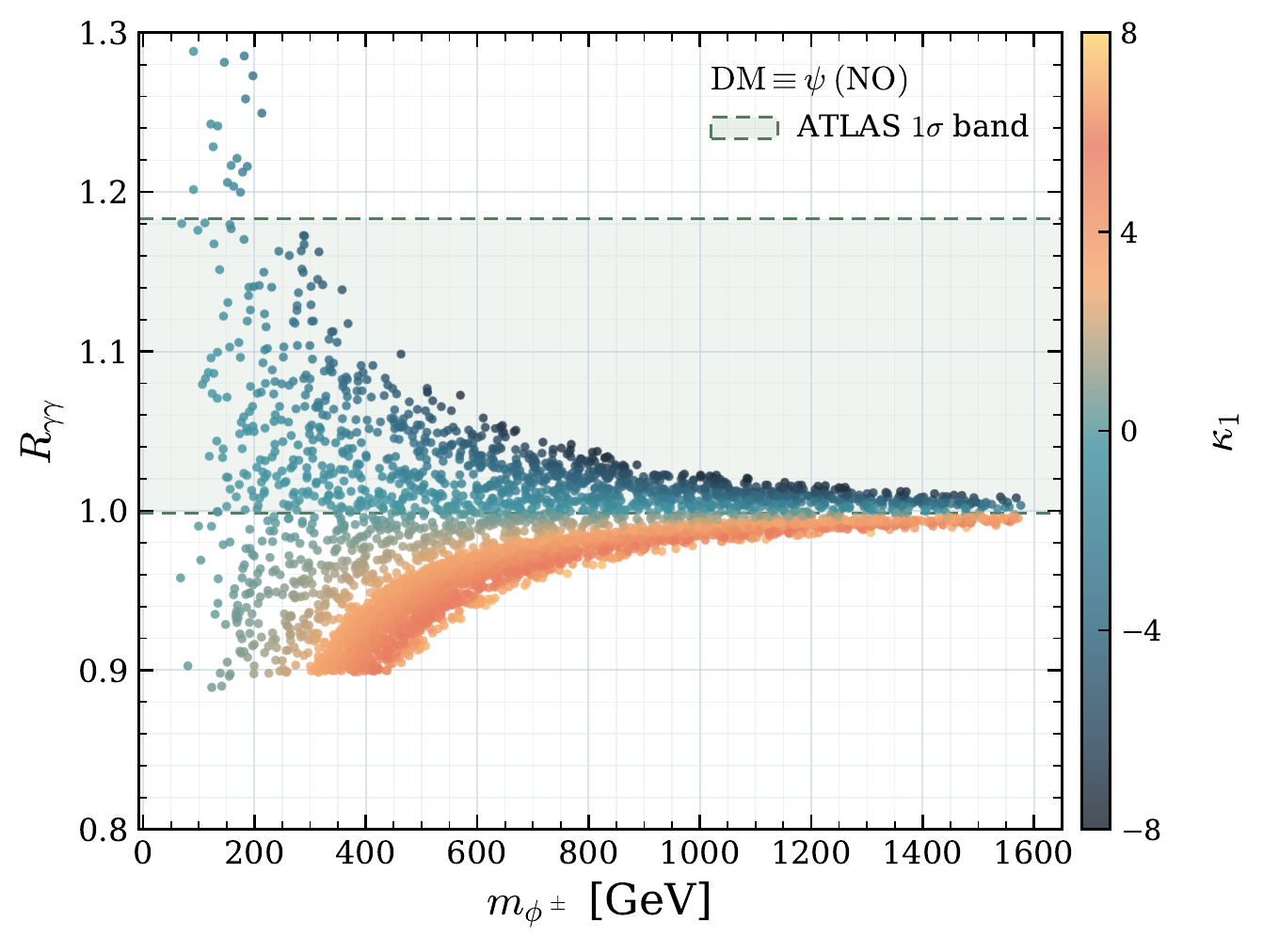}
\includegraphics[width=0.33\linewidth]{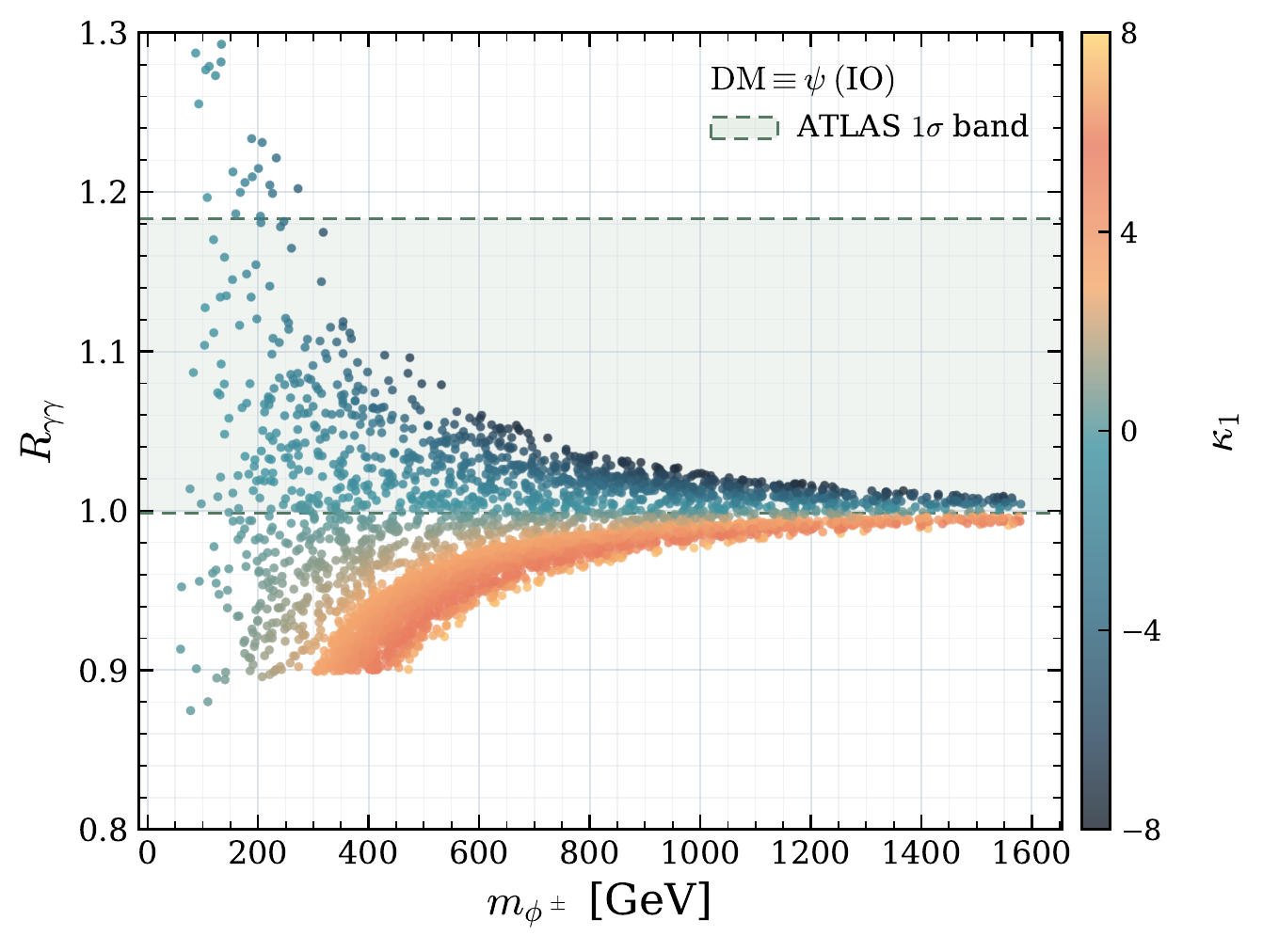}
\includegraphics[width=0.33\linewidth]{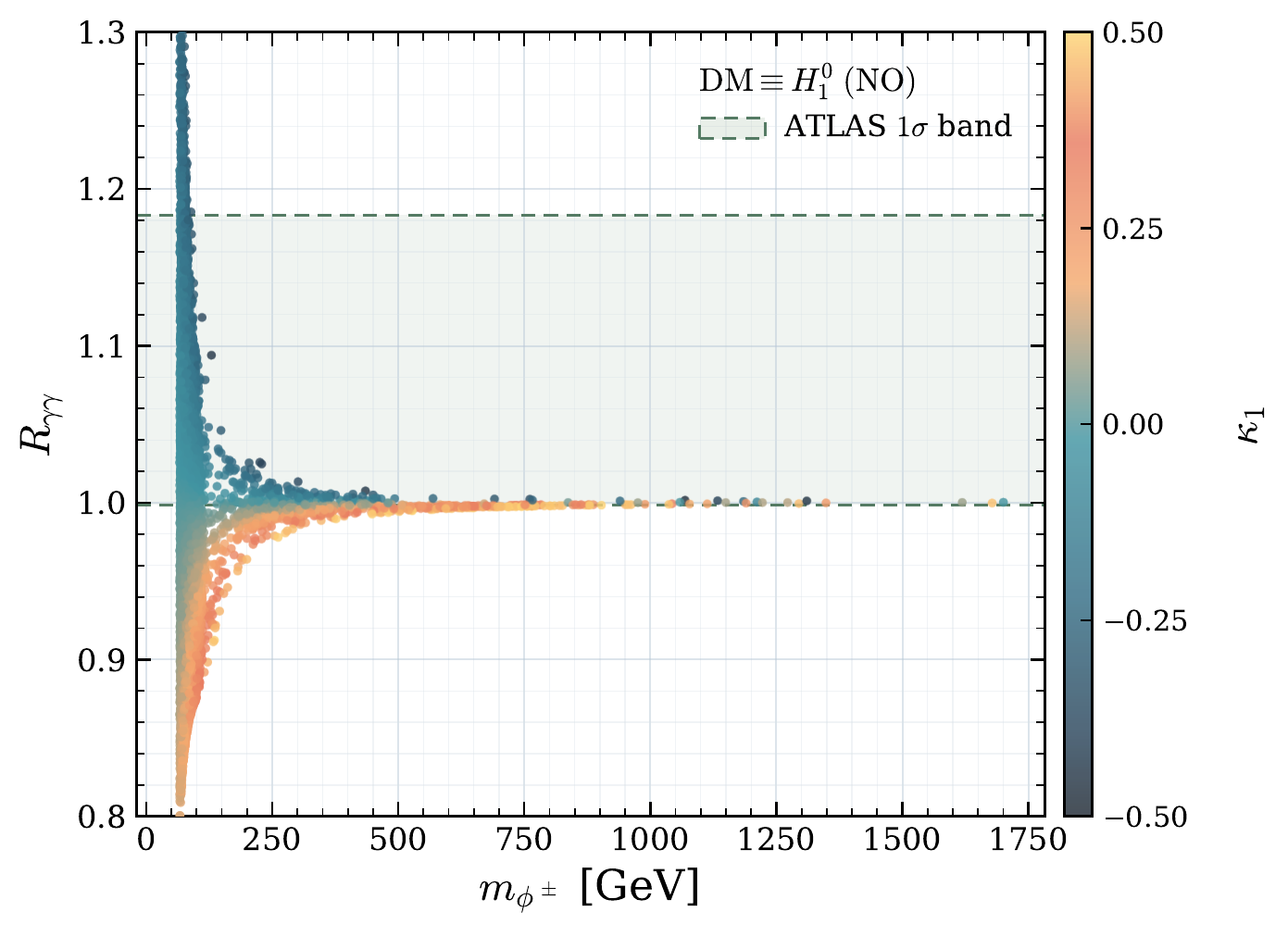}
\includegraphics[width=0.33\linewidth]{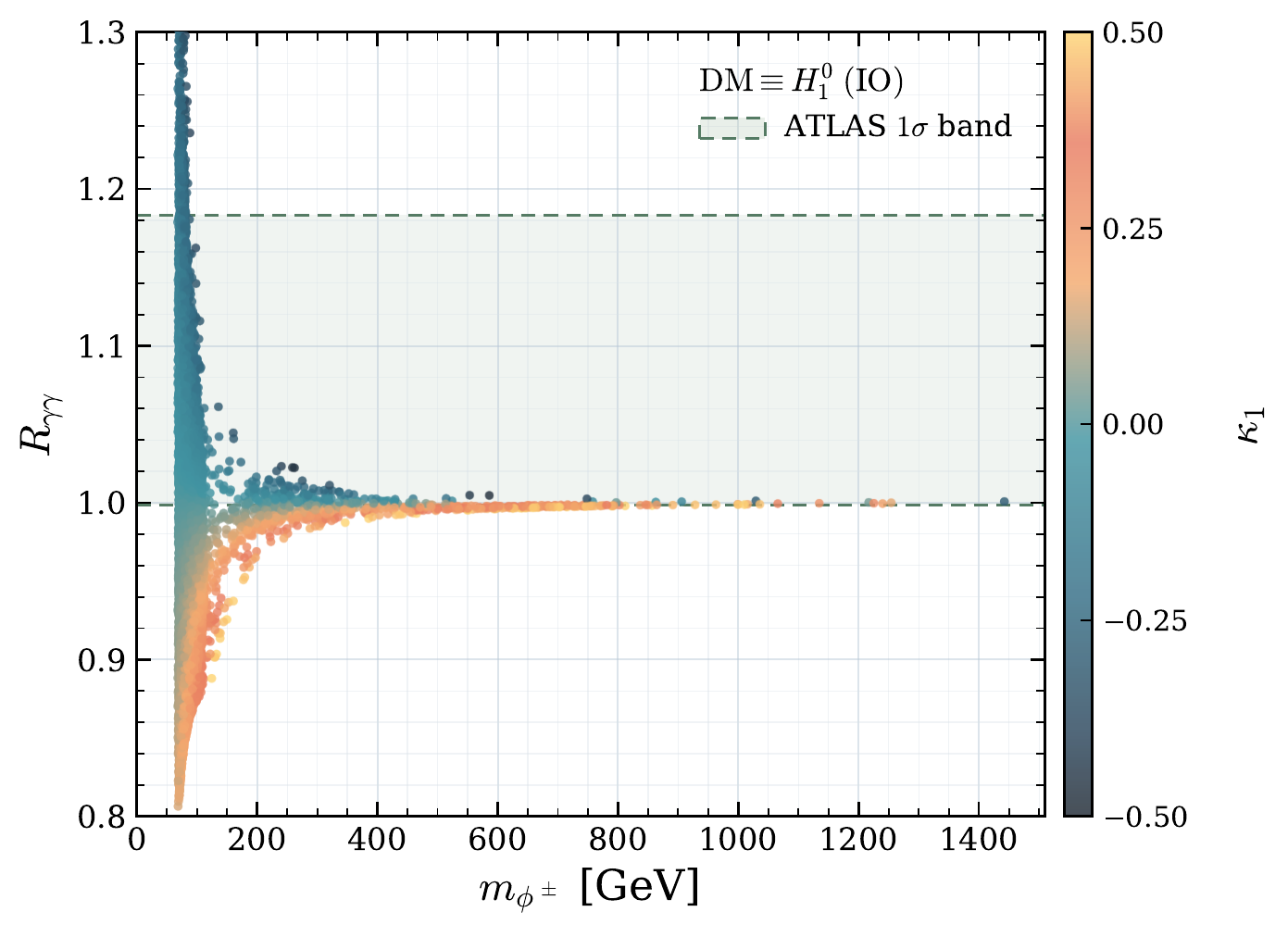}
\caption{Higgs diphoton constraint in the four scans. Top-left panel: fermionic DM with NO. Top-right panel: fermionic DM with IO. Bottom-left panel: scalar DM with NO. Bottom-right panel: scalar DM with IO. Each panel shows $R_{\gamma\gamma}$ as a function of $m_{\phi^\pm}$, with the color scale indicating $\kappa_1$.}
\label{f14}
\end{figure}
Figure~\ref{f14} displays the Higgs diphoton ratio $R_{\gamma\gamma}$ as a function of the charged-scalar mass $m_{\phi^\pm}$ in the four scans, with the quartic coupling $\kappa_1$ shown by the color scale. The common behavior is the expected decoupling of the charged-scalar loop. For large $m_{\phi^\pm}$, the new contribution is suppressed and $R_{\gamma\gamma}$ approaches the SM value. For lighter charged scalars, the correction becomes more visible, and its sign and size are controlled mainly by the combination of $\kappa_1$ and $m_{\phi^\pm}$. The ATLAS band therefore constrains the charged-scalar sector directly. This constraint is relevant for the rest of the phenomenology because $\kappa_1$ also enters the charged-scalar mass relation and the Higgs interactions of the $Z_2$-odd sector, including the loop-induced Higgs coupling of fermionic DM and the Higgs-portal interaction of scalar DM.

The best benchmark points used in the channel discussion are summarized in Tables~\ref{tab:bp-spectrum},~\ref{tab:bp-yukawas} and~\ref{tab:bp-observables}. They are chosen as the accepted points with the smallest total fit $\chi^2$ in each of the four scans.
\begin{table}[H]
\centering
\begin{tabular}{l|c|c|c|c|c|c|c|c|c|c|c|c}
\hline
Case & $M_\psi$ & $M_N$ & $m_{H_1^0}$ & $m_{H_2^0}$ & $m_{A_1^0}$ & $m_{A_2^0}$ & $m_{\phi^\pm}$ & $s_H$ & $\kappa_1$ & $\kappa_4$ & $\kappa_5$ & $\mu_\phi$ \\
\hline
$\psi$ NO & 685.03 & 1169.22 & 721.29 & 727.94 & 712.12 & 727.51 & 764.53 & 0.108 & 3.567 & 0.435 & 0.488 & $-17.20$ \\
$\psi$ IO & 546.80 & 1871.81 & 560.29 & 626.78 & 612.34 & 557.20 & 565.95 & 0.875 & 0.475 & 0.376 & $-0.204$ & $-149.90$ \\
$H_1^0$ NO & 135.82 & 2795.41 & 63.56 & 107.20 & 70.93 & 107.59 & 96.94 & 0.068 & 0.159 & 0.243 & 0.165 & $-10.77$ \\
$H_1^0$ IO & 237.62 & 2467.18 & 68.97 & 73.53 & 74.71 & 69.49 & 99.75 & 0.628 & 0.015 & 0.094 & 0.386 & $-1.81$ \\
\hline
\end{tabular}
\caption{Mass spectra and scalar sector inputs for the four best benchmark points. All masses and $\mu_\phi$ are in GeV.}
\label{tab:bp-spectrum}
\end{table}
\begin{table}[H]
\centering
\scriptsize
\begin{adjustbox}{width=\textwidth}
\begin{tabular}{l|c|c|c|c}
\hline
Coupling & $\psi$ NO & $\psi$ IO & $H_1^0$ NO & $H_1^0$ IO \\
\hline
$Y_e$ & $(1.61\times10^{-3}-2.45\times10^{-3}i)$ & $(7.23\times10^{-1}+5.76\times10^{-1}i)$ & $(8.54\times10^{-6}-2.38\times10^{-5}i)$ & $(1.64\times10^{-6}+3.70\times10^{-7}i)$ \\
$Y_\mu$ & $(-3.54\times10^{-3}+4.43\times10^{-4}i)$ & $(-6.76\times10^{-1}+6.53\times10^{-1}i)$ & $(-1.03\times10^{-5}+2.72\times10^{-5}i)$ & $(-5.54\times10^{-6}+5.65\times10^{-7}i)$ \\
$Y_\tau$ & $(-6.01\times10^{-3}+3.63\times10^{-3}i)$ & $(5.96\times10^{-1}-5.99\times10^{-1}i)$ & $(-3.18\times10^{-5}+5.84\times10^{-5}i)$ & $(6.15\times10^{-6}-2.48\times10^{-7}i)$ \\
$y_e$ & $(8.75\times10^{-3}-8.14\times10^{-3}i)$ & $(2.26\times10^{-4}-1.19\times10^{-4}i)$ & $(2.85\times10^{-4}-7.19\times10^{-5}i)$ & $(2.14\times10^{-2}-9.66\times10^{-4}i)$ \\
$y_\mu$ & $(1.20\times10^{-1}-7.22\times10^{-2}i)$ & $(-3.63\times10^{-5}-9.31\times10^{-5}i)$ & $(9.55\times10^{-4}-1.62\times10^{-3}i)$ & $(3.51\times10^{-3}+1.34\times10^{-3}i)$ \\
$y_\tau$ & $(8.24\times10^{-2}-1.47\times10^{-2}i)$ & $(5.63\times10^{-5}+1.60\times10^{-4}i)$ & $(4.35\times10^{-4}-9.00\times10^{-4}i)$ & $(-2.72\times10^{-3}+3.19\times10^{-3}i)$ \\
$y'_{11}$ & $5.36\times10^{-6}$ & $3.58\times10^{-6}$ & $1.36\times10^{-2}$ & $1.26\times10^{-1}$ \\
\hline
\end{tabular}
\end{adjustbox}
\caption{Complex Yukawa couplings for the four best benchmark points. The notation follows Eqs.~\eqref{eq:Mnu_compact} and \ref{eq:Mnu_explicit}: $Y_\alpha$ and $y_\alpha$ are the two flavor vectors entering the one-loop neutrino mass matrix, while $y'_{11}$ couples $\psi$ to $N$ and $\phi'$. The displayed benchmarks are written in a phase convention where $y'_{11}$ is real.}
\label{tab:bp-yukawas}
\end{table}
\begin{table}[H]
\centering
\scriptsize
\begin{adjustbox}{width=\textwidth}
\begin{tabular}{l|c|c|c|c|c|c|c|c|c|c|c|c}
\hline
Case & $\chi^2_{\rm tot}$ & $\Omega h^2$ & $x_F$ & $\sigma_{\rm SI}$ & $R_{\gamma\gamma}$ & $\Delta S$ & $\Delta T$ & $\sum_i m_i$ & $m_\beta$ & $m_{\beta\beta}$ & ${\rm BR}(\mu\to e\gamma)$ & $CR(\mu{\rm Al}\to e{\rm Al})$ \\
\hline
$\psi$ NO & 3.65 & 0.12017 & 27.05 & $1.00\times10^{-50}$ & 0.981 & $-3.41\times10^{-3}$ & 0.0408 & 0.0586 & $8.88\times10^{-3}$ & $3.43\times10^{-3}$ & $5.76\times10^{-14}$ & $3.61\times10^{-19}$ \\
$\psi$ IO & 4.69 & 0.11869 & 26.90 & $7.63\times10^{-63}$ & 0.995 & $3.93\times10^{-3}$ & 0.0338 & 0.0993 & $4.89\times10^{-2}$ & $4.63\times10^{-2}$ & $3.98\times10^{-23}$ & $2.63\times10^{-29}$ \\
$H_1^0$ NO & 4.55 & 0.12044 & 23.95 & $4.84\times10^{-50}$ & 0.933 & $-2.10\times10^{-2}$ & 0.0126 & 0.0589 & $8.93\times10^{-3}$ & $2.00\times10^{-3}$ & $1.11\times10^{-17}$ & $9.90\times10^{-22}$ \\
$H_1^0$ IO & 3.80 & 0.12154 & 24.10 & $2.02\times10^{-48}$ & 0.994 & $-2.02\times10^{-2}$ & 0.0134 & 0.0991 & $4.89\times10^{-2}$ & $1.91\times10^{-2}$ & $5.96\times10^{-14}$ & $5.57\times10^{-17}$ \\
\hline
\end{tabular}
\end{adjustbox}
\caption{Main observables for the four best benchmark points. The SI cross section is given in ${\rm cm}^2$, while the neutrino masses are in eV.}
\label{tab:bp-observables}
\end{table}
The benchmark spectra also motivate a dedicated collider study. Multi-TeV muon colliders have been shown to probe lepton-portal DM, mono-Higgs DM production and charged scalar sectors in radiative neutrino-mass models~\cite{Jueid:2023zxx,Jueid:2023qcf,Belfkir:2023vpo}. The collider phenomenology of the present \texttt{T4-3-i-B1} model is not included in this fit and is left for future work.
%%%%%%%%%%%%%%%%%%%%%%%%%%%%%%%%%%%%%%%%%%%%%%%%%%%%%%%%%%%%%%%%%
\section{Summary and Conclusions}
\label{SecVII}
%%%%%%%%%%%%%%%%%%%%%%%%%%%%%%%%%%%%%%%%%%%%%%%%%%%%%%%%%%%%%%%%%
In this work we have built the first complete model based on the genuine \texttt{T4-3-i} one-loop topology for Majorana neutrino mass. In this topology, the internal fermion connected to the external lepton and Higgs fields would generate the usual type-I or type-III seesaw if it were Majorana. We avoid this tree-level contribution by taking this fermion to be a Dirac field $N$ and by placing lepton number violation in a different Majorana fermion $\psi$ inside the loop. A $Z_2$ symmetry is imposed to keep the new scalars inert and stabilizes the lightest odd particle. We first classified the possible models that follow from this topology in the case of a singlet Dirac fermion $N$ by using the hypercharge parameter $\alpha$ and the gauge representations of the particles in the loop. This gives the variants \texttt{T4-3-i-A}, \texttt{T4-3-i-B}, \texttt{T4-3-i-C} and \texttt{T4-3-i-D}, with several allowed choices of $\alpha$ that contain a neutral DM candidate. We then focused on the simplest case, \texttt{T4-3-i-B1}, which contains one Dirac fermion, one Majorana fermion, one inert scalar doublet and one inert scalar singlet. This choice keeps the new sector small while allowing either $\psi$ or $H_i^0$ to play the role of DM.

The one-loop neutrino mass matrix has the form $M_\nu=\Lambda(Yy^T+yY^T)$. Therefore, with one Dirac fermion $N$ and one Majorana fermion $\psi$, the matrix has rank-two for generic flavor vectors. This predicts one massless neutrino at leading order. The complex Yukawa couplings were scanned directly through their real and imaginary parts, and the resulting mass matrix was diagonalized numerically using the Takagi factorization. The same Yukawa structures that generate the PMNS pattern also control cLFV, so the oscillation fit is strongly correlated with $\mu\to e\gamma$, $\mu\to3e$ and coherent $\mu-e$ conversion. The numerical analysis combined neutrino oscillation likelihoods from \texttt{NuFIT 6.0}, the cosmological bound on the neutrino mass sum, the Planck relic abundance, DD limits, cLFV bounds, oblique parameters and the Higgs diphoton measurement. We performed four independent scans: fermionic DM with NO, fermionic DM with IO, scalar DM with NO and scalar DM with IO. The results show that the model can satisfy all imposed constraints in both neutrino mass orderings and for both DM assignments.

For fermionic DM, the lightest odd state is the Majorana fermion $\psi$. Its Higgs interaction is absent at tree level and is generated only through a loop-induced effective $h\psi\psi$ coupling involving $\kappa_1$, the charged scalar and the lepton sector Yukawa couplings. As a result, the SI DD cross section is naturally suppressed. The viable fermionic samples lie below the present LZ and XENONnT bounds and also below the xenon neutrino floor reference. The relic abundance is not obtained through pure $\psi\psi$ annihilation alone; instead, the dominant freeze-out channels are controlled by annihilation and coannihilation of nearby inert scalars. This coannihilation behavior is a direct consequence of the compressed odd spectrum required by the relic density constraint. For scalar DM, we chose $H_1^0$ as the lightest odd state. In this case the diagonal $h H_1^0 H_1^0$ coupling is present at tree level, so DD is parametrically less suppressed than in the fermionic case. The accepted $H_1^0$-DM samples populate the region below the LZ curve, with part of the viable parameter space lying between the LZ bound and the neutrino floor reference. This makes the scalar realization more accessible to next-generation liquid xenon experiments. The relic abundance is mainly shaped by the Higgs-resonance region near $m_{H_1^0}\simeq m_h/2$ and by coannihilation with nearby neutral and charged inert scalars at larger masses.

The neutrino sector predictions are characteristic of a rank-two Majorana mass matrix. In NO, the mass sum is around $0.058$--$0.059~{\rm eV}$, $m_\beta$ is at the $9~{\rm meV}$ level and $m_{\beta\beta}$ remains in the few-meV range. These values are below the reach of the direct and neutrinoless-double-beta searches shown in our plots. In IO, the mass sum is around $0.098$--$0.100~{\rm eV}$, $m_\beta$ is close to the Project~8 benchmark sensitivity and $m_{\beta\beta}$ lies in the interval targeted by next-generation neutrinoless-double-beta decay experiments. The phenomenology is therefore sharply ordered: cLFV observables test the Yukawa structure, direct detection separates the fermionic and scalar DM assignments, Higgs diphoton and oblique observables constrain the scalar sector, and lepton-number-violation searches are most powerful for IO. The framework thus provides a compact and predictive connection between radiative neutrino mass, dark matter and precision flavor phenomenology.

Several directions remain open. The other variants found in the $\alpha$ classification (see Sec. \ref{SecII-A}) deserve a separate study, since different gauge charges can change the scalar spectrum, the coannihilation channels, the cLFV rates and the collider signals. Larger versions of the model, with more odd fermions or extra scalar copies, could also give three nonzero neutrino masses while keeping the same loop mechanism. Another important step is to study the production and decay of the charged and neutral odd scalars at colliders, especially in compressed spectra where the final states contain soft leptons and missing energy. It would also be useful to include indirect detection, future cLFV searches, future DD data and the running of the scalar potential to higher scales. These studies would show how far the present construction can be extended, and whether the genuine \texttt{T4-3-i} topology points to a wider class of testable radiative models. Some of these directions are already under study, while the others are left for future work.
%%%%%%%%%%%%%%%%%%%%%%%%%%%%%%%%%%%%%%%%%%%%%%%%%%%%%%%%%%%%%%%%%
\acknowledgments
%%%%%%%%%%%%%%%%%%%%%%%%%%%%%%%%%%%%%%%%%%%%%%%%%%%%%%%%%%%%%%%%%
The work of M.A.L and S.N. is supported by the United Arab Emirates University (UAEU) under UPAR Grant No. 12S162. We thank Dr. Adil Jueid for reading the manuscript and for his useful feedback and valuable discussions.
%%%%%%%%%%%%%%%%%%%%%%%%%%%%%%%%%%%%%%%%%%%%%%%%%%%%%%%%%%%%%%%%%
\appendix
%%%%%%%%%%%%%%%%%%%%%%%%%%%%%%%%%%%%%%%%%%%%%%%%%%%%%%%%%%%%%%%%%
\section{Unitarity amplitude matrices}
\label{app.A}
%%%%%%%%%%%%%%%%%%%%%%%%%%%%%%%%%%%%%%%%%%%%%%%%%%%%%%%%%%%%%%%%%
As discussed in subsection \ref{sub.A}, the presence of exact symmetries--electric charge conservation, {\it CP}, and a global $Z_2$ symmetry--allows the full scattering amplitude matrix to be block-diagonalized into six independent sub-matrices. Each block corresponds to a distinct symmetry sector defined by quantum numbers of the initial and final states. These sectors and their associated sub-matrices are:\\
\textbf{Neutral {\it CP}-even and $Z_2$-even 8 by 8 matrix in the basis:} \{$h h$, $G^0 G^0$, $h_1 h_1$, $h_2 h_2$, $\omega_1 \omega_1$, $\omega_2 \omega_2$, $G^+ G^-$, $\phi^+ \phi^-$\}.
\begin{equation}
\left(
\begin{array}{cccccccc}
 6\lambda_1 & 2\lambda_1 & \kappa_1 + \kappa_2 + \kappa_3 & \kappa_4 & \kappa_1 + \kappa_2 - \kappa_3 & \kappa_4 & 2\lambda_1 & \kappa_1 \\
 2\lambda_1 & 6\lambda_1 & \kappa_1 + \kappa_2 - \kappa_3 & \kappa_4 & \kappa_1 + \kappa_2 + \kappa_3 & \kappa_4 & 2\lambda_1 & \kappa_1 \\
 \kappa_1 + \kappa_2 + \kappa_3 & \kappa_1 + \kappa_2 - \kappa_3 & 6\lambda_2 & \kappa_5 & 2\lambda_2 & \kappa_5 & \kappa_1 & 2\lambda_2 \\
 \kappa_4 & \kappa_4 & \kappa_5 & 6\lambda_3 & \kappa_5 & 2\lambda_3 & \kappa_4 & \kappa_5 \\
 \kappa_1 + \kappa_2 - \kappa_3 & \kappa_1 + \kappa_2 + \kappa_3 & 2\lambda_2 & \kappa_5 & 6\lambda_2 & \kappa_5 & \kappa_1 & 2\lambda_2 \\
 \kappa_4 & \kappa_4 & \kappa_5 & 2\lambda_3 & \kappa_5 & 6\lambda_3 & \kappa_4 & \kappa_5 \\
 2\lambda_1 & 2\lambda_1 & \kappa_1 & \kappa_4 & \kappa_1 & \kappa_4 & 4\lambda_1 & \kappa_1 + \kappa_2 \\
 \kappa_1 & \kappa_1 & 2\lambda_2 & \kappa_5 & 2\lambda_2 & \kappa_5 & \kappa_1 + \kappa_2 & 4\lambda_2
\end{array}
\right)
\end{equation}
\textbf{Neutral {\it CP}-even and $Z_2$-odd 5 by 5 matrix in the basis:} \{$h h_1$, $h h_2$, $G^0 \omega_1$, $G^0 \omega_2$, $G^+ \phi^-$\}.
\begin{equation}
\left(
\begin{array}{ccccc}
\kappa_1 & 0 & \kappa_3 & 0 & \frac{\kappa_1}{2} + \frac{\kappa_2}{2} \\
0 & \kappa_4 & 0 & 0 & 0 \\
\kappa_3 & 0 & \frac{\kappa_1}{4} + \frac{\kappa_2}{4} + \frac{\kappa_3}{4} & 0 & \frac{\kappa_2}{2} + \frac{\kappa_3}{2} \\
0 & 0 & 0 & \frac{\kappa_4}{2} & 0 \\
\frac{\kappa_2}{2} + \frac{\kappa_3}{2} & 0 & \frac{\kappa_2}{2} + \frac{\kappa_3}{2} & 0 & \frac{\kappa_3}{2}
\end{array}
\right)
\end{equation}
\textbf{Neutral {\it CP}-odd and $Z_2$-even 7 by 7 matrix in the basis:} \{$h G^0$, $h_1 \omega_1$, $h_1 \omega_2$, $h_2 \omega_1$, $h_2 \omega_2$, $G^+ G^-$, $\phi^+ \phi^-$\}.
\begin{equation}
\left(
\begin{array}{ccccccc}
2\lambda_1 & \kappa_3 & 0 & 0 & 0 & 0 & 0 \\
\kappa_3 & 2\lambda_2 & 0 & 0 & 0 & 0 & 0 \\
0 & 0 & \kappa_5 & 0 & 0 & 0 & 0 \\
0 & 0 & 0 & \kappa_5 & 0 & 0 & 0 \\
0 & 0 & 0 & 0 & 2\lambda_3 & 0 & 0 \\
0 & 0 & 0 & 0 & 0 & 4\lambda_1 & \kappa_1 + \kappa_2 \\
0 & 0 & 0 & \kappa_5 & 0 & \kappa_1 + \kappa_2 & 4\lambda_2
\end{array}
\right)    
\end{equation}
\textbf{Neutral {\it CP}-odd and $Z_2$-odd 5 by 5 matrix in the basis:} \{$h \omega_1$, $h \omega_2$, $G^0 h_1$, $G^0 h_2$, $G^+ \phi^-$\}.
\begin{equation}
\left(
\begin{array}{ccccc}
\kappa_1 + \kappa_2 - \kappa_3 & 0 & \kappa_3 & 0 & \frac{\kappa_2}{2} - \frac{\kappa_3}{2} \\
0 & \kappa_4 & 0 & 0 & 0 \\
\kappa_3 & 0 & \kappa_1 + \kappa_2 - \kappa_3 & 0 & -\frac{\kappa_2}{2} + \frac{\kappa_3}{2} \\
0 & 0 & 0 & \kappa_4 & 0 \\
\frac{\kappa_2}{2} - \frac{\kappa_3}{2} & 0 & -\frac{\kappa_2}{2} + \frac{\kappa_3}{2} & 0 & \frac{\kappa_3}{2}
\end{array}
\right)    
\end{equation}
\textbf{Charged and $Z_2$-even 6 by 6 matrix in the basis:} \{$h G^+$, $G^0 G^+$, $h_1 \phi^+$, $h_2 \phi^+$, $\omega_1 \phi^+$, $\omega_2 \phi^+$\}.
\begin{equation}
\left(
\begin{array}{cccccc}
2\lambda_1 & 0 & \frac{\kappa_2}{2} + \frac{\kappa_3}{2} & 0 & \frac{\kappa_2}{2} - \frac{\kappa_3}{2} & 0 \\
0 & 2\lambda_1 & -\frac{\kappa_2}{2} + \frac{\kappa_3}{2} & 0 & \frac{\kappa_2}{2} + \frac{\kappa_3}{2} & 0 \\
\frac{\kappa_2}{2} + \frac{\kappa_3}{2} & -\frac{\kappa_2}{2} + \frac{\kappa_3}{2} & 2\lambda_2 & 0 & 0 & 0 \\
0 & 0 & 0 & \kappa_5 & 0 & 0 \\
\frac{\kappa_2}{2} - \frac{\kappa_3}{2} & \frac{\kappa_2}{2} + \frac{\kappa_3}{2} & 0 & 0 & 2\lambda_2 & 0 \\
0 & 0 & 0 & 0 & 0 & \kappa_5
\end{array}
\right)      
\end{equation}
\textbf{Charged and $Z_2$-odd 6 by 6 matrix in the basis:} \{$h \phi^+$, $G^0 \phi^+$, $h_1 G^+$, $h_2 G^+$, $\omega_1 G^+$, $\omega_2 G^+$\}
\begin{equation}
\left(
\begin{array}{cccccc}
\kappa_1 & 0 & \frac{\kappa_2}{2} + \frac{\kappa_3}{2} & 0 & \frac{\kappa_2}{2} - \frac{\kappa_3}{2} & 0 \\
0 & \kappa_1 & -\frac{\kappa_2}{2} + \frac{\kappa_3}{2} & 0 & \frac{\kappa_2}{2} + \frac{\kappa_3}{2} & 0 \\
\frac{\kappa_2}{2} + \frac{\kappa_3}{2} & -\frac{\kappa_2}{2} + \frac{\kappa_3}{2} & \kappa_1 & 0 & 0 & 0 \\
0 & 0 & 0 & \kappa_4 & 0 & 0 \\
\frac{\kappa_2}{2} - \frac{\kappa_3}{2} & \frac{\kappa_2}{2} + \frac{\kappa_3}{2} & 0 & 0 & \kappa_1 & 0 \\
0 & 0 & 0 & 0 & 0 & \kappa_4
\end{array}
\right)      
\end{equation}

\bibliographystyle{JHEP}
\bibliography{bibliography.bib}

@article{Bonnet:2012kz,
    author = "Bonnet, Florian and Hirsch, Martin and Ota, Toshihiko and Winter, Walter",
    title = "{Systematic study of the d=5 Weinberg operator at one-loop order}",
    eprint = "1204.5862",
    archivePrefix = "arXiv",
    primaryClass = "hep-ph",
    reportNumber = "IFIC-12-26, MPP-2012-76",
    doi = "10.1007/JHEP07(2012)153",
    journal = "JHEP",
    volume = "07",
    pages = "153",
    year = "2012"
}

@article{AristizabalSierra:2014wal,
    author = "Aristizabal Sierra, D. and Degee, A. and Dorame, L. and Hirsch, M.",
    title = "{Systematic classification of two-loop realizations of the Weinberg operator}",
    eprint = "1411.7038",
    archivePrefix = "arXiv",
    primaryClass = "hep-ph",
    doi = "10.1007/JHEP03(2015)040",
    journal = "JHEP",
    volume = "03",
    pages = "040",
    year = "2015"
}

@article{Cao:2017xgk,
    author = "Cao, Qing-Hong and Chen, Shao-Long and Ma, Ernest and Yan, Bin and Zhang, Dong-Ming",
    title = "{New Class of Two-Loop Neutrino Mass Models with Distinguishable Phenomenology}",
    eprint = "1707.05896",
    archivePrefix = "arXiv",
    primaryClass = "hep-ph",
    reportNumber = "UCRHEP-T580",
    doi = "10.1016/j.physletb.2018.02.038",
    journal = "Phys. Lett. B",
    volume = "779",
    pages = "430--435",
    year = "2018"
}

@article{Cepedello:2018rfh,
    author = "Cepedello, Ricardo and Fonseca, Renato M. and Hirsch, Martin",
    title = "{Systematic classification of three-loop realizations of the Weinberg operator}",
    eprint = "1807.00629",
    archivePrefix = "arXiv",
    primaryClass = "hep-ph",
    reportNumber = "IFIC/18-20, IFIC-18-20",
    doi = "10.1007/JHEP10(2018)197",
    journal = "JHEP",
    volume = "10",
    pages = "197",
    year = "2018",
    note = "[Erratum: JHEP 06, 034 (2019)]"
}

@article{Cai:2017jrq,
    author = "Cai, Yi and Herrero-Garc{\'\i}a, Juan and Schmidt, Michael A. and Vicente, Avelino and Volkas, Raymond R.",
    title = "{From the trees to the forest: a review of radiative neutrino mass models}",
    eprint = "1706.08524",
    archivePrefix = "arXiv",
    primaryClass = "hep-ph",
    reportNumber = "ADP-17-29-T1035",
    doi = "10.3389/fphy.2017.00063",
    journal = "Front. in Phys.",
    volume = "5",
    pages = "63",
    year = "2017"
}

@article{Ma:2006km,
    author = "Ma, Ernest",
    title = "{Verifiable radiative seesaw mechanism of neutrino mass and dark matter}",
    eprint = "hep-ph/0601225",
    archivePrefix = "arXiv",
    reportNumber = "UCRHEP-T403",
    doi = "10.1103/PhysRevD.73.077301",
    journal = "Phys. Rev. D",
    volume = "73",
    pages = "077301",
    year = "2006"
}

@article{Budhi:2014gxa,
    author = "Budhi, Romy H. S. and Kashiwase, Shoichi and Suematsu, Daijiro",
    title = "{Inflation in a modified radiative seesaw model}",
    eprint = "1409.6889",
    archivePrefix = "arXiv",
    primaryClass = "hep-ph",
    reportNumber = "KANAZAWA-14-08",
    doi = "10.1103/PhysRevD.90.113013",
    journal = "Phys. Rev. D",
    volume = "90",
    number = "11",
    pages = "113013",
    year = "2014"
}

@article{Kashiwase:2015yla,
    author = "Kashiwase, Shoichi and Suematsu, Daijiro",
    title = "{Lepton number asymmetry via inflaton decay in a modified radiative seesaw model}",
    eprint = "1507.06782",
    archivePrefix = "arXiv",
    primaryClass = "hep-ph",
    reportNumber = "KANAZAWA-15-13",
    doi = "10.1016/j.physletb.2015.08.062",
    journal = "Phys. Lett. B",
    volume = "749",
    pages = "603--612",
    year = "2015"
}

@article{Budhi:2015sha,
    author = "Budhi, Romy H. S. and Kashiwase, Shoichi and Suematsu, Daijiro",
    title = "{Inflation due to a nonminimal coupling of singlet scalars in the radiative seesaw model}",
    eprint = "1509.05841",
    archivePrefix = "arXiv",
    primaryClass = "hep-ph",
    reportNumber = "KANAZAWA-15-14",
    doi = "10.1103/PhysRevD.93.013022",
    journal = "Phys. Rev. D",
    volume = "93",
    number = "1",
    pages = "013022",
    year = "2016"
}

@article{Zee:1980ai,
    author = "Zee, A.",
    title = "{A Theory of Lepton Number Violation, Neutrino Majorana Mass, and Oscillation}",
    reportNumber = "UPR-0150T",
    doi = "10.1016/0370-2693(80)90349-4",
    journal = "Phys. Lett. B",
    volume = "93",
    pages = "389",
    year = "1980",
    note = "[Erratum: Phys.Lett.B 95, 461 (1980)]"
}

@article{Ma:1998dn,
    author = "Ma, Ernest",
    title = "{Pathways to naturally small neutrino masses}",
    eprint = "hep-ph/9805219",
    archivePrefix = "arXiv",
    reportNumber = "UCRHEP-T-222",
    doi = "10.1103/PhysRevLett.81.1171",
    journal = "Phys. Rev. Lett.",
    volume = "81",
    pages = "1171--1174",
    year = "1998"
}

@article{Restrepo:2013aga,
    author = "Restrepo, Diego and Zapata, Oscar and Yaguna, Carlos E.",
    title = "{Models with radiative neutrino masses and viable dark matter candidates}",
    eprint = "1308.3655",
    archivePrefix = "arXiv",
    primaryClass = "hep-ph",
    doi = "10.1007/JHEP11(2013)011",
    journal = "JHEP",
    volume = "11",
    pages = "011",
    year = "2013"
}

@article{Avila:2025qsc,
    author = "{\'A}vila, Ivania M. and Karan, Anirban and Mandal, Sanjoy and Sadhukhan, Soumya and Valle, Jos{\'e} W. F.",
    title = "{Dark matter as the source of neutrino mass: Theory overview and experimental prospects}",
    eprint = "2506.24027",
    archivePrefix = "arXiv",
    primaryClass = "hep-ph",
    doi = "10.1016/j.physrep.2026.02.003",
    journal = "Phys. Rept.",
    volume = "1173",
    pages = "1--81",
    year = "2026"
}

@article{Darricau:2026iwg,
    author = "Darricau, A. and Teixeira, A. M.",
    title = "{Revisiting cLFV in ''T1-2-A'' scotogenic models: asymmetries in three-body lepton decays}",
    eprint = "2607.07484",
    archivePrefix = "arXiv",
    primaryClass = "hep-ph",
    month = "7",
    year = "2026"
}

@article{Loualidi:2020jlj,
    author = "Loualidi, M. A. and Miskaoui, M.",
    title = "{One-loop Type II Seesaw Neutrino Model with Stable Dark Matter Candidates}",
    eprint = "2003.11434",
    archivePrefix = "arXiv",
    primaryClass = "hep-ph",
    doi = "10.1016/j.nuclphysb.2020.115219",
    journal = "Nucl. Phys. B",
    volume = "961",
    pages = "115219",
    year = "2020"
}

@article{Kashav:2022kpk,
    author = "Kashav, Monal and Verma, Surender",
    title = "{On minimal realization of topological Lorentz structures with one-loop seesaw extensions in A$_{4}$ modular symmetry}",
    eprint = "2205.06545",
    archivePrefix = "arXiv",
    primaryClass = "hep-ph",
    doi = "10.1088/1475-7516/2023/03/010",
    journal = "JCAP",
    volume = "03",
    pages = "010",
    year = "2023"
}

@article{Loualidi:2026pld,
    author = "Loualidi, Mohamed Amin and Miskaoui, Mohamed and Nasri, Salah",
    title = "{Radiative Neutrino Mass in a Nonholomorphic $T'$ Modular Invariant Model}",
    eprint = "2606.11346",
    archivePrefix = "arXiv",
    primaryClass = "hep-ph",
    month = "6",
    year = "2026"
}

@article{Wang:2015saa,
    author = "Wang, Weijian and Han, Zhi-Long",
    title = "{Radiative linear seesaw model, dark matter, and $U(1)_{B-L}$}",
    eprint = "1508.00706",
    archivePrefix = "arXiv",
    primaryClass = "hep-ph",
    doi = "10.1103/PhysRevD.92.095001",
    journal = "Phys. Rev. D",
    volume = "92",
    pages = "095001",
    year = "2015"
}

@book{autonne1915matrices,
  title={Sur les matrices hypohermitiennes et sur les matrices unitaires},
  author={Autonne, L{\'e}on},
  year={1915},
  publisher={A. Rey}
}

@inproceedings{takagi1924algebraic,
  title={On an algebraic problem reluted to an analytic theorem of carath{\'e}odory and fej{\'e}r and on an allied theorem of landau},
  author={Takagi, Teiji},
  booktitle={Japanese journal of mathematics: transactions and abstracts},
  volume={1},
  pages={83--93},
  year={1924},
  organization={The Mathematical Society of Japan}
}

@article{Kannike:2012pe,
    author = "Kannike, Kristjan",
    title = "{Vacuum Stability Conditions From Copositivity Criteria}",
    eprint = "1205.3781",
    archivePrefix = "arXiv",
    primaryClass = "hep-ph",
    doi = "10.1140/epjc/s10052-012-2093-z",
    journal = "Eur. Phys. J. C",
    volume = "72",
    pages = "2093",
    year = "2012"
}

@article{ParticleDataGroup:2024cfk,
    author = "Navas, S. and others",
    collaboration = "Particle Data Group",
    title = "{Review of particle physics}",
    doi = "10.1103/PhysRevD.110.030001",
    journal = "Phys. Rev. D",
    volume = "110",
    number = "3",
    pages = "030001",
    year = "2024"
}

@article{deGouvea:2013zba,
    author = "de Gouvea, Andre and Vogel, Petr",
    title = "{Lepton Flavor and Number Conservation, and Physics Beyond the Standard Model}",
    eprint = "1303.4097",
    archivePrefix = "arXiv",
    primaryClass = "hep-ph",
    doi = "10.1016/j.ppnp.2013.03.006",
    journal = "Prog. Part. Nucl. Phys.",
    volume = "71",
    pages = "75--92",
    year = "2013"
}

@article{Bernstein:2013hba,
    author = "Bernstein, Robert H. and Cooper, Peter S.",
    title = "{Charged Lepton Flavor Violation: An Experimenter's Guide}",
    eprint = "1307.5787",
    archivePrefix = "arXiv",
    primaryClass = "hep-ex",
    reportNumber = "FERMILAB-PUB-13-259-PPD",
    doi = "10.1016/j.physrep.2013.07.002",
    journal = "Phys. Rept.",
    volume = "532",
    pages = "27--64",
    year = "2013"
}

@article{Calibbi:2017uvl,
    author = "Calibbi, Lorenzo and Signorelli, Giovanni",
    title = "{Charged Lepton Flavour Violation: An Experimental and Theoretical Introduction}",
    eprint = "1709.00294",
    archivePrefix = "arXiv",
    primaryClass = "hep-ph",
    doi = "10.1393/ncr/i2018-10144-0",
    journal = "Riv. Nuovo Cim.",
    volume = "41",
    number = "2",
    pages = "71--174",
    year = "2018"
}

@article{Ardu:2022sbt,
    author = "Ardu, Marco and Pezzullo, Gianantonio",
    title = "{Introduction to Charged Lepton Flavor Violation}",
    eprint = "2204.08220",
    archivePrefix = "arXiv",
    primaryClass = "hep-ph",
    doi = "10.3390/universe8060299",
    journal = "Universe",
    volume = "8",
    number = "6",
    pages = "299",
    year = "2022"
}

@article{Davidson:2022jai,
    author = "Davidson, Sacha and Echenard, Bertrand and Bernstein, Robert H. and Heeck, Julian and Hitlin, David G.",
    title = "{Charged Lepton Flavor Violation}",
    eprint = "2209.00142",
    archivePrefix = "arXiv",
    primaryClass = "hep-ex",
    month = "8",
    year = "2022"
}

@article{MEGII:2025muegamma,
    collaboration = "MEG II",
    title = "{New limit on the $\mu^+\to e^+\gamma$ decay with the MEG II experiment}",
    year = "2025",
    eprint = "2504.15711",
    archivePrefix = "arXiv",
    primaryClass = "hep-ex"
}

@article{Baldini:2018uhj,
    author = "Baldini, A. M. and others",
    collaboration = "MEG II",
    title = "{The design of the MEG II experiment}",
    eprint = "1801.04688",
    archivePrefix = "arXiv",
    primaryClass = "physics.ins-det",
    doi = "10.1140/epjc/s10052-018-5845-6",
    journal = "Eur. Phys. J. C",
    volume = "78",
    number = "5",
    pages = "380",
    year = "2018"
}

@article{Blondel:2021fji,
    author = "Blondel, Alain and others",
    collaboration = "Mu3e",
    title = "{Technical design of the phase I Mu3e experiment}",
    eprint = "2009.11690",
    archivePrefix = "arXiv",
    primaryClass = "physics.ins-det",
    doi = "10.1016/j.nima.2021.165679",
    journal = "Nucl. Instrum. Meth. A",
    volume = "1014",
    pages = "165679",
    year = "2021"
}

@article{Belle-II:2018jsg,
    author = "Kou, E. and others",
    collaboration = "Belle-II",
    title = "{The Belle II Physics Book}",
    eprint = "1808.10567",
    archivePrefix = "arXiv",
    primaryClass = "hep-ex",
    doi = "10.1093/ptep/ptz106",
    journal = "PTEP",
    volume = "2019",
    number = "12",
    pages = "123C01",
    year = "2019",
    note = "[Erratum: PTEP 2020, 029201 (2020)]"
}

@article{BaBar:2010taulgamma,
    collaboration = "BaBar",
    title = "{Searches for Lepton Flavor Violating Decays $\tau^\pm\to e^\pm\gamma$ and $\tau^\pm\to \mu^\pm\gamma$}",
    journal = "Phys. Rev. Lett.",
    volume = "104",
    pages = "021802",
    year = "2010",
    eprint = "0908.2381",
    archivePrefix = "arXiv",
    primaryClass = "hep-ex"
}

@article{Belle:2021taugamma,
    collaboration = "Belle",
    title = "{Search for lepton-flavor-violating $\tau$ decays into a lepton and a photon at Belle}",
    journal = "JHEP",
    volume = "10",
    pages = "019",
    year = "2021",
    eprint = "2103.12994",
    archivePrefix = "arXiv",
    primaryClass = "hep-ex"
}

@article{SINDRUM:1988mu3e,
    collaboration = "SINDRUM",
    title = "{Search for the Decay $\mu^+\to e^+e^+e^-$}",
    journal = "Nucl. Phys. B",
    volume = "299",
    pages = "1",
    year = "1988"
}

@article{Belle:2010tau3l,
    author = "Hayasaka, K. and others",
    collaboration = "Belle",
    title = "{Search for Lepton Flavor Violating Tau Decays into Three Leptons with 719 Million Produced Tau+Tau- Pairs}",
    journal = "Phys. Lett. B",
    volume = "687",
    pages = "139-143",
    year = "2010",
    eprint = "1001.3221",
    archivePrefix = "arXiv",
    primaryClass = "hep-ex"
}

@article{Aoki:2025clfv,
    author = "Aoki, M. and Baldini, A. M. and Bernstein, R. H. and Carloganu, C. and Mihara, S. and Miscetti, S. and Mori, T. and Ootani, W. and Renga, F. and Ritt, S. and Sch{\"o}ning, A.",
    title = "{Charged Lepton Flavour Violations searches with muons: present and future}",
    year = "2025",
    eprint = "2503.22461",
    archivePrefix = "arXiv",
    primaryClass = "hep-ex"
}

@article{Toma:2013zsa,
    author = "Toma, Takashi and Vicente, Avelino",
    title = "{Lepton Flavor Violation in the Scotogenic Model}",
    eprint = "1312.2840",
    archivePrefix = "arXiv",
    primaryClass = "hep-ph",
    reportNumber = "DCPT-13-198",
    doi = "10.1007/JHEP01(2014)160",
    journal = "JHEP",
    volume = "01",
    pages = "160",
    year = "2014"
}

@article{Kuno:1999jp,
    author = "Kuno, Yoshitaka and Okada, Yasuhiro",
    title = "{Muon decay and physics beyond the standard model}",
    eprint = "hep-ph/9909265",
    archivePrefix = "arXiv",
    doi = "10.1103/RevModPhys.73.151",
    journal = "Rev. Mod. Phys.",
    volume = "73",
    pages = "151--202",
    year = "2001"
}

@article{Abada:2014kba,
    author = "Abada, A. and Krauss, M. E. and Porod, W. and Staub, F. and Vicente, A. and Weiland, C.",
    title = "{Lepton flavor violation in low-scale seesaw models: SUSY and non-SUSY contributions}",
    eprint = "1408.0138",
    archivePrefix = "arXiv",
    primaryClass = "hep-ph",
    doi = "10.1007/JHEP11(2014)048",
    journal = "JHEP",
    volume = "11",
    pages = "048",
    year = "2014"
}

@article{Lindner:2016bgg,
    author = "Lindner, Manfred and Platscher, Moritz and Queiroz, Farinaldo S.",
    title = "{A Call for New Physics: The Muon Anomalous Magnetic Moment and Lepton Flavor Violation}",
    eprint = "1610.06587",
    archivePrefix = "arXiv",
    primaryClass = "hep-ph",
    doi = "10.1016/j.physrep.2017.12.001",
    journal = "Phys. Rept.",
    volume = "731",
    pages = "1--82",
    year = "2018"
}

@article{Kitano:2002mt,
    author = "Kitano, Ryuichiro and Koike, Masafumi and Okada, Yasuhiro",
    title = "{Detailed calculation of lepton flavor violating muon electron conversion rate for various nuclei}",
    eprint = "hep-ph/0203110",
    archivePrefix = "arXiv",
    reportNumber = "KEK-TH-808",
    doi = "10.1103/PhysRevD.66.096002",
    journal = "Phys. Rev. D",
    volume = "66",
    pages = "096002",
    year = "2002",
    note = "[Erratum: Phys. Rev. D 76, 059902 (2007)]"
}

@article{Akeroyd:2000wc,
    author = "Akeroyd, Andrew G. and Arhrib, Abdesslam and Naimi, El-Mokhtar",
    title = "{Note on tree level unitarity in the general two Higgs doublet model}",
    eprint = "hep-ph/0006035",
    archivePrefix = "arXiv",
    reportNumber = "UFR-HEP-00-06, KEK-TH-00-699, KEK-TH-699",
    doi = "10.1016/S0370-2693(00)00962-X",
    journal = "Phys. Lett. B",
    volume = "490",
    pages = "119--124",
    year = "2000"
}

@article{Lee:1977yc,
    author = "Lee, Benjamin W. and Quigg, C. and Thacker, H. B.",
    title = "{The Strength of Weak Interactions at Very High-Energies and the Higgs Boson Mass}",
    reportNumber = "FERMILAB-PUB-77-022-T",
    doi = "10.1103/PhysRevLett.38.883",
    journal = "Phys. Rev. Lett.",
    volume = "38",
    pages = "883--885",
    year = "1977"
}

@inproceedings{Arhrib:2000is,
    author = "Arhrib, Abdesslam",
    title = "{Unitarity constraints on scalar parameters of the standard and two Higgs doublets model}",
    booktitle = "{Workshop on Noncommutative Geometry, Superstrings and Particle Physics}",
    eprint = "hep-ph/0012353",
    archivePrefix = "arXiv",
    month = "12",
    year = "2000"
}

@article{Arhrib:2012ia,
    author = "Arhrib, Abdesslam and Benbrik, Rachid and Gaur, Naveen",
    title = "{$H\to \gamma \gamma$ in Inert Higgs Doublet Model}",
    eprint = "1201.2644",
    archivePrefix = "arXiv",
    primaryClass = "hep-ph",
    doi = "10.1103/PhysRevD.85.095021",
    journal = "Phys. Rev. D",
    volume = "85",
    pages = "095021",
    year = "2012"
}

@article{Arhrib:2013ela,
    author = "Arhrib, Abdesslam and Tsai, Yue-Lin Sming and Yuan, Qiang and Yuan, Tzu-Chiang",
    title = "{An Updated Analysis of Inert Higgs Doublet Model in light of the Recent Results from LUX, PLANCK, AMS-02 and LHC}",
    eprint = "1310.0358",
    archivePrefix = "arXiv",
    primaryClass = "hep-ph",
    doi = "10.1088/1475-7516/2014/06/030",
    journal = "JCAP",
    volume = "06",
    pages = "030",
    year = "2014"
}

@article{Belyaev:2016lok,
    author = "Belyaev, Alexander and Cacciapaglia, Giacomo and Ivanov, Igor P. and Rojas-Abatte, Felipe and Thomas, Marc",
    title = "{Anatomy of the Inert Two Higgs Doublet Model in the light of the LHC and non-LHC Dark Matter Searches}",
    eprint = "1612.00511",
    archivePrefix = "arXiv",
    primaryClass = "hep-ph",
    doi = "10.1103/PhysRevD.97.035011",
    journal = "Phys. Rev. D",
    volume = "97",
    number = "3",
    pages = "035011",
    year = "2018"
}

@book{Gunion:1989we,
    author = "Gunion, John F. and Haber, Howard E. and Kane, Gordon L. and Dawson, Sally",
    title = "{The Higgs Hunter's Guide}",
    reportNumber = "SCIPP-89/13, UCD-89-4, BNL-41644",
    doi = "10.1201/9780429496448",
    isbn = "978-0-429-49644-8",
    volume = "80",
    year = "2000"
}

@article{Djouadi:2005gi,
    author = "Djouadi, Abdelhak",
    title = "{The Anatomy of electro-weak symmetry breaking. I: The Higgs boson in the standard model}",
    eprint = "hep-ph/0503172",
    archivePrefix = "arXiv",
    reportNumber = "LPT-ORSAY-05-17",
    doi = "10.1016/j.physrep.2007.10.004",
    journal = "Phys. Rept.",
    volume = "457",
    pages = "1--216",
    year = "2008"
}

@article{Chen:2013vi,
    author = "Chen, Chian-Shu and Geng, Chao-Qiang and Huang, Da and Tsai, Lu-Hsing",
    title = "{New Scalar Contributions to $h\to Z\gamma$}",
    eprint = "1301.4694",
    archivePrefix = "arXiv",
    primaryClass = "hep-ph",
    doi = "10.1103/PhysRevD.87.075019",
    journal = "Phys. Rev. D",
    volume = "87",
    pages = "075019",
    year = "2013"
}

@article{ATLAS:2022vkf,
    author = "Aad, Georges and others",
    collaboration = "ATLAS",
    title = "{A detailed map of Higgs boson interactions by the ATLAS experiment ten years after the discovery}",
    eprint = "2207.00092",
    archivePrefix = "arXiv",
    primaryClass = "hep-ex",
    reportNumber = "CERN-EP-2022-057",
    doi = "10.1038/s41586-022-04893-w",
    journal = "Nature",
    volume = "607",
    number = "7917",
    pages = "52--59",
    year = "2022",
    note = "[Erratum: Nature 612, E24 (2022)]"
}

@article{Peskin:1990zt,
    author = "Peskin, Michael E. and Takeuchi, Tatsu",
    title = "{A New constraint on a strongly interacting Higgs sector}",
    reportNumber = "SLAC-PUB-5272",
    doi = "10.1103/PhysRevLett.65.964",
    journal = "Phys. Rev. Lett.",
    volume = "65",
    pages = "964--967",
    year = "1990"
}

@article{Peskin:1991sw,
    author = "Peskin, Michael E. and Takeuchi, Tatsu",
    title = "{Estimation of oblique electroweak corrections}",
    reportNumber = "SLAC-PUB-5618",
    doi = "10.1103/PhysRevD.46.381",
    journal = "Phys. Rev. D",
    volume = "46",
    pages = "381--409",
    year = "1992"
}

@article{Grimus:2008nb,
    author = "Grimus, W. and Lavoura, L. and Ogreid, O. M. and Osland, P.",
    title = "{The Oblique parameters in multi-Higgs-doublet models}",
    eprint = "0802.4353",
    archivePrefix = "arXiv",
    primaryClass = "hep-ph",
    reportNumber = "UWTHPH-2008-4",
    doi = "10.1016/j.nuclphysb.2008.04.019",
    journal = "Nucl. Phys. B",
    volume = "801",
    pages = "81--96",
    year = "2008"
}

@article{Gondolo:1990dk,
    author = "Gondolo, Paolo and Gelmini, Graciela",
    title = "{Cosmic abundances of stable particles: Improved analysis}",
    reportNumber = "UCLA-90-TEP-68",
    doi = "10.1016/0550-3213(91)90438-4",
    journal = "Nucl. Phys. B",
    volume = "360",
    pages = "145--179",
    year = "1991"
}

@article{Griest:1990kh,
    author = "Griest, Kim and Seckel, David",
    title = "{Three exceptions in the calculation of relic abundances}",
    reportNumber = "CFPA-TH-90-001A, BA-90-79",
    doi = "10.1103/PhysRevD.43.3191",
    journal = "Phys. Rev. D",
    volume = "43",
    pages = "3191--3203",
    year = "1991"
}

@article{Planck:2018vyg,
    author = "Aghanim, N. and others",
    collaboration = "Planck",
    title = "{Planck 2018 results. VI. Cosmological parameters}",
    eprint = "1807.06209",
    archivePrefix = "arXiv",
    primaryClass = "astro-ph.CO",
    doi = "10.1051/0004-6361/201833910",
    journal = "Astron. Astrophys.",
    volume = "641",
    pages = "A6",
    year = "2020",
    note = "[Erratum: Astron.Astrophys. 652, C4 (2021)]"
}

@article{Vicente:2014wga,
    author = "Vicente, Avelino and Yaguna, Carlos E.",
    title = "{Probing the scotogenic model with lepton flavor violating processes}",
    eprint = "1412.2545",
    archivePrefix = "arXiv",
    primaryClass = "hep-ph",
    reportNumber = "MS-TP-14-37",
    doi = "10.1007/JHEP02(2015)144",
    journal = "JHEP",
    volume = "02",
    pages = "144",
    year = "2015"
}

@article{Hagedorn:2018spx,
    author = "Hagedorn, Claudia and Herrero-Garc{\'i}a, Juan and Molinaro, Emiliano and Schmidt, Michael A.",
    title = "{Phenomenology of the Generalised Scotogenic Model with Fermionic Dark Matter}",
    eprint = "1804.04117",
    archivePrefix = "arXiv",
    primaryClass = "hep-ph",
    reportNumber = "ADP-18-9-T1057, CP3-ORIGINS-2018-012, ADP-18-9/T1057, CP3-Origins-2018-012-DNRF90",
    doi = "10.1007/JHEP11(2018)103",
    journal = "JHEP",
    volume = "11",
    pages = "103",
    year = "2018"
}

@article{Karan:2023adm,
    author = "Karan, Anirban and Sadhukhan, Soumya and Valle, Jos{\'e} W. F.",
    title = "{Phenomenological profile of scotogenic fermionic dark matter}",
    eprint = "2308.09135",
    archivePrefix = "arXiv",
    primaryClass = "hep-ph",
    doi = "10.1007/JHEP12(2023)185",
    journal = "JHEP",
    volume = "12",
    pages = "185",
    year = "2023"
}

@article{Esteban:2024eli,
    author = "Esteban, Ivan and Gonzalez-Garcia, M. C. and Maltoni, Michele and Martinez-Soler, Ivan and Pinheiro, Jo{\~a}o Paulo and Schwetz, Thomas",
    title = "{NuFit-6.0: updated global analysis of three-flavor neutrino oscillations}",
    eprint = "2410.05380",
    archivePrefix = "arXiv",
    primaryClass = "hep-ph",
    reportNumber = "IFT-UAM/CSIC-24-140, YITP-SB-2024-24, IPPP/24/64, IPPP/24/64, IFT-UAM/CSIC-24-140, YITP-SB-2024-24",
    doi = "10.1007/JHEP12(2024)216",
    journal = "JHEP",
    volume = "12",
    pages = "216",
    year = "2024"
}

@article{LZ:2024zvo,
    author = "Aalbers, J. and others",
    collaboration = "LZ",
    title = "{Dark Matter Search Results from 4.2{\,}{\,}Tonne-Years of Exposure of the LUX-ZEPLIN (LZ) Experiment}",
    eprint = "2410.17036",
    archivePrefix = "arXiv",
    primaryClass = "hep-ex",
    reportNumber = "FERMILAB-PUB-24-0796-V",
    doi = "10.1103/4dyc-z8zf",
    journal = "Phys. Rev. Lett.",
    volume = "135",
    number = "1",
    pages = "011802",
    year = "2025"
}

@article{XENON:2023cxc,
    author = "Aprile, E. and others",
    collaboration = "XENON",
    title = "{First Dark Matter Search with Nuclear Recoils from the XENONnT Experiment}",
    eprint = "2303.14729",
    archivePrefix = "arXiv",
    primaryClass = "hep-ex",
    doi = "10.1103/PhysRevLett.131.041003",
    journal = "Phys. Rev. Lett.",
    volume = "131",
    number = "4",
    pages = "041003",
    year = "2023"
}

@article{JUNO:2025gmd,
    author = "Abusleme, Angel and others",
    collaboration = "JUNO",
    title = "{First measurement of reactor neutrino oscillations at JUNO}",
    eprint = "2511.14593",
    archivePrefix = "arXiv",
    primaryClass = "hep-ex",
    month = "11",
    year = "2025"
}

@article{KamLAND-Zen:2016pfg,
    author = "Gando, A. and others",
    collaboration = "KamLAND-Zen",
    title = "{Search for Majorana Neutrinos near the Inverted Mass Hierarchy Region with KamLAND-Zen}",
    journal = "Phys. Rev. Lett.",
    volume = "117",
    pages = "082503",
    year = "2016",
    eprint = "1605.02889",
    archivePrefix = "arXiv",
    primaryClass = "hep-ex",
    doi = "10.1103/PhysRevLett.117.082503"
}

@article{GERDA:2020xhi,
    author = "Agostini, M. and others",
    collaboration = "GERDA",
    title = "{Final Results of GERDA on the Search for Neutrinoless Double-$\beta$ Decay}",
    journal = "Phys. Rev. Lett.",
    volume = "125",
    pages = "252502",
    year = "2020",
    eprint = "2009.06079",
    archivePrefix = "arXiv",
    primaryClass = "nucl-ex",
    doi = "10.1103/PhysRevLett.125.252502"
}

@article{CUORE:2022fgx,
    author = "Adams, D. Q. and others",
    collaboration = "CUORE",
    title = "{Search for Majorana neutrinos exploiting millikelvin cryogenics with CUORE}",
    journal = "Nature",
    volume = "604",
    pages = "53--58",
    year = "2022",
    eprint = "2104.06906",
    archivePrefix = "arXiv",
    primaryClass = "nucl-ex",
    doi = "10.1038/s41586-022-04497-4"
}

@article{LEGEND:2021bnm,
    collaboration = "LEGEND",
    title = "{LEGEND-1000 Preconceptual Design Report}",
    year = "2021",
    eprint = "2107.11462",
    archivePrefix = "arXiv",
    primaryClass = "physics.ins-det"
}

@article{nEXO:2021ujk,
    collaboration = "nEXO",
    title = "{nEXO: neutrinoless double beta decay search beyond $10^{28}$ year half-life sensitivity}",
    journal = "Phys. Rev. C",
    volume = "105",
    pages = "065503",
    year = "2022",
    eprint = "2106.16243",
    archivePrefix = "arXiv",
    primaryClass = "physics.ins-det",
    doi = "10.1103/PhysRevC.105.065503"
}

@article{CUPID:2022olt,
    author = "Armstrong, W. and others",
    collaboration = "CUPID",
    title = "{CUPID: The Next-Generation Neutrinoless Double Beta Decay Experiment}",
    journal = "J. Low Temp. Phys.",
    volume = "209",
    pages = "1079--1086",
    year = "2022",
    eprint = "1907.09376",
    archivePrefix = "arXiv",
    primaryClass = "physics.ins-det",
    doi = "10.1007/s10909-022-02909-3"
}

@article{KATRIN:2024mass,
    collaboration = "KATRIN",
    title = "{Direct neutrino-mass measurement based on 259 days of KATRIN data}",
    year = "2024",
    eprint = "2406.13516",
    archivePrefix = "arXiv",
    primaryClass = "hep-ex"
}

@article{Project8:2022bla,
    collaboration = "Project 8",
    title = "{The Project 8 Neutrino Mass Experiment}",
    year = "2022",
    eprint = "2203.07349",
    archivePrefix = "arXiv",
    primaryClass = "physics.ins-det"
}

@article{OHare:2021utq,
    author = "O'Hare, Ciaran A. J.",
    title = "{New Definition of the Neutrino Floor for Direct Dark Matter Searches}",
    eprint = "2109.03116",
    archivePrefix = "arXiv",
    primaryClass = "hep-ph",
    doi = "10.1103/PhysRevLett.127.251802",
    journal = "Phys. Rev. Lett.",
    volume = "127",
    number = "25",
    pages = "251802",
    year = "2021"
}

@article{Mu2e:2014fns,
    author = "Bartoszek, L. and others",
    collaboration = "Mu2e",
    title = "{Mu2e Technical Design Report}",
    eprint = "1501.05241",
    archivePrefix = "arXiv",
    primaryClass = "physics.ins-det",
    reportNumber = "FERMILAB-TM-2594, FERMILAB-DESIGN-2014-01",
    doi = "10.2172/1172555",
    month = "10",
    year = "2014"
}

@article{Kurup:2011zza,
    author = "Kurup, A.",
    collaboration = "COMET",
    title = "{The COMET Experiment}",
    journal = "Nucl. Phys. B Proc. Suppl.",
    volume = "218",
    pages = "38--43",
    year = "2011",
    note = "\url{http://j-parc.jp/researcher/Hadron/en/pac_1203/pdf/COMET-PhaseI-LoI.pdf}"
}

@article{DARWIN:2016hyl,
    author = "Aalbers, J. and others",
    collaboration = "DARWIN",
    title = "{DARWIN: towards the ultimate dark matter detector}",
    eprint = "1606.07001",
    archivePrefix = "arXiv",
    primaryClass = "astro-ph.IM",
    doi = "10.1088/1475-7516/2016/11/017",
    journal = "JCAP",
    volume = "11",
    pages = "017",
    year = "2016"
}

@article{XLZD:2024nsu,
    author = "Aalbers, J. and others",
    collaboration = "XLZD",
    title = "{The XLZD Design Book: towards the next-generation liquid xenon observatory for dark matter and neutrino physics}",
    eprint = "2410.17137",
    archivePrefix = "arXiv",
    primaryClass = "hep-ex",
    doi = "10.1140/epjc/s10052-025-14810-w",
    journal = "Eur. Phys. J. C",
    volume = "85",
    number = "10",
    pages = "1192",
    year = "2025"
}

@article{Jueid:2023zxx,
    author = "Jueid, Adil and Nasri, Salah",
    title = "{Lepton portal dark matter at muon colliders: Total rates and generic features for phenomenologically viable scenarios}",
    eprint = "2301.12524",
    archivePrefix = "arXiv",
    primaryClass = "hep-ph",
    reportNumber = "CTPU-PTC-2023-02",
    doi = "10.1103/PhysRevD.107.115027",
    journal = "Phys. Rev. D",
    volume = "107",
    number = "11",
    pages = "115027",
    year = "2023"
}

@article{Jueid:2023qcf,
    author = "Jueid, Adil and Chowdhury, Talal Ahmed and Nasri, Salah and Saad, Shaikh",
    title = "{Probing Zee-Babu states at muon colliders}",
    eprint = "2306.01255",
    archivePrefix = "arXiv",
    primaryClass = "hep-ph",
    reportNumber = "CTPU-PTC-23-23",
    doi = "10.1103/PhysRevD.109.075011",
    journal = "Phys. Rev. D",
    volume = "109",
    number = "7",
    pages = "075011",
    year = "2024"
}

@article{Belfkir:2023vpo,
    author = "Belfkir, Mohamed and Jueid, Adil and Nasri, Salah",
    title = "{Boosting dark matter searches at muon colliders with machine learning: The mono-Higgs channel as a case study}",
    eprint = "2309.11241",
    archivePrefix = "arXiv",
    primaryClass = "hep-ph",
    reportNumber = "CTPU-PTC-23-37",
    doi = "10.1093/ptep/ptad144",
    journal = "PTEP",
    volume = "2023",
    number = "12",
    pages = "123B03",
    year = "2023"
}

\end{document}